\pdfoutput=1
\RequirePackage{fix-cm}
\documentclass[prl,aps,reprint,twocolumn,amsmath,superscriptaddress,notitlepage,longbibliography,floatfix]{revtex4-2}

\usepackage[T1,OT1]{fontenc}
\DeclareTextSymbolDefault{\DJ}{T1}
\DeclareTextSymbolDefault{\dj}{T1}

\usepackage{amssymb,mathtools,amsmath}
\usepackage{physics}
\usepackage{bbold}
\usepackage{dsfont}
\usepackage{graphicx}
\usepackage[dvipsnames]{xcolor}
\usepackage{hyperref}
\hypersetup{
	colorlinks,
	citecolor=NavyBlue,
	linkcolor=purple,
	urlcolor=NavyBlue}

\renewcommand{\figurename}{\textbf{Fig.}}

\newcommand{\FigTwoCaptionSize}{8.5}
\newcommand{\FigTwoCaptionSkip}{10}
\newcommand{\FigTwoCaptionFont}{\fontsize{\FigTwoCaptionSize}{\FigTwoCaptionSkip}\selectfont}

\makeatletter
\newcommand{\SIhideTOC}{\let\SIsaved@sec\l@section\let\SIsaved@subsec\l@subsection\let\SIsaved@subsubsec\l@subsubsection\let\l@section\@gobbletwo\let\l@subsection\@gobbletwo\let\l@subsubsection\@gobbletwo}
\newcommand{\SIshowTOC}{\let\l@section\SIsaved@sec\let\l@subsection\SIsaved@subsec\let\l@subsubsection\SIsaved@subsubsec}
\makeatother

\newcommand{\SIpart}[1]{\par\bigskip\begin{center}\large\bfseries #1\end{center}\phantomsection\addcontentsline{toc}{part}{#1}\par\medskip}

\begin{document}

	\title{Ultrahigh-Q resonances from stabilized exceptional points}

	\author{Haydar~Sahin}
	\email{sahinhaydar@u.nus.edu}
	\affiliation{Department of Physics, National University of Singapore, Singapore 117551, Republic of Singapore}
	\affiliation{Department of Electrical and Computer Engineering, National University of Singapore, Singapore 117583, Republic of Singapore}
	\affiliation{Institute of High Performance Computing, Agency for Science, Technology and Research (A*STAR), Singapore 138632, Republic of Singapore}

	\author{Fatih~Ozaydin}
	\affiliation{Institute for International Strategy and Emerging Technologies, Tokyo International University, Tokyo 170-0013, Japan}
	\affiliation{Nanoelectronics Research Center, Istanbul, T\"urkiye}

	\author{Zhuo~Bin~Siu}
	\affiliation{Department of Electrical and Computer Engineering, National University of Singapore, Singapore 117583, Republic of Singapore}

	\author{Shuhan~Yang$^\diamond$}
	\affiliation{Department of Electrical and Computer Engineering, National University of Singapore, Singapore 117583, Republic of Singapore}

	\author{Md.~Saddam~Hossain~Razo$^\diamond$}
	\affiliation{Department of Electrical and Computer Engineering, National University of Singapore, Singapore 117583, Republic of Singapore}

	\author{S.~M.~\mbox{Rafi-Ul-Islam}}
	\affiliation{Department of Electrical and Computer Engineering, National University of Singapore, Singapore 117583, Republic of Singapore}

	\author{Ramy~\mbox{El-Ganainy}}
	\affiliation{Department of Electrical and Computer Engineering, Saint Louis University, St.~Louis, MO 63103, USA}

	\author{Hyunsoo~Yang}
	\affiliation{Department of Electrical and Computer Engineering, National University of Singapore, Singapore 117583, Republic of Singapore}

	\author{Jian~Feng~Kong}
	\affiliation{Institute of High Performance Computing, Agency for Science, Technology and Research (A*STAR), Singapore 138632, Republic of Singapore}

	\author{Ronny~Thomale}
	\affiliation{Institut f\"ur Theoretische Physik und Astrophysik, Universit\"at W\"urzburg, W\"urzburg, Germany}

	\author{\c{S}ahin~K.~\"Ozdemir}
	\affiliation{Department of Electrical and Computer Engineering, Saint Louis University, St.~Louis, MO 63103, USA}

	\author{Mansoor~B.~A.~Jalil}
	\email{elembaj@nus.edu.sg}
	\affiliation{Department of Electrical and Computer Engineering, National University of Singapore, Singapore 117583, Republic of Singapore}

	\author{Ching~Hua~Lee}
	\email{phylch@nus.edu.sg}
	\affiliation{Department of Physics, National University of Singapore, Singapore 117551, Republic of Singapore}

	\collaboration{\small$^\diamond$ These authors contributed equally to this work}

\begin{abstract}
	Resonances with high quality factors (Q) underpin precision clocks, lasers, filters and sensors~\cite{ludlow_optical_2015,vahala_optical_2003}. Because loss, disorder and noise broaden resonances, ultrahigh-Q resonances have mainly been pursued through tightly controlled ultrapure materials, cryogenic cooling or extreme shielding~\cite{megrant_planar_2012,romanenko_three-dimensional_2020,armani_ultra-high-q_2003,ghadimi_elastic_2018}. Here, we instead uncover a universal mechanism that increases Q by orders of magnitude despite unavoidable noise and disorder. This is accomplished through a unique symbiosis between exceptional points (EPs)~\cite{bender_real_1998,el-ganainy_non-hermitian_2018,ozdemir_paritytime_2019} and band topology, by coupling two topological lattices, one with gain and one with loss, in a parity--time (PT)-symmetric arrangement. Through its characteristic EP structure, the topological band gap suppresses linewidth broadening induced by dynamical gain fluctuations, diverting their effects into resonance-frequency shifts and thereby enabling an unprecedented degree of linewidth stability. Exploiting this protection, a room-temperature circuit of standard discrete components reaches $\mathrm{Q}=3.2\times10^5$, the highest Q reported, to our knowledge, for a macroscale discrete-element network, compared with $\mathrm{Q}=88$ for a passive resonator built from the same type of components. This makes ultrahigh-Q resonances a matter of network band structure design rather than material perfection or environmental control, with topology protecting the sharpness of a resonance rather than a boundary state. The mechanism applies to electrical, acoustic, mechanical and photonic networks alike and offers a route towards integrated ultrahigh-Q resonators operating at GHz and higher frequencies.
\end{abstract}

	\maketitle
	\addtocontents{toc}{\protect\SIhideTOC}

	High-quality resonance phenomena, in which a specific mode of a system is selectively enhanced in response to an external excitation, manifest across a wide range of physical media. Examples include Fano resonances in photonic and nanoscale structures~\cite{limonov_fano_2017,miroshnichenko_fano_2010}, electron paramagnetic resonances in atomic physics~\cite{baumann_electron_2015}, magnon resonances in magnetic materials~\cite{makihara_ultrastrong_2021,shiota_tunable_2020}, whispering-gallery mode resonances in optical microcavities~\cite{peng_paritytime-symmetric_2014,he_detecting_2011}, optical resonances in photonic crystals~\cite{lee_observation_2012} and impedance resonances in electrical circuits~\cite{imhof_topolectrical-circuit_2018,zhang_anomalous_2023,sahin_impedance_2023,sahin_topolectrical_2025,hohmann_observation_2023,stegmaier_topological_2025,li_emergence_2019,rafi-ul-islam_unconventional_2022,lenggenhager_simulating_2022,zhang_topolectrical_2025}. Across all such systems, the quality factor (Q) serves as the universal figure of merit, conventionally quantifying the efficiency of energy storage relative to dissipation. A higher Q gives a narrower resonance line, which makes small shifts of the resonance easier to resolve. The pursuit of ever-higher Q has therefore been a persistent and cross-cutting goal in modern physics, driving advances in optical, mechanical, magnetic and electronic systems alike, from precision clocks and lasers to filters and sensors~\cite{kumar_topological_2022,ye_micro_2026}.

	High-Q resonances require precise control of wave--medium interactions and minimal fabrication imperfections, so they are primarily associated with finely engineered devices~\cite{yang_bridging_2018}. For example, crystalline quartz resonators reach Q of $10^4$ to $10^5$ at room temperature and above $10^9$ at millikelvin temperatures~\cite{friedt_introduction_2007,goryachev_extremely_2012}, and superconducting resonators reach $10^6$ to beyond $10^{10}$ at cryogenic temperatures~\cite{megrant_planar_2012,romanenko_three-dimensional_2020}. Optical microcavities and nanomechanical resonators exceed $10^8$ through strong confinement, nanoscale fabrication and careful control of radiation and clamping losses~\cite{armani_ultra-high-q_2003,ghadimi_elastic_2018}. Bound states in the continuum in photonic crystals~\cite{hsu_observation_2013,hsu_bound_2016,jin_topologically_2019,kang_applications_2023,reva_bound_2026} and metasurfaces~\cite{koshelev_asymmetric_2018} bring high-Q resonances to open structures by exploiting destructive interference to suppress radiative losses. Each of these advances is remarkable, and every rise in Q has marked a milestone for its platform. In each case, however, high-Q operation comes from removing loss at its source, and it is tied to a specific material, temperature or fabrication process. Consequently, conventional efforts towards ultrahigh-Q resonances have largely focused on platform-specific designs and continual refinements in fabrication quality. In this work, we pursue a different question, namely whether a universal mechanism can enable ultrahigh-Q resonances across distinct physical platforms and remain intrinsically robust against disorder. To address this question, we combine the extreme parametric sensitivity of near-divergent resonant modes with a spectral stabilization mechanism that suppresses disorder-induced eigenvalue fluctuations.

	\begin{figure}[t!]
	\centering
	\includegraphics[width=\columnwidth]{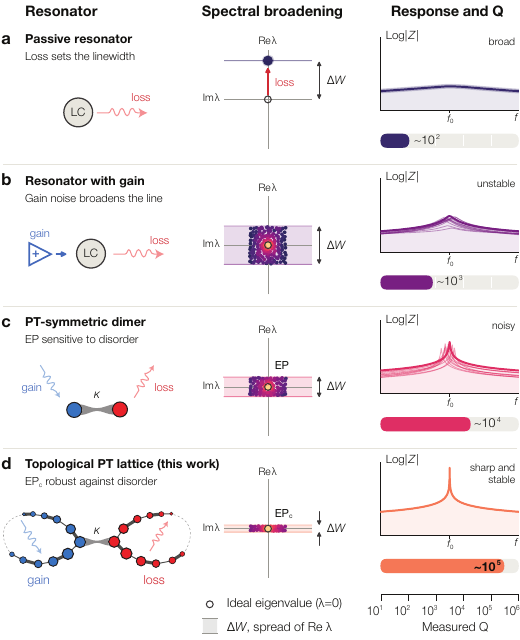}
	\caption{\textbf{Route to stable ultrahigh-Q resonances.} 
		While an eigenvalue tuned to the spectral origin gives rise to a sharp resonance in the clean limit (without disorder), under disorder the eigenvalue of a resonance spreads along $\mathrm{Re}\,\lambda$, and this spectral width $\Delta W$ limits its Q (Eq.~\eqref{EqWidthQ}). The four resonators are built from the same type of discrete components (left), and the sketches show the eigenvalue closest to the spectral origin over many disorder realizations (middle, dots) together with driven responses obtained in repeated realizations (right, thin lines). Here, a line is called unstable when it fluctuates in time and its height and width change from sweep to sweep, not when the circuit self-oscillates.
		\textbf{a,} In a passive LC resonator, loss holds the eigenvalue away from the origin, so the line is broad.
		\textbf{b,} Gain compensates the real-axis offset and brings the eigenvalue to the origin on average, but every gain fluctuation enters $\mathrm{Re}\,\lambda$ in full, so the line is unstable and broad.
		\textbf{c,} In a PT-symmetric dimer, an exceptional point (EP) steers the eigenvalue to the origin owing to the two-dimensional spectral control, but half of each gain fluctuation still enters $\mathrm{Re}\,\lambda$ (Eq.~\eqref{EqDimerShare}), so the line remains noisy.
		\textbf{d,} In the topological PT lattice, the band gap diverts the fluctuations near the central EP, EP$_\mathrm{c}$, into shifts of the tunable resonance frequency, so the line stays sharp and stable and Q reaches the order of $10^5$. This stabilization gives access to Q values at which any further increase is otherwise extremely challenging.}
	\label{figMain_Overview}
	\end{figure}

	A general alternative to removing loss is to compensate it with gain, as in regenerative receivers~\cite{shang_high-sensitivity_2014}, Q-enhanced filters~\cite{kuhn_200_1996} and lasers operated below threshold~\cite{monifi_tunable_2013,kullig_exceptional_2025,alizadeh_lasing_2025}. Gain can compensate the loss on average and, in principle, increase Q without limit, independently of the individual quality of the components (Fig.~\ref{figMain_Overview}a). In practice, however, the active elements used to compensate loss inevitably amplify noise, introduce frequency drift and produce instabilities. We collectively refer to these dynamical effects, together with environmental and thermal perturbations, as \textit{disorder}. In a single LC resonator with gain, every gain fluctuation enters the linewidth in full (Fig.~\ref{figMain_Overview}b), and this \textit{spectral broadening} limits practical gain-compensated (Q-enhanced) LC filters to Q of $10^2$ to $10^3$~\cite{kuhn_200_1996}. Since such fluctuations are largely unavoidable in real devices at room temperature, we follow a different strategy. Instead of removing them, we divert them into the resonance frequency, where they do not broaden the line. Consequently, achieving ultrahigh-Q resonances with gain requires spectral control that (i) steers the resonance towards zero net loss and (ii) keeps disorder out of its linewidth.

	A prominent route towards the first requirement is the engineering of eigenvalue trajectories in parity--time (PT)-symmetric systems~\cite{el-ganainy_non-hermitian_2018,ozdemir_paritytime_2019,miri_exceptional_2019}, which can support a real eigenfrequency spectrum despite the presence of gain and loss~\cite{bender_real_1998,meng_exceptional_2024}. This principle has been realized in optical systems~\cite{guo_observation_2009,ruter_observation_2010,regensburger_paritytime_2012}, electronic circuits~\cite{schindler_experimental_2011,assawaworrarit_robust_2017,sahin_topolectrical_2025,fernandez-alcazar_paritytime_2026,chen_generalized_2018,dong_sensitive_2019,zhang_observation_2024}, microcavities~\cite{peng_paritytime-symmetric_2014} and lasers~\cite{feng_single-mode_2014,hodaei_parity-timesymmetric_2014}. In the elementary PT dimer of two resonators with balanced gain and loss, increasing the coupling $\kappa$ makes their eigenvalues and eigenvectors coalesce at an exceptional point (EP)~\cite{heiss_physics_2012,chen_exceptional_2017,hodaei_enhanced_2017,meng_exceptional_2024,zhao_exceptional_2024,lai_observation_2019,hokmabadi_non-hermitian_2019,bai_recent_2026,yang_beyond_2026}, which can bring a mode to zero net loss. However, real-world EPs are prone to avoided crossings~\cite{haus_coupled-mode_1991,el-ganainy_exceptional_2014,peng_what_2014}, and their incomplete eigenbases amplify perturbations~\cite{wiersig_review_2020,bai_nonlinear_2023,bai_observation_2024} and noise~\cite{wang_petermann-factor_2020,lau_fundamental_2018,langbein_no_2018,zhang_quantum_2019,wiersig_prospects_2020,suntharalingam_noise_2023,duggan_limitations_2022,kononchuk_exceptional-point-based_2022}. Moreover, as we show below, even on the lossless branch of an ideal PT dimer, half of each gain fluctuation still enters the linewidth (Fig.~\ref{figMain_Overview}c). Proximity to an EP in PT systems therefore does not by itself ensure a narrow, stable resonance.

	For the second requirement, we here introduce a mechanism that suppresses disorder-induced spectral broadening near zero net loss and thereby enables orders-of-magnitude improvements in resonance quality. We couple two topological Su--Schrieffer--Heeger (SSH) lattices~\cite{hasan_colloquium_2010,ozawa_topological_2019,wang_non-hermitian_2022}, one with gain and one with loss, through a single coupling. Gain and loss in topological lattices are a central theme of non-Hermitian physics~\cite{bergholtz_exceptional_2021,ding_non-hermitian_2022,okuma_non-hermitian_2023,zhu_hybrid_2022,zhao_feedback_2025}, with experiments in photonic, electrical and acoustic systems~\cite{weimann_topologically_2017,zhao_non-hermitian_2019,weidemann_topological_2020,hofmann_reciprocal_2020}, and non-Hermitian topology has been used to enhance the performance of sensors~\cite{budich_non-hermitian_2020,yuan_nonhermitian_2023,chen_ultra-sensitivity_2024,deng_ultrasensitive_2024,parto_enhanced_2025}. The topological band gap supports interface modes whose coupling creates three EPs in the regime studied here. The central exceptional point, EP$_\mathrm{c}$, lies at the spectral origin in the middle of the band gap. Near this EP, the strong response to local gain fluctuations is directed mainly into resonance-frequency shifts, while the portion that enters the linewidth drops to about a quarter and, at an optimal gain--loss strength, vanishes to first order (Fig.~\ref{figMain_Overview}d). Neither the conventional PT dimer nor a gapless PT lattice has this property. In a circuit built from standard discrete components, we measure $\mathrm{Q}=3.2\times10^5$, compared with $\mathrm{Q}=1.8\times10^4$ for a PT dimer and $\mathrm{Q}=88$ for a passive resonator built from the same type of components. To our knowledge, this is the highest Q reported for a resonator assembled from discrete lumped elements at room temperature. Because this Q is a property of the spectrum of a network of lossy components rather than of a material, the mechanism offers a route to ultrahigh Q in discrete-element networks and metamaterials, and can, in principle, extend to photonic, acoustic and mechanical resonators with gain, loss, a band gap and a tunable coupling.

	\section{Relating the Q factor to the spectral width}
	\label{SecBroadening}

	To find what a design must control, we first relate the Q factor to the spectrum of a resonator. The Q factor of a driven resonator is set by how close its resonant eigenvalue comes to the origin of the complex plane, and in particular by the real part of this eigenvalue. We describe the response through the complex eigenvalues $\lambda_n$ of the operator that governs it, the admittance matrix (circuit Laplacian) $J(\omega)$ of an electrical circuit or the effective Hamiltonian of an optical or mechanical resonator. The impedance between two nodes $a$ and $b$ of a circuit is a sum over the eigenmodes~\cite{lee_topolectrical_2018,imhof_topolectrical-circuit_2018,sahin_topolectrical_2025,sahin_impedance_2023,jiang_engineering_2026},
	\begin{equation}
		Z_{ab} = \sum_{n} \frac{\left(\psi^{R}_{n,a} - \psi^{R}_{n,b}\right)\left(\psi^{L}_{n,a} - \psi^{L}_{n,b}\right)}{\lambda_n},
		\label{EqImpEig}
	\end{equation}
	where $\psi^{R}_n$ and $\psi^{L}_n$ are the right and left eigenvectors of $J$, normalized such that $\langle\psi^{L}_m|\psi^{R}_n\rangle=\delta_{mn}$ ($\psi^{L}_n$ is the transpose of $\psi^{R}_n$ for the reciprocal circuits studied here). In this admittance picture, $\mathrm{Im}\,\lambda$ measures the detuning between the drive and the mode, and $\mathrm{Re}\,\lambda$ measures its net loss (positive) or net gain (negative), the reverse of the familiar Hamiltonian convention. The response diverges when an eigenvalue reaches the spectral origin. Tuning the drive frequency moves all eigenvalues along the imaginary axis and removes the detuning, so the attainable Q is set by $\mathrm{Re}\,\lambda$.

	We use the spectral Q of the driven response,
	\begin{equation}
		\mathrm{Q} = \frac{f_0}{\Delta f},
		\label{EqQfactor}
	\end{equation}
	where $f_0$ is the peak frequency and $\Delta f$ is the full width at half power~\cite{gregory_q-factor_2022}. Near zero net loss, disorder changes $\mathrm{Re}\,\lambda$, and with it $\Delta f$, by a large fraction, whereas it shifts $f_0$ by only a tiny fraction of its value, even when this shift exceeds the linewidth. The Q of a resonance therefore changes through its linewidth rather than its frequency, and a higher Q means a narrower line. Near the peak, the eigenvalue closest to the origin dominates Eq.~\eqref{EqImpEig} and gives a Lorentzian line. In a circuit, the onsite admittance $i\omega c_\Sigma+1/(i\omega L)$ of each node, where $c_\Sigma$ is the total capacitance of a node, sets the frequency dependence of this eigenvalue, which gives (Methods)
	\begin{equation}
		\mathrm{Q}\simeq\frac{\omega_0 c_\Sigma}{|\mathrm{Re}\,\lambda|},
		\label{EqQLinewidth}
	\end{equation}
	with $\omega_0=2\pi f_0$. For a resonator with the complex frequency $\omega_0-i\gamma$, the same steps give the familiar $\mathrm{Q}=\omega_0/2\gamma$, so the link between Q and $\mathrm{Re}\,\lambda$ holds for any platform described by Eq.~\eqref{EqImpEig}.

	Disorder turns $\mathrm{Re}\,\lambda$ from a single value into a distribution. In many realistic systems, disorder is not uniform in space. The couplings and the active elements that steer a resonance are the most sensitive parts of a device, and the near-divergent modes that carry the resonance concentrate around the couplings, so the dominant disorder is expected at the coupling nodes (Supplementary Section~\ref{SuppMatDisorderProfile}). Active elements, in turn, depend dynamically on their operating temperature, supply, and environmental conditions, which makes them the main source of time-dependent disorder~\cite{li_stochastic_2023,li_synergetic_2023,sun_optimal_2020}. Static deviations of the components only shift the eigenvalue, and we remove this shift by tuning the drive frequency, the coupling and the drive amplitude. Fluctuating disorder, by contrast, scatters the eigenvalue into a cloud around the origin. We call the spread of $\mathrm{Re}\,\lambda$ in this cloud the spectral width $\Delta W$. When the cloud is centred on the origin, the largest net loss in it is $\Delta W/2$, and Eq.~\eqref{EqQLinewidth} gives
	\begin{equation}
		\mathrm{Q}\geq\frac{2\,\omega_0c_\Sigma}{\Delta W}.
		\label{EqWidthQ}
	\end{equation}
	A narrower $\Delta W$ therefore increases the Q that a resonance keeps under disorder by the same proportion. The spread along the imaginary axis does not change the linewidth, but it makes the resonance frequency jitter.

	Figure~\ref{figMain_Overview} illustrates this criterion for four resonators built from the same type of components. The middle column shows the distribution of the eigenvalue closest to the origin over many disorder realizations, where each dot is one sample, and the right column shows the corresponding driven response. The shape of each distribution is set by the resonator itself. In a passive LC resonator, loss keeps the eigenvalue away from the origin, $\Delta W$ is the loss itself, and the line is broad (Q $\sim10^2$, Fig.~\ref{figMain_Overview}a). Gain moves the eigenvalue to the origin on average. Its fluctuations and those of the reactive elements, however, act independently, so the samples fill a rectangle whose width along $\mathrm{Re}\,\lambda$ is the full gain fluctuation, and the line fluctuates strongly from one measurement to the next (Q $\sim10^3$, Fig.~\ref{figMain_Overview}b). In a PT dimer, the coupling $\kappa$ steers the eigenvalue to the origin through an EP, but the samples still spread widely along $\mathrm{Re}\,\lambda$ near the origin, and the line is noisy (Q $\sim10^4$, Fig.~\ref{figMain_Overview}c). In the coupled topological lattices, the samples near the central EP stay close to the imaginary axis, $\Delta W$ is small, and the line is sharp and stable (Q $\sim10^5$, Fig.~\ref{figMain_Overview}d). These orders of magnitude agree with those measured in our experiment, which we present later.

	What decides between these designs is how much of a gain fluctuation reaches $\mathrm{Re}\,\lambda$. A small change $\delta$ of the gain admittance of a coupling node shifts the resonant eigenvalue by $\delta\lambda$, and we write
	\begin{equation}
		\mathrm{Re}\,\delta\lambda=\eta\,\delta,\qquad \Delta W=|\eta|\,\Delta W_\mathrm{g},
		\label{EqGainShare}
	\end{equation}
	where $\Delta W_\mathrm{g}$ is the spread of the gain admittance itself. We call $\eta$ the broadening coefficient. It measures how much of a gain fluctuation broadens the line, and it follows from the weight of the resonant mode on the fluctuating node (Methods). The smaller $|\eta|$ is, the narrower the line stays, and for $|\eta|=1$ the full gain fluctuation reaches the linewidth. The imaginary part of $\delta\lambda$ only shifts the resonance frequency. An ultrahigh-Q design must therefore steer the eigenvalue to the origin and, at the same time, make $|\eta|$ small.

	\section{Circuit resonators and their clean-limit responses}
	\label{SecThreeEPs}

	\begin{figure*}[t!]
		\centering
		\includegraphics[width=\textwidth,height=0.80\textheight,keepaspectratio]{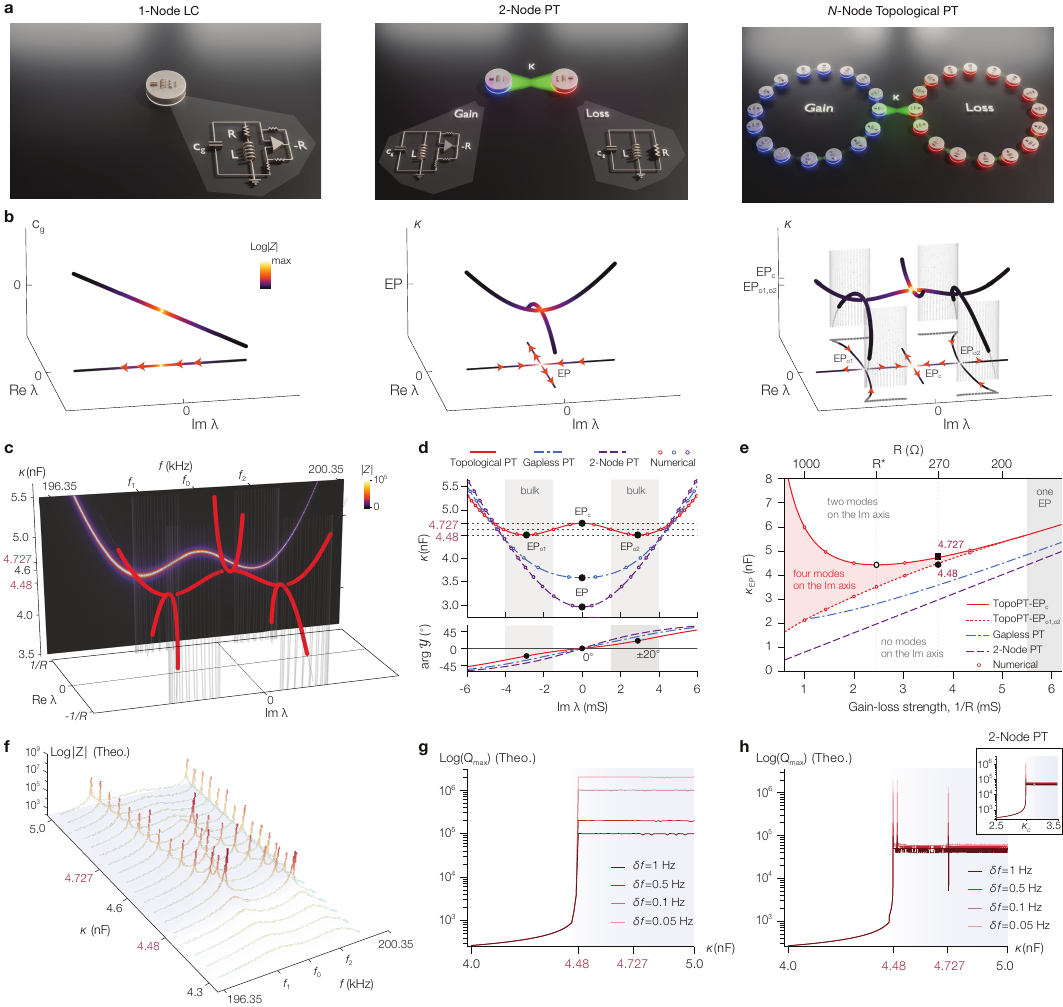}
		\let\small\FigTwoCaptionFont
		\caption{\textbf{Topological origin of the ultrahigh-Q resonances in the clean limit.} 
			Arising from the PT transitions, exceptional points mark the onset of divergent modes, which give rise to ultrahigh-Q resonances through the unique spectral profile of the topological configuration.
			\textbf{a,} The three models are built from the same LC resonator, and a variable coupling $\kappa$ couples the gain (blue) and loss (red) parts of the dimer (middle) and of two SSH rings of $N$ nodes (right).
			\textbf{b,} Sweeping a circuit parameter such as $c_g$ or $\kappa$ drives the eigenvalues in the complex plane and gives rise to a resonance (warm colours) once an eigenvalue reaches the spectral origin. The dimer reaches the origin through a single EP, whereas the topological PT lattice has three EPs, of which the central one, EP$_\mathrm{c}$, passes through the origin.
			\textbf{c,} Increasing $\kappa$ detaches four eigenvalues (red) from the bulk bands (grey). They are complex below 4.48\,nF, purely imaginary up to 4.727\,nF, and two real and two imaginary beyond. Since the drive frequency shifts the spectrum along the imaginary axis, each lossless eigenvalue gives a ridge of the impedance $|Z(f,\kappa)|$ between the coupling nodes (background map, Eq.~\eqref{EqMethodsFinalZ}).
			\textbf{d,} The extrema of the coupling needed for a lossless mode (Eq.~\eqref{EqHump}) are the EPs (black dots). Since $\mathcal{Y}$ is real at EP$_\mathrm{c}$ (bottom), the strong EP response to a gain fluctuation only shifts the frequency there, whereas the phase of about $\pm20^\circ$ at EP$_\mathrm{o1,o2}$ lets part of it broaden the line.
			\textbf{e,} Curves in the parameter space of the gain--loss strength and coupling that place each EP at the spectral origin show that the three EPs persist over a wide range of gain--loss strengths. The flat bottom of the EP$_\mathrm{c}$ curve at $R^*$ marks the point where EP$_\mathrm{c}$ is least sensitive to gain fluctuations.
			\textbf{f,} Up to four sharp impedance peaks appear between 4.48 and 4.727\,nF, where the outer EPs reach the origin at $f_1$ and $f_2$ and the inner pair merges into EP$_\mathrm{c}$ at $f_0$.
			\textbf{g,} With balanced gain and loss, the Q above 4.48\,nF is limited only by the frequency step $\delta f$, since the true linewidth vanishes.
			\textbf{h,} With a small excess of gain, the eigenvalues acquire a small real part and Q becomes finite, but one coupling near each EP still reaches the resolution limit, whereas the dimer has only one, $\kappa_c$ (inset).
				The parameters are $c_1=1.005$\,nF, $c_2=2.212$\,nF, $c_g=333$\,nF, $L=1.915\,\mu$H, $R=270\,\Omega$ and $N=16$ nodes per ring.}
		\label{figMain_CleanLimit}
	\end{figure*}

	To find how the broadening coefficient $\eta$ can be made small while the eigenvalue is still steered to the origin, we examine three models in the clean limit, in which every component takes its design value and neither disorder nor a mismatch of the resonators is present, from the fundamental LC resonator to our topological PT lattice (Fig.~\ref{figMain_CleanLimit}). Each model adds one ingredient to the previous one, first a coupling and then a topological band gap, and the comparison shows what each ingredient brings. We use the language of circuits, but the equations apply to any coupled resonators described by Eq.~\eqref{EqImpEig}.

	\subsection{Single LC resonator with gain}

	The simplest resonator is a single LC node, a capacitor and an inductor with a parallel loss resistance. A gain element can, in principle, cancel this loss and move the eigenvalue to the origin (Fig.~\ref{figMain_CleanLimit}a, left). Gain compensation alone, however, does not give a robust ultrahigh-Q resonance under disorder, for two reasons. The admittance of the node, which is also its only eigenvalue, is $\lambda=i\omega c_\Sigma+1/(i\omega L)+1/R_p-1/R$, where $1/R_p$ is the parasitic loss and $-1/R$ is the gain supplied by a negative-impedance converter~\cite{hofmann_chiral_2019,helbig_generalized_2020}. First, a gain fluctuation $\delta$ reaches $\mathrm{Re}\,\lambda$ in full, so that $\eta=1$. Second, apart from the gain itself, no element of the resonator moves its eigenvalue along the real axis. Sweeping the capacitance or the drive frequency only changes the detuning along the imaginary axis and leaves the net loss unchanged (Fig.~\ref{figMain_CleanLimit}b, left). Gain can thus reduce the loss and increase Q, but it cannot hold the eigenvalue at the origin against its own fluctuations. Improving Q under disorder therefore requires two kinds of control, one that steers the eigenvalue in both directions of the complex plane and one that reduces its response to fluctuations.

	\subsection{Exceptional points in PT systems and their linewidth limit}

	Coupling two resonators provides the first of these controls. In a PT dimer, a lossy resonator and a gain resonator are coupled through a capacitor $\kappa$ (Fig.~\ref{figMain_CleanLimit}a, middle). For identical resonators with the onsite admittance $Y_0=i\omega c_\Sigma+1/(i\omega L)$, the circuit Laplacian and its eigenvalues are
	\begin{equation}
		\begin{aligned}
			J_\text{2-Node}&=
			\begin{pmatrix}
				Y_0+\dfrac{1}{R} & -i\omega\kappa\\[6pt]
				-i\omega\kappa & Y_0-\dfrac{1}{R}
			\end{pmatrix},\\[4pt]
			\lambda_\pm&=Y_0\pm\sqrt{\frac{1}{R^2}-\omega^2\kappa^2}.
		\end{aligned}
		\label{Eq2NodePTLaplacian}
	\end{equation}
	The two eigenvalues merge at an EP at $\omega\kappa=1/R$. At the resonance frequency, $Y_0=0$, this EP lies exactly at the spectral origin (Fig.~\ref{figMain_CleanLimit}b, middle). As $\kappa$ increases and the system moves through the EP, the pair changes from purely real, with one mode of net loss and one of net gain, to purely imaginary, with two lossless modes. In the clean limit, the coupling thus steers the eigenvalues along the EP trajectories to the origin, which a single resonator cannot do.

	In a real dimer, two effects limit this route. To include both, we add a small difference $\pm i\zeta$ to the onsite admittances of the two resonators, which describes a mismatch of their natural frequencies, and replace $-1/R$ by $-1/R+\delta$ at the gain node. Equation~\eqref{Eq2NodePTLaplacian} then gives, without approximation,
	\begin{equation}
		\lambda_\pm=Y_0+\frac{\delta}{2}\pm\sqrt{\left(\frac{1}{R}-\frac{\delta}{2}+i\zeta\right)^2-\omega^2\kappa^2}.
		\label{EqDimerShare}
	\end{equation}
	The first effect is the mismatch of the resonators, which is mostly unavoidable in practice. For $\zeta\neq0$, the square root vanishes at no real $\kappa$, so the EP turns into an avoided crossing~\cite{haus_coupled-mode_1991,el-ganainy_exceptional_2014}. It lifts the degeneracy at the origin and pushes the two trajectories apart, and neither of them reaches the origin through the EP. A slight excess of gain, however, shifts the spectrum along the real axis and lets one trajectory cross the imaginary axis. Tuning the drive frequency then shifts the whole spectrum along the imaginary axis and brings this crossing to the origin (Fig.~\ref{figMain_SpectralBroadening}a,b). The coupling, a slight excess of gain and the drive frequency together can therefore still steer one eigenvalue through the origin.

	The second effect is disorder. Dynamical fluctuations of the gain broaden the trajectories, so that a trajectory that passes through the origin still spreads along the real axis, and this spread limits the Q. For $\zeta=0$, Eq.~\eqref{EqDimerShare} gives $\mathrm{Re}\,\lambda_\pm=\delta/2$ at every coupling above the EP, so exactly half of every gain fluctuation reaches the linewidth, $\eta=1/2$, whatever $\kappa$ is chosen. Near the EP, the eigenspace becomes defective and adds a frequency jitter~\cite{bai_nonlinear_2023,wang_petermann-factor_2020}, since the imaginary part of Eq.~\eqref{EqDimerShare} changes with $\delta$ at a rate that diverges at the EP, a form of the known noise enhancement at EPs~\cite{lau_fundamental_2018,li_synergetic_2023,kononchuk_exceptional-point-based_2022,li_stochastic_2023,sun_noise-assisted_2026}. The spectral width of the dimer therefore stays broad near its EP (Supplementary Section~\ref{SuppMat2NodePT}).

	The dimer therefore provides the first control but not the second. By Eq.~\eqref{EqDimerShare}, $\eta$ stays at $1/2$ on the lossless branch in the clean limit, whatever $\kappa$ is chosen, so the coupling alone cannot reduce it. To lower $|\eta|$, the response of the coupling nodes themselves must change. Embedding each resonator in a topological lattice achieves this, because the response of each coupling node then depends on the gapped spectrum of the whole topological lattice, an additional degree of spectral control that the dimer lacks.

	\subsection{Topological PT lattice and its input admittance}

	A topological lattice supplies this control through its band gap and the interface modes that the gap hosts. We embed each resonator in a closed SSH ring of $N$ nodes, in which identical LC resonators are coupled through alternating capacitors $c_1$ and $c_2$ (Fig.~\ref{figMain_CleanLimit}a, right). Measured from the onsite admittance, its bulk modes form two bands at $\omega(c_2-c_1)<|\mathrm{Im}\,\lambda|<\omega(c_1+c_2)$, separated by a band gap centred at zero (Methods). We couple a lossy ring and a gain ring through a single capacitor $\kappa$ between node $a$ of the lossy ring and node $b$ of the gain ring. Since an isolated ring is translationally invariant, its winding number depends only on the choice of the unit cell, and both choices are equally valid. The coupling, however, breaks this symmetry and provides a natural reference point, which fixes the unit cell at the coupling node, as is standard for SSH domain walls and topological interfaces~\cite{su_solitons_1979}. With this choice, each ring is in the topological phase of the SSH model for $c_2>c_1$~\cite{sahin_protected_2025} (Methods), and exchanging $c_1$ and $c_2$ only mirrors each ring about its coupling node and leaves the spectrum unchanged. In both cases, each gapped ring binds one interface mode at its coupling node, so that the topological distinction relevant here is between gapped ($c_1\neq c_2$) and gapless ($c_1=c_2$) rings (Methods and Supplementary Section~\ref{SuppMatCleanLimit}). The resulting $2N\times2N$ circuit Laplacian is
	\begin{equation}
		J_\text{Topo-PT} =
		\begin{pmatrix}
			J_\mathrm{SSH} + \dfrac{1}{R}\,\mathds{I}_N & K \\[8pt]
			K^\mathsf{T} & J_\mathrm{SSH} - \dfrac{1}{R}\,\mathds{I}_N
		\end{pmatrix},
		\label{EqMainTopoPT}
	\end{equation}
	where $J_\mathrm{SSH}$ is the Laplacian of an isolated ring with the onsite admittance $Y_0$ (Methods, Eq.~\eqref{EqMethodsSSH}), and $K$ has a single nonzero entry, $-i\omega\kappa$, which couples $a$ and $b$. The strong coupling acts as a local defect in each closed ring, and the topological rings host modes localized at this interface (Supplementary Section~\ref{SuppMatCleanLimit}).

	The influence of the coupling on the entire ring is encoded in its input admittance, the inverse of its one-point impedance. Eliminating all other nodes gives an exact condition for the coupling-induced modes, which contains only the input admittances of the two finite rings (Methods). At the resonance frequency, where $Y_0=0$, the eigenvalues of $J_\text{Topo-PT}$ satisfy
	\begin{equation}
		\mathcal{Y}\!\left(\tfrac{1}{R}-\lambda\right)\mathcal{Y}\!\left(-\tfrac{1}{R}-\lambda\right)+\omega^2\kappa^2=0,
		\label{EqDressedDimer}
	\end{equation}
	where, for a long ring,
	\begin{equation}
		\mathcal{Y}(y)=\frac{1}{y}\sqrt{\left[y^2+\omega^2(c_2-c_1)^2\right]\left[y^2+\omega^2(c_1+c_2)^2\right]}
		\label{EqDressedY}
	\end{equation}
	is the input admittance of a ring whose nodes have the admittance $y$ (Supplementary Section~\ref{SuppMatTwoNodeReduction}). Here, $y$ is the onsite admittance of a node shifted by the eigenvalue, $y=Y_0+1/R-\lambda$ for the lossy ring, where $+1/R$ is the loss, and $y=Y_0-1/R-\lambda$ for the gain ring, where $-1/R$ is the gain. At resonance, these reduce to the two arguments of Eq.~\eqref{EqDressedDimer}. The input admittance carries the band structure of the topological ring through two terms, the gap edge $\omega(c_2-c_1)$ and the band edge $\omega(c_1+c_2)$. For a single node, the input admittance is the node admittance itself, $\mathcal{Y}(y)=y$, so every difference between the topological PT lattice and the dimer comes from these two band terms. A gain fluctuation at node $b$ enters as $\mathcal{Y}\rightarrow\mathcal{Y}+\delta$, and a detuning of the drive frequency shifts $\lambda$ along the imaginary axis. For the ring size used in Fig.~\ref{figMain_CleanLimit} ($N=16$), the solutions of Eq.~\eqref{EqDressedDimer} agree with the eigenvalues of the coupling-induced modes of the full Laplacian to four digits (Supplementary Section~\ref{SuppMatTwoNodeReduction}), so this compact description captures the relevant trajectories accurately in this parameter regime.

	Figure~\ref{figMain_CleanLimit}b shows that the band terms change the spectrum qualitatively. As $\kappa$ increases, four modes detach from the bulk bands, localize at the coupling nodes and reach the imaginary axis, and the topological PT lattice forms three EPs instead of the single EP of the dimer, EP$_\mathrm{o1}$ and EP$_\mathrm{o2}$ at $\kappa\approx4.48$\,nF and EP$_\mathrm{c}$ at $4.727$\,nF (Fig.~\ref{figMain_CleanLimit}b, right). The corresponding impedance map in Fig.~\ref{figMain_CleanLimit}c connects these trajectories to the resonant response. Bright ridges occur where a mode with vanishing real admittance reaches the origin as the drive frequency changes. The topological PT lattice therefore provides several routes to divergent clean-limit Q, whose different sensitivity to disorder must be understood from the band structure.

	\subsection{Three exceptional points from the topological band gap}

	It is the specific configuration that we have described by Eq.~\eqref{EqMainTopoPT} and shown in Fig.~\ref{figMain_CleanLimit}a, right. Its interface mode trajectories can be understood from the competition between the gap response and the bulk-band response of each topological ring. Equation~\eqref{EqDressedY} can be written as $\mathcal{Y}^2(y)=y^2+2\omega^2(c_1^2+c_2^2)+\omega^4(c_2-c_1)^2(c_1+c_2)^2/y^2$. The last term contains the product of the gap edge $\omega(c_2-c_1)$ and the band edge $\omega(c_1+c_2)$, and it exists only for a nonzero gap. It dominates near the gap centre for weak gain and loss and disappears when the gap closes, so it changes how the local coupling competes with the response of the surrounding ring. Since $\mathcal{Y}$ is odd and real for real arguments, an eigenvalue $\lambda$ on the imaginary axis, a lossless mode, requires
	\begin{equation}
		\omega\kappa=\left|\mathcal{Y}\!\left(\tfrac{1}{R}+\lambda\right)\right|.
		\label{EqHump}
	\end{equation}
	The right-hand side is the coupling admittance needed to hold a lossless mode at the detuning $\mathrm{Im}\,\lambda$, and Fig.~\ref{figMain_CleanLimit}d plots it in units of capacitance. For the dimer, $\mathcal{Y}=y$, and a detuning only adds reactive admittance, so the required coupling, $\sqrt{1/R^2+(\mathrm{Im}\,\lambda)^2}/\omega$, grows monotonically away from the centre. Its single minimum gives the single EP of the dimer (purple in Fig.~\ref{figMain_CleanLimit}d).

	The topological PT model exhibits distinct properties inside and outside its band gap. In the middle of the gap, the ring has no bulk modes to respond with. It acts as a stiff element with a large input admittance, $\mathcal{Y}(y)\simeq\omega^2(c_2-c_1)(c_1+c_2)/y$ for weak gain and loss, which is set entirely by the gap. Near the bulk bands, the ring can respond through its bulk modes, its input admittance drops, and less coupling is needed. Far from the bands, $\mathcal{Y}\simeq y$ grows again. The required coupling therefore has a maximum at the gap centre between two minima near the bulk bands (red in Fig.~\ref{figMain_CleanLimit}d). At each extremum, two lossless modes merge and their eigenvectors coalesce into one, which forms an EP. The two minima give the outer EPs, EP$_\mathrm{o1}$ and EP$_\mathrm{o2}$, at $\kappa\approx4.48$\,nF and $\mathrm{Im}\,\lambda\approx\pm2.93$\,mS, inside the frequency range of the bulk bands (grey). The maximum gives the central EP, EP$_\mathrm{c}$, at the spectral origin, with the coupling
	\begin{equation}
		\begin{aligned}
			\kappa_{\mathrm{EP_c}}&=\frac{\mathcal{Y}(1/R)}{\omega}\\
			&=\frac{1}{\omega R}\sqrt{\left[1+\omega^2(c_2-c_1)^2R^2\right]\left[1+\omega^2(c_1+c_2)^2R^2\right]},
		\end{aligned}
		\label{EqMainKappaEPc}
	\end{equation}
	which is $4.727$\,nF for our parameters. The gap edge and the band edge enter this coupling directly. At $\kappa_{\mathrm{EP_c}}$, the $2\times2$ matrix of the coupling nodes becomes proportional to $\begin{psmallmatrix}1&-i\\-i&-1\end{psmallmatrix}$, which squares to zero and has a single eigenvector, and the eigenvectors of the full Laplacian confirm the coalescence at all three EPs (Methods).

	\subsection{Distinctive features of the central exceptional point}

	The central EP differs physically from the outer ones. It forms on a maximum rather than a minimum, so its transition is reversed. The pair approaches the origin along the imaginary axis and leaves it along the real axis as $\kappa$ increases (Im$\lambda$-to-Re$\lambda$), whereas the outer EPs, like the dimer EP, show Re$\lambda$-to-Im$\lambda$ transitions (Fig.~\ref{figMain_CleanLimit}b, right). More importantly, EP$_\mathrm{c}$ sits at the centre of the band gap, isolated from all bulk modes and formed by modes localized at the coupling interface, whereas the outer EPs sit within the frequency range of the bulk bands, where the interface modes hybridize with bulk states. Beyond EP$_\mathrm{c}$, the pair becomes purely real and moves into the band gaps of the two rings, where its modes decay exponentially from the interface on alternating sites, the hallmark of the boundary states of an open topological SSH chain (Supplementary Section~\ref{SuppMatCleanLimit}). The strong coupling thus turns each closed ring into an effectively open topological chain, and EP$_\mathrm{c}$ is an EP of its topological interface modes.

	The phase of $\mathcal{Y}$ along the curve (Fig.~\ref{figMain_CleanLimit}d, bottom) reveals a further difference, which decides how each EP responds to disorder. At EP$_\mathrm{c}$, the bulk bands of each ring lie symmetrically above and below in frequency, their contributions to $\mathcal{Y}$ balance, and $\mathcal{Y}$ is real, as at the dimer EP. At the outer EPs, inside the frequency range of the bulk bands, this balance is lost, and $\mathcal{Y}$ acquires a phase of about $\pm20^\circ$. Near any EP, the eigenvectors nearly coalesce, and the response of the eigenvalue to a perturbation grows without bound~\cite{wiersig_review_2020,wang_petermann-factor_2020}. For a gain fluctuation at the coupling node, the phase of $\mathcal{Y}$ at the EP sets the direction of this response in the complex plane (Supplementary Section~\ref{SuppMatGainFluctuations}). At EP$_\mathrm{c}$, the growing response is purely imaginary and only shifts the resonance frequency. At the outer EPs, the phase tilts it towards the real axis, so that a part $\tan20^\circ\approx0.36$ of a response that diverges at the EP broadens the line. The band gap thus keeps the strong EP response out of the linewidth at EP$_\mathrm{c}$, but not at the outer EPs.

	Figure~\ref{figMain_CleanLimit}e shows that the band gap keeps the three EPs over a wide range of gain--loss strengths. In the shaded window between the curves of EP$_\mathrm{c}$ (solid) and EP$_\mathrm{o1,o2}$ (dashed), four modes lie on the imaginary axis, above it two remain, and below it none. The window stays open up to $1/R=5.50$\,mS ($R=182\,\Omega$), where the three EPs merge into one (grey), and the EP$_\mathrm{c}$ curve has a flat bottom at $R^*=407\,\Omega$ (open circle), where $\mathcal{Y}$ is smallest. We use $R=270\,\Omega$, which places EP$_\mathrm{c}$ within the 4.0 to 5.1\,nF tuning range of the coupling capacitor of our circuit, due to its experimental feasibility. In this operating regime, a gapless PT lattice ($c_1=c_2$) has only a single EP, like the dimer (blue in Fig.~\ref{figMain_CleanLimit}d,e and Supplementary Section~\ref{SuppMatThreeEPs}).

	\subsection{Divergent Q factors in the clean limit}

	To connect the three-EP spectrum to Q, we follow the impedance resonances as the coupling increases in the clean limit (Fig.~\ref{figMain_CleanLimit}f--h). Because the drive frequency shifts the spectrum along the imaginary axis (Supplementary Section~\ref{SuppMatFreqShift}), each lossless eigenvalue reaches the origin at its own drive frequency and gives a sharp impedance peak. Figure~\ref{figMain_CleanLimit}f follows these peaks as $\kappa$ increases. Below 4.48\,nF, no eigenvalue lies on the imaginary axis, and the impedance shows only broad and weak maxima. At 4.48\,nF, sharp peaks appear at $f_1$ and $f_2$, about 0.7\,kHz below and above $f_0$, where EP$_\mathrm{o1}$ and EP$_\mathrm{o2}$ reach the origin. Each of them then splits into two peaks, which gives four sharp peaks between 4.48 and 4.727\,nF, compared with at most two peaks for the dimer. The two inner peaks move towards $f_0$ and merge there at EP$_\mathrm{c}$. Beyond EP$_\mathrm{c}$, this pair becomes real and its peak vanishes, whereas the two outer peaks remain and move apart.

	With gain and loss exactly balanced, the extracted Q jumps at 4.48\,nF onto a plateau that rises as the frequency step $\delta f$ of the calculation shrinks (Fig.~\ref{figMain_CleanLimit}g). The plateau is the resolution limit $f_0/(2\delta f)$, because the true linewidth vanishes, so the plotted values are lower bounds (Methods).

	A small excess of gain gives every eigenvalue on the imaginary axis a finite real part, and the Q becomes finite (Fig.~\ref{figMain_CleanLimit}h). In the clean limit, a uniform imbalance of all gain elements shifts the real part of every such eigenvalue by half of the imbalance, in every model (Supplementary Section~\ref{SuppMatGainFluctuations}), so the topological PT lattice and the dimer have the same Q plateau (Fig.~\ref{figMain_CleanLimit}h and inset). The EP trajectories, however, still cross the imaginary axis. Near an EP, the real part of an eigenvalue changes rapidly with $\kappa$ and passes through zero at a coupling close to the EP, where the eigenvalue reaches the origin at its drive frequency and Q rises to the resolution limit. The topological PT lattice has three such crossings, one near each EP, whereas the dimer has only one (inset). Every EP of the topological PT lattice thus offers its own ultrahigh-Q coupling, even with a gain--loss imbalance. Under disorder, however, these EPs behave very differently, as we show next.

	\section{Band-gap stabilization of the central exceptional point}
	\label{SecSuppression}

	\begin{figure*}[t!]
		\centering
		\includegraphics[width=\linewidth]{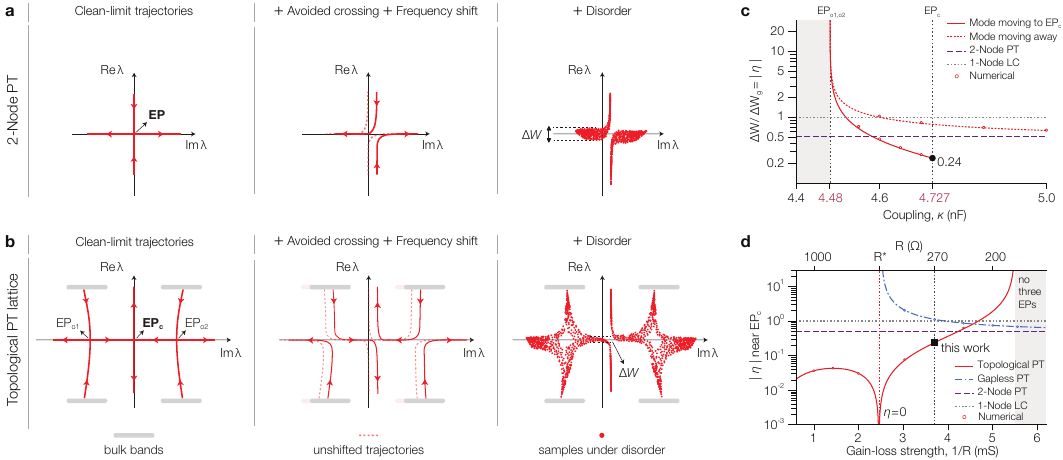}
		\caption{\textbf{Spectral evolution from the clean limit to the disordered regime.}
		In a real circuit, avoided crossings make the spectral origin hard to reach, and disorder makes the eigenvalue hard to keep there. The central EP of the topological PT lattice keeps it best, with the smallest broadening coefficient $\eta$ of all models.
			\textbf{a,} In the clean limit, the EP of the 2-Node PT dimer lies exactly at the origin (left, arrows for increasing $\kappa$). A mismatch of the two resonators turns it into an avoided crossing with neither branch passing through the origin (middle). Tuning the drive frequency, which shifts the entire spectrum along the imaginary axis, nevertheless brings one trajectory through the origin (middle and Supplementary Fig.~\ref{FigSuppMatDisorderAnalyses}). Disorder, however, broadens the trajectories into a cloud with spectral width $\Delta W$ at the origin (right), which translates into fluctuations of the peak in the response spectrum.
			\textbf{b,} The same steps apply to the topological PT lattice, which has three EPs. EP$_\mathrm{c}$ is approached along the imaginary axis and left along the real axis (Im$\lambda$-to-Re$\lambda$ transition), the reverse of the outer EPs and the dimer EP. Under disorder, the clouds near the outer EPs broaden, whereas the cloud at EP$_\mathrm{c}$ stays narrow, which is the key to the ultrahigh-Q resonance.
			\textbf{c,} On the branch towards EP$_\mathrm{c}$, the broadening coefficient $|\eta|=\Delta W/\Delta W_\mathrm{g}$ (Eq.~\eqref{EqGainShare}) falls below the dimer value of 1/2 and reaches 0.24 at EP$_\mathrm{c}$ (a smaller coefficient means less spectral broadening). The single resonator, with $|\eta|=1$, has no suppression at all, whereas the critical coefficient $|\eta_\mathrm{c}|$ (Eq.~\eqref{EqEPcShare}) of the topological PT lattice is the smallest of all models.
			\textbf{d,} The broadening coefficient $|\eta|$ near EP$_\mathrm{c}$ vanishes at $R^*$ and diverges where the three EPs merge. This shows the flexibility of the topological PT lattice, whose broadening coefficient can be tuned over orders of magnitude, unlike those of the dimer and the single resonator, which remain constant.
			}
		\label{figMain_SpectralBroadening}
	\end{figure*}

	\subsection{From the clean limit to disorder}

	The clean-limit resonances identify where Q can become large. Under disorder, the decisive question is how each EP converts gain fluctuations into linewidth, that is, how large its broadening coefficient $\eta$ is, since a smaller $|\eta|$ keeps the line narrower (Eq.~\eqref{EqGainShare}). We find that the central EP converts them almost entirely into shifts of the resonance frequency, so that only a small part reaches the linewidth. Figure~\ref{figMain_SpectralBroadening}a,b sketches how the trajectories of the two PT systems change from the clean limit to a real circuit. In the clean limit, the EP of the dimer lies at the origin (Fig.~\ref{figMain_SpectralBroadening}a, left). A mismatch of the two resonators turns it into an avoided crossing (Fig.~\ref{figMain_SpectralBroadening}a, middle). At the resonance frequency, the avoided crossing stays centred at the origin, but neither trajectory passes through it. The drive frequency shifts the entire spectrum along the imaginary axis, and a slight imbalance of gain and loss shifts it along the real axis (Eq.~\eqref{EqDimerShare}), so that one trajectory can still be placed at the origin. In our simulations, the resonance then lies at a drive frequency slightly above the resonance frequency of a single node (Supplementary Fig.~\ref{FigSuppMatDisorderAnalyses}). Finally, disorder broadens the trajectories into clouds with a large $\Delta W$ at the origin (Fig.~\ref{figMain_SpectralBroadening}a, right). In the topological PT lattice, avoided crossings form at all three EPs, and the trajectory towards EP$_\mathrm{c}$ can be placed at the origin in the same way (Fig.~\ref{figMain_SpectralBroadening}b, middle).

	\subsection{Robustness of the central exceptional point to disorder}

	Under disorder, the clouds near the outer EPs broaden as in the dimer, whereas the trajectory towards EP$_\mathrm{c}$ stays concentrated, and the cloud at the origin has a much smaller $\Delta W$ (Fig.~\ref{figMain_SpectralBroadening}b, right). Two properties of Eq.~\eqref{EqDressedDimer} explain this difference.

	First, as the phase of $\mathcal{Y}$ in Fig.~\ref{figMain_CleanLimit}d shows, the band gap keeps the strong response near the EPs out of the linewidth at EP$_\mathrm{c}$, whereas a part $\tan20^\circ\approx0.36$ of it broadens the line at the outer EPs. Figure~\ref{figMain_SpectralBroadening}c confirms this along $\kappa$ for the two modes that are created at the outer EPs. Their $|\eta|$ diverges at 4.48\,nF, as the $20^\circ$ phase predicts. The mode that moves towards EP$_\mathrm{c}$ (solid) then falls below the dimer value near 4.6\,nF and reaches $|\eta|=0.24$ at EP$_\mathrm{c}$, whereas the mode that moves away (dotted) stays above it. The ultrahigh-Q resonance therefore lives on the branch towards EP$_\mathrm{c}$.

	Second, the band terms of $\mathcal{Y}$ set the finite value of $\eta$ that remains at EP$_\mathrm{c}$. A gain fluctuation at the coupling node acts like a small change of $1/R$. It shifts EP$_\mathrm{c}$ along its curve in Fig.~\ref{figMain_CleanLimit}e, while the gap holds the lossless modes at the maximum of the curve in Fig.~\ref{figMain_CleanLimit}d. Comparing these two effects gives (Supplementary Section~\ref{SuppMatGainFluctuations})
	\begin{equation}
		\eta_\mathrm{c}=\frac{d\kappa_{\mathrm{EP_c}}/d(1/R)}{2\,\omega\,\kappa_{\mathrm{EP_c}}\,d^2\kappa/d(\mathrm{Im}\,\lambda)^2}.
		\label{EqEPcShare}
	\end{equation}
	The numerator is the slope of the EP$_\mathrm{c}$ curve in Fig.~\ref{figMain_CleanLimit}e, and the denominator contains the curvature of the curve in Fig.~\ref{figMain_CleanLimit}d at its maximum. A flatter EP$_\mathrm{c}$ curve or a sharper maximum therefore gives a smaller $|\eta_\mathrm{c}|$ and a narrower line. For the dimer, the same expression returns $\eta=1/2$. For our parameters, it gives $\eta_\mathrm{c}=-0.24$, about half of the dimer value and a quarter of that of the single resonator (dot in Fig.~\ref{figMain_SpectralBroadening}c).

	The band gap also gives a design rule for an even smaller $|\eta|$ (Fig.~\ref{figMain_SpectralBroadening}d). By Eq.~\eqref{EqEPcShare}, $\eta_\mathrm{c}$ vanishes where the EP$_\mathrm{c}$ curve of Fig.~\ref{figMain_CleanLimit}e is flat, that is, at the minimum of $\mathcal{Y}$. This occurs at $1/R^*=\omega\sqrt{(c_2-c_1)(c_1+c_2)}$, the geometric mean of the gap edge and the band edge, which gives $R^*=407\,\Omega$ for our capacitors. At $R^*$, a gain fluctuation at the coupling node moves the resonant eigenvalue only along the imaginary axis, to first order. For small fluctuations, this first-order term dominates the broadening and already explains the differences between the models, and we therefore discuss only the first-order effects. At $R^*$, however, where this term vanishes, the second-order terms set the remaining broadening, so that disorder also enters through these terms and limits the Q when they are relevant beyond the first-order limitations. Towards strong gain and loss, $|\eta_\mathrm{c}|$ grows and diverges at $R=182\,\Omega$, where the three EPs merge (grey). Our parameters, with $R=270\,\Omega$, lie between these limits (square), and approaching $R^*$ would lower $|\eta|$ further. We chose $R=270\,\Omega$ for the experimental feasibility rather than $R^*$ because the parasitic dissipation of the nodes, about 2\,mS (Methods), is comparable to $1/R^*\approx2.5$\,mS and known only approximately, so a design at $R^*$ would leave the effective gain--loss strength poorly defined. At $270\,\Omega$, the designed loss of 3.7\,mS clearly exceeds it, and $|\eta_\mathrm{c}|$ is already about half of the dimer value, so the measured Q is not the limit of the mechanism. Platforms with continuously tunable gain, such as optically pumped resonators~\cite{peng_paritytime-symmetric_2014,hodaei_parity-timesymmetric_2014,bandres_topological_2018}, can approach $R^*$ in situ. Without a gap, the gap term vanishes, $\mathcal{Y}$ has no minimum, and $\eta_\mathrm{c}$ has no zero. At our gain--loss strength, the gapless PT lattice has $|\eta|\approx1.1$, more than even a single resonator (blue).

	The suppression of spectral broadening at EP$_\mathrm{c}$ is therefore a property of the topological band gap. The gap places EP$_\mathrm{c}$ at the centre of a spectral region free of bulk modes, where the topological interface modes of the coupled rings form it, and its edges set $\eta_\mathrm{c}$ through Eq.~\eqref{EqEPcShare} and fix $R^*$. This suppression acts on fluctuations at the coupling nodes, where the dominant disorder is expected.

	\section{Simulated spectra and responses under disorder}
	\label{SecSimulations}

	\begin{figure*}[t!]
		\centering
		\includegraphics[width=\textwidth]{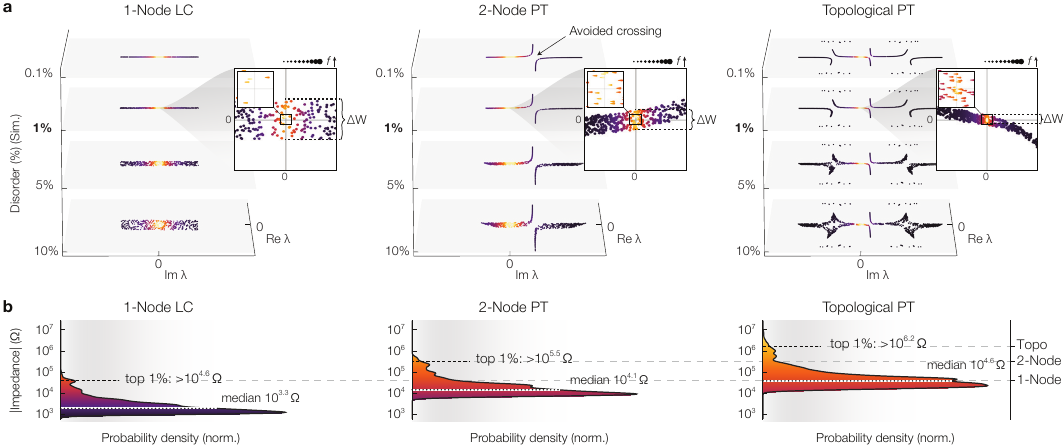}
		\caption{\textbf{Simulated spectra and responses under disorder.}
			The interplay of topology and PT symmetry keeps the eigenvalues of the topological PT lattice least scattered along their trajectory, with the smallest $\Delta W$ and the highest response.
			\textbf{a,} Spectra of the three models for gain disorder from 0.1\% to 10\%, the main source of dynamical jitter. Each level overlays 400 realizations with a random variation of $\kappa$ within $\pm10\%$ (or of $c_g$, Methods). Insets, the origin at 1\% disorder, with each realization also evaluated at 50 drive frequencies within 2\,Hz. Remarkably, the trajectory towards EP$_\mathrm{c}$ stays concentrated near the origin with a narrow $\Delta W$, whereas the single resonator and the dimer broaden strongly along $\mathrm{Re}\,\lambda$.
			\textbf{b,} Probability density of $|Z|$ at 1\% disorder for 500 realizations, each evaluated at 50 drive frequencies, with all models initially tuned to a comparable clean-limit peak of about $10^7\,\Omega$ near zero admittance. The topological PT lattice retains the highest median impedance and top-1\% threshold, a manifestation of its narrower $\Delta W$ and smaller broadening coefficient $\eta$.
			}
		\label{figMain_Simulations}
	\end{figure*}

	The small-fluctuation analysis predicts that the central trajectory should retain the narrowest spectral width. To test whether the suppression survives realistic disorder of increasing strength, we simulate the three circuits with the component values of our experiment (Fig.~\ref{figMain_Simulations}). Saliently, although disorder broadens the trajectories of all three models, the cloud at EP$_\mathrm{c}$ of the topological PT lattice stays the narrowest, and its response stays the highest.

	All simulations start from the same clean-limit condition. For each model, we choose the coupling and the drive frequency at which one eigenvalue sits at the origin, which gives a clean-limit impedance peak of about $10^7\,\Omega$ in every model (Supplementary Fig.~\ref{FigSuppMatDisorderAnalyses}). For the topological PT lattice, this eigenvalue lies on the branch towards EP$_\mathrm{c}$, and for the dimer on the branch above its EP (Methods). We then apply a local disorder of the gain admittance, from 0.1\% to 10\%, at the gain-side coupling node, motivated by the measured spatial concentration of fluctuations near the interface (Supplementary Section~\ref{SuppMatDisorderProfile}). Disorder turns the EPs into avoided crossings, but each trajectory still passes close to the origin at a drive frequency that depends on $\kappa$. Each realization therefore also carries a random variation of $\kappa$ within $\pm10\%$, which samples the eigenvalue along its trajectory in the complex plane. The single resonator has no coupling, so its capacitor $c_g$ varies instead. Both $\kappa$ and $c_g$ are reactive elements, which move the eigenvalues mainly along the imaginary axis, so a fair comparison is set by their admittance in siemens rather than their capacitance in farads. We choose a variation of $c_g$ within $\pm0.5\%$, which spreads $\mathrm{Im}\,\lambda$ over about 4\,mS, as the variation of $\kappa$ does near the origin for the dimer (Methods).

	Despite the broadening of all trajectories, the trajectory towards EP$_\mathrm{c}$ remains concentrated near the origin up to the highest disorder level (Fig.~\ref{figMain_Simulations}a). Each panel overlays the eigenvalues of 400 realizations per level, coloured by the weight of each eigenvalue in the impedance between the coupling nodes (Methods). For the single resonator, the cloud grows into a rectangle that spans both axes, and its spread along the real axis stays wide, as $\eta=1$ predicts. For the dimer, the cloud near the origin spreads quickly as the disorder grows. For the topological PT lattice, the clouds near EP$_\mathrm{o1}$ and EP$_\mathrm{o2}$ broaden into star-like shapes, as their divergent $\eta$ predicts, whereas the trajectory towards EP$_\mathrm{c}$ stays narrow. The insets zoom into the origin at 1\% disorder, where each realization is evaluated at 50 random drive frequencies within a 2\,Hz window, comparable to the linewidth of an ultrahigh-Q resonance. Tuning the drive frequency moves the eigenvalues along the imaginary axis and brings some of them close to the origin in every model, but only the topological PT lattice keeps a small $\Delta W$ there. A gapless PT lattice of the same geometry broadens like the dimer or more (Supplementary Section~\ref{SuppMatDisorderSpectra}). The order of the clouds thus follows the order of $\eta$ in Fig.~\ref{figMain_SpectralBroadening}c,d.

	To connect this spectral confinement to the driven response, we use a second ensemble (Fig.~\ref{figMain_Simulations}b). At 1\% disorder, we evaluate $|Z|$ between the coupling nodes, or between the node and ground for the single resonator, for 500 realizations, each at 50 random drive frequencies within the 2\,Hz window. Disorder lowers the response of all models from the clean-limit peak of about $10^7\,\Omega$, but by very different amounts. The median falls to $10^{3.3}\,\Omega$ for the single resonator and to $10^{4.1}\,\Omega$ for the dimer, whereas the topological PT lattice keeps $10^{4.6}\,\Omega$. The upper tail, which measures how often a high response survives, separates the models more clearly. The top 1\% of the values exceed $10^{4.6}$, $10^{5.5}$ and $10^{6.2}\,\Omega$ for the single resonator, the dimer and the topological PT lattice, respectively, so the topological PT lattice keeps its top responses within a decade of the clean-limit peak. Across the whole distribution, the response is largest for the topological PT lattice, intermediate for the dimer and smallest for the single resonator.

	\section{Experimental realization of ultrahigh-Q resonances}
	\label{SecExperiment}

	\begin{figure*}[t!]
	\centering
	\includegraphics[width=\textwidth]{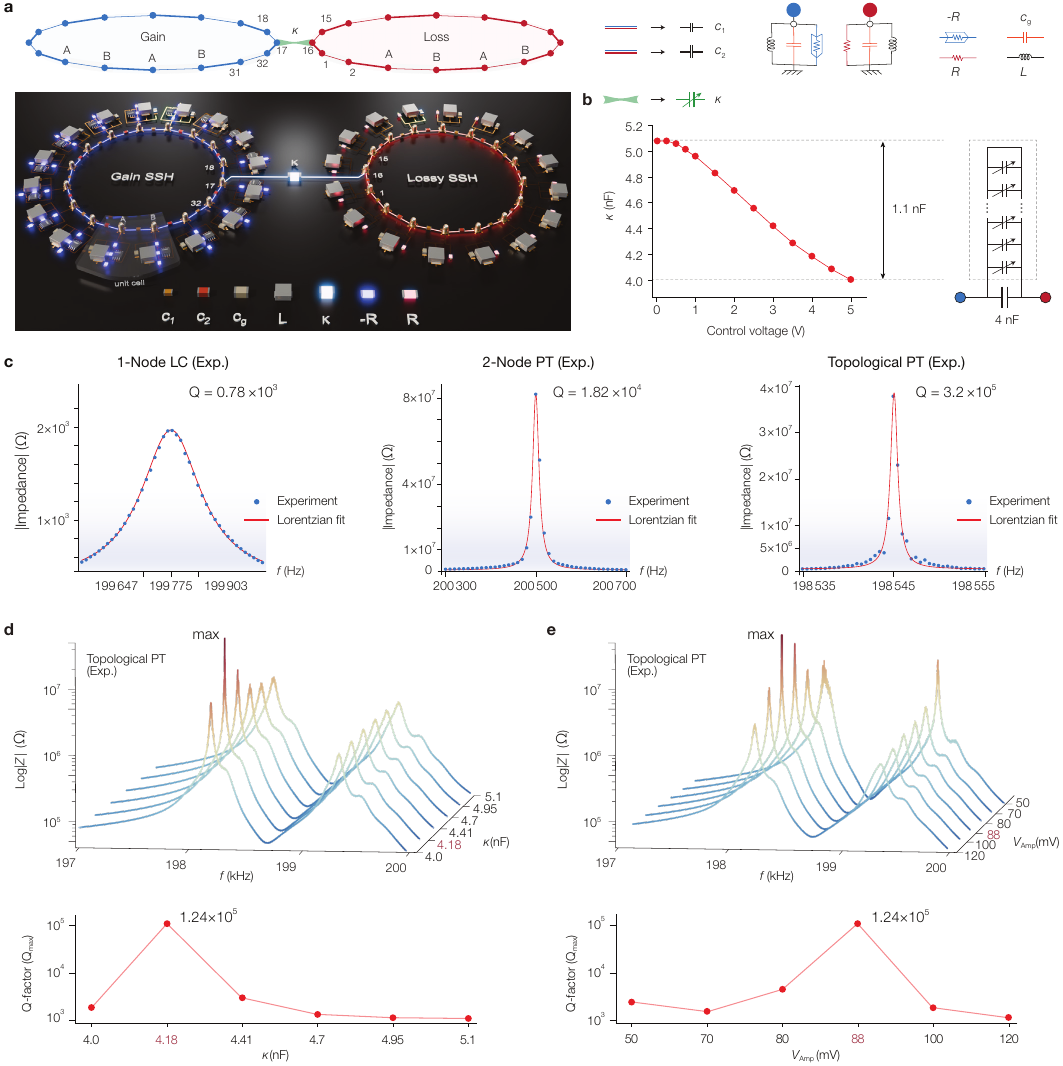}
	\caption{\textbf{Experimental ultrahigh-Q enhancement through a topologically stabilized EP.}
		Built from the same type of components, the topological PT circuit reaches resonances with Q of the order of $10^5$, the highest among the three circuits, together with a high parametric sensitivity.
		\textbf{a,} Topological PT circuit of two 16-node SSH rings, one with gain (blue) and one with loss (red), coupled through a tunable capacitor $\kappa$ between node 16 of the lossy ring and node 17 of the gain ring. It is built from standard surface-mount capacitors $c_1$, $c_2$ and $c_g$ and inductors $L$, with a negative resistance $-R$ from a negative-impedance converter at each gain node and a resistor $R$ at each loss node (Methods).
		\textbf{b,} The variable coupling is realized by eleven voltage-controlled capacitors in parallel with a fixed 4\,nF capacitor, which tune $\kappa$ continuously from 4.0 to 5.1\,nF, a range that covers $\kappa_{\mathrm{EP_c}}$ (Methods).
		\textbf{c,} In the measured impedance spectra, the 1-Node LC resonator with gain reaches Q = $0.78\times10^3$ and the 2-Node PT dimer Q = $1.82\times10^4$, whereas the topological PT lattice reaches Q = $3.2\times10^5$, to our knowledge the highest Q reported for a macroscale network of discrete elements.
		\textbf{d,} The ultrahigh-Q resonance depends sharply on the coupling. At a fixed $V_\mathrm{Amp}=88$\,mV, the sharpest peak, $\mathrm{Q}_\mathrm{max}=1.24\times10^5$, appears at $\kappa_c=4.18$\,nF (top), while Q falls to the order of $10^3$ about 0.2\,nF away (bottom).
		\textbf{e,} The resonance depends just as sharply on the drive amplitude. At $\kappa_c$, the same $\mathrm{Q}_\mathrm{max}$ appears at $V_\mathrm{Amp}=88$\,mV and falls to about $10^3$ some 10\,mV away, so $\kappa$ and $V_\mathrm{Amp}$ together bring the circuit to its ultrahigh-Q point.
	}
	\label{figMain_Experiment}
	\end{figure*}

	We now test the predictions of the theory and the simulations in experiment (Fig.~\ref{figMain_Experiment}). We build the three models of Fig.~\ref{figMain_CleanLimit}a from standard discrete components, with the same component values and the same type of operational amplifiers, and measure their Q under ambient conditions.

	\subsection{Implementation of the topological PT circuit}

	All three circuits are built from the same LC node (Fig.~\ref{figMain_Experiment}a,b). Each node is grounded through a capacitor $c_g=333$\,nF and an inductor $L=1.915\,\mu$H, all standard surface-mount components with tolerances of 1\% to 20\% (Methods). After assembly, we tune each node with ferrite sheets and magnets, which reduces the spread of the node resonance frequencies to 0.05\% around $f_0\approx198.35$\,kHz (Supplementary Section~\ref{SuppMatNodeTuning}). The gain nodes (blue) use negative-impedance converters built with AD8057 operational amplifiers~\cite{hofmann_chiral_2019,helbig_generalized_2020}, and the loss nodes (red) use $270\,\Omega$ resistors. To compensate the parasitic dissipation of the nodes, the converters use $180\,\Omega$, which brings gain and loss close to balance with a slight excess of gain (Methods). The 1-Node LC resonator combines the loss and the gain in a single node. The 2-Node PT dimer couples a loss node and a gain node through the same voltage-controlled coupling module as the topological PT lattice (Supplementary Section~\ref{SuppMat2NodePT}). The topological PT lattice realizes Eq.~\eqref{EqMainTopoPT} with two closed 16-node SSH rings with $c_1=1.005$\,nF and $c_2=2.212$\,nF. Its coupling module, eleven voltage-controlled capacitors in parallel with a fixed capacitor, couples node 16 of the lossy ring to node 17 of the gain ring and tunes $\kappa$ continuously from 4.0 to 5.1\,nF, a range that covers $\kappa_{\mathrm{EP_c}}$ (Fig.~\ref{figMain_Experiment}b). The slight excess of gain helps one eigenvalue branch pass the residual avoided crossing near EP$_\mathrm{c}$, as in Fig.~\ref{figMain_CleanLimit}h. Auxiliary $-\kappa$ elements compensate the extra admittance at the coupling nodes, and we measure the impedance between nodes 16 and 17 in a floating configuration (Methods).

	\subsection{Measured ultrahigh-Q resonance}

	The topological PT lattice gives by far the highest Q of the three resonators (Fig.~\ref{figMain_Experiment}c). We obtain Q from Lorentzian fits of the measured impedance with Eq.~\eqref{EqQfactor} (Supplementary Sections~\ref{SuppMat1NodeLC} and~\ref{SuppMat2NodePT}). The LC resonator has Q = 88 without gain and Q = $0.78\times10^3$ with gain. The PT dimer reaches Q = $1.82\times10^4$, limited by the strong fluctuations of its eigenvalue near the EP (Supplementary Section~\ref{SuppMat2NodePT}). The topological PT lattice reaches Q = $3.2\times10^5$ near 198.55\,kHz. Through Eq.~\eqref{EqQLinewidth}, this Q corresponds to a net loss of the resonant eigenvalue of about $3.5\times10^{-4}$ of the loss $1/R$ of a single node, so the network compensates 99.96\% of the dissipation of its components while its line stays stable (Extended Data Fig.~\ref{FigMethods_Noise}). The linewidth of about 0.6\,Hz is only $3\times10^{-6}$ of the resonance frequency. At this sub-hertz level, small noise and jitter of the circuit and of the instruments shift the peak by amounts comparable to its width, which makes the measurement of such narrow lines challenging (Methods). The measured admittance spectra of the circuit, which show the interface modes approaching the origin and the Im$\lambda$-to-Re$\lambda$ transition of EP$_\mathrm{c}$, are presented in Extended Data Fig.~\ref{FigMethods_Spectra} and Methods.

	\subsection{Parametric sensitivity and stability of the ultrahigh-Q resonance}

	Strikingly, this ultrahigh-Q resonance can be switched on and off by very small changes of the coupling or of the drive amplitude, which shows how precisely the coupling steers the eigenvalue to the origin. A detuning of the drive by a few hertz, about $10^{-5}$ of the resonance frequency, already suppresses the response (Fig.~\ref{figMain_Experiment}c). The coupling moves the eigenvalue along its trajectory near EP$_\mathrm{c}$. Figure~\ref{figMain_Experiment}d shows the impedance spectra between nodes 16 and 17 as $\kappa$ is swept from 4.0 to 5.1\,nF at the fixed drive amplitude $V_\mathrm{Amp}=88$\,mV. Over most of this range, the spectra show only weak and broad peaks, with Q of the order of $10^3$. At $\kappa_c=4.18$\,nF, a single sharp peak rises above all others and reaches $\mathrm{Q}_\mathrm{max}=1.24\times10^5$. Notably, a change of $\kappa$ by only 0.2\,nF, less than 0.1\% of the node capacitance of about 336\,nF, brings Q back to the order of $10^3$, consistent with the narrow high-Q regions of Fig.~\ref{figMain_CleanLimit}h. The coupling of the sharpest peak lies below $\kappa_{\mathrm{EP_c}}=4.727$\,nF, mainly because the gain bias, the parasitic resistances and the residual negative resistance of the $-\kappa$ elements shift the EP trajectories in the real circuit (Methods and Supplementary Section~\ref{SuppMatSpectraOuterEPs}).

	The drive amplitude provides a second, independent handle, because the inductors and the negative-impedance converters depend weakly on it, which changes the effective gain and detuning (Methods). Figure~\ref{figMain_Experiment}e shows the spectra at the fixed coupling $\kappa_c=4.18$\,nF for $V_\mathrm{Amp}$ from 50 to 120\,mV. Again, Q stays of the order of $10^3$ over most of the range, and a single sharp peak appears at $V_\mathrm{Amp}=88$\,mV, with the same $\mathrm{Q}_\mathrm{max}=1.24\times10^5$. A change of the amplitude by about 10\,mV suppresses it as strongly as 0.2\,nF of coupling. Both sweeps pass through the same operating point, and both find their maximum exactly there. The ultrahigh-Q resonance is thus a sharp optimum along both directions of the parameter plane, and two independent sweeps reproduce the same $\mathrm{Q}_\mathrm{max}$.

	Significantly, the ultrahigh-Q resonance is also stable. Long recordings show that the measured impedance is least noisy exactly at $\kappa_c$, where the response to the coupling is steepest (Extended Data Fig.~\ref{FigMethods_Noise} and Methods), in contrast to the dimer, whose resonance fluctuates strongly between measurements near its EP (Supplementary Section~\ref{SuppMat2NodePT}). Without a drive, the circuit does not oscillate, and with a drive the waveforms stay sinusoidal (Supplementary Section~\ref{SuppMatTemporalWaveforms}). All measurements are therefore taken below the oscillation threshold, where the ultrahigh-Q resonance is a stable property of the driven circuit.

	\section{Discussion}

	In this work, we introduced a mechanism for ultrahigh-Q enhancement through the suppression of disorder-induced spectral broadening. In the topological PT lattice, gain compensates the dissipation of the components, while the topological band gap directs the strong response of the central EP (EP$_\mathrm{c}$, at the spectral origin in the middle of the band gap) to gain fluctuations into the resonance frequency rather than into the linewidth. The attainable Q is therefore governed by the spectral structure of the coupled network as well as by the losses of its constituent resonators. A circuit built from resonators with Q = 88 thereby reaches Q = $3.2\times10^5$ at room temperature, whereas a PT dimer built from the same type of components reaches Q = $1.82\times10^4$ and a gapless PT lattice with the same geometry shows no such robustness (Supplementary Section~\ref{SuppMatDisorderSpectra}). This identifies the topological band gap as the origin of the effect. Achieving Q above $10^5$ near 200\,kHz is particularly significant for a discrete-element platform, where resistive parasitics and gain-induced fluctuations strongly influence the resonance and have so far limited gain-compensated LC resonators to Q of $10^2$ to $10^3$~\cite{kuhn_200_1996}.

	Ultrahigh-Q resonances have so far been closely tied to the perfection of a material or a device. Superconducting, optical and nanomechanical resonators reach their remarkable Q by reducing loss at its source through materials, cryogenic cooling or nanofabrication~\cite{megrant_planar_2012,romanenko_three-dimensional_2020,armani_ultra-high-q_2003,ghadimi_elastic_2018}. Our results open a different route. In our circuit, the individual components retain finite dissipation and fabrication tolerances, and gain compensates their loss, yet their coupled spectral response supports an ultrahigh-Q resonance near EP$_\mathrm{c}$. The resonance can be switched on and off by a small change of the coupling $\kappa$ or the drive amplitude $V_\mathrm{Amp}$ (Fig.~\ref{figMain_Experiment}d,e), and its frequency is set by standard inductors and capacitors. An ultrahigh-Q resonance thus becomes a reconfigurable property of a network rather than a fixed property of a material.

	The Q reported here is the spectral Q of a driven, gain-compensated resonator. It quantifies the resonance linewidth relative to the operating frequency, a relevant figure of merit for filtering and sensing~\cite{gregory_q-factor_2022,vahala_optical_2003}, while the coupling and amplitude sweeps characterize the associated parametric response. In active resonators, gain compensation can narrow the linewidth but also amplify fluctuations~\cite{kuhn_200_1996,chapeau-blondeau_input-output_1997}, and near conventional EPs, noise enhancement can likewise offset the benefit of the increased response~\cite{wang_petermann-factor_2020,lau_fundamental_2018,langbein_no_2018,duggan_limitations_2022}. In the topological PT lattice, by contrast, the measured impedance is both most sensitive and least noisy at the same coupling (Methods). The band gap thus protects the linewidth of an EP resonance against the fluctuations of its own gain, while preserving the steep parametric response that makes EPs attractive.

	Previous efforts to improve resonators with topological concepts have targeted other quantities. Topological charges of bound states in the continuum make a radiation-limited Q robust against fabrication imperfections~\cite{jin_topologically_2019}, and lattices with gain have been used to select robust lasing modes~\cite{schomerus_topologically_2013,weimann_topologically_2017,st-jean_lasing_2017,parto_edge-mode_2018,stegmaier_topological_2021}. Here, the topological band gap instead acts on an EP of a driven resonator below the oscillation threshold and suppresses the contribution of local gain fluctuations to the linewidth. The protected quantity is therefore the linewidth of the resonance rather than its radiation loss or the profile of a mode. Because Eq.~\eqref{EqDressedDimer} holds for any two lattices coupled at a single site, this protection is not tied to our circuit. Its strength is set by the gap edge $\omega(c_2-c_1)$ and the band edge $\omega(c_1+c_2)$ of the SSH rings, where $c_1$ and $c_2$ are their alternating coupling capacitors. These two edges fix the broadening coefficient at EP$_\mathrm{c}$, $\eta_\mathrm{c}$, the part of a gain fluctuation that reaches the linewidth, and its zero at the gain--loss strength $1/R^*=\omega\sqrt{(c_2-c_1)(c_1+c_2)}$, so that the band gap itself becomes a design parameter for the linewidth of a resonance.

	Two points qualify our result. First, the Q reported here measures the linewidth of a driven, gain-compensated resonance below the oscillation threshold, rather than the free decay time of a passive resonator, and the circuit does not oscillate without a drive (Supplementary Section~\ref{SuppMatTemporalWaveforms}). Second, its sub-hertz linewidth approaches the level at which small noise and jitter affect the measurement, and resolving these contributions will show how close the reported Q is to the intrinsic limit of the driven response. A wider band gap, which moves $R^*$ towards stronger gain and loss where the parasitic dissipation matters less, and more uniform active elements offer clear routes to even higher Q, and feedback control of $\kappa$ or $V_\mathrm{Amp}$ could hold the circuit at its optimal operating point.

	\section{Outlook}

	For macroscopic resonators, this network-based approach complements the established material-based routes. Quartz crystals, for example, reach $Q$ values of $10^4$ to $10^6$ at room temperature and remain the standard for frequency control~\cite{friedt_introduction_2007,goryachev_extremely_2012}. In photonic crystals and metasurfaces, engineered interference can suppress radiation loss through bound states in the continuum~\cite{hsu_observation_2013,koshelev_asymmetric_2018}. What sets our approach apart is that it keeps the resonance narrow even as the individual components fluctuate. It therefore brings ultrahigh $Q$ within reach of discrete-element networks and metamaterials built from lossy, imperfect components.

	In particular, the approach may benefit integrated electronics, where resistive and substrate losses typically limit the $Q$ of on-chip inductors to a few tens~\cite{murphy_implicit_2017,yang_high-performance_2016,washio_02-m_2002}. Frequency references and narrowband filters therefore still rely on separate quartz and acoustic resonators~\cite{hagelauer_microwave_2023}. Our circuit already uses transistor-based negative resistances, as in on-chip LC oscillators~\cite{jamil_tunable_2007}, and coupled inductors and capacitors that standard chip processes can fabricate. Since the mechanism depends on the spectral arrangement of the coupled modes rather than on a specific frequency, GHz resonators for wireless communication are a promising extension. Such an extension, however, remains to be demonstrated, as it requires active elements that remain stable and low in noise over the bandwidth of the lattice, as well as a layout that keeps parasitic couplings and signal delays small compared with the intended couplings.

	At higher frequencies, gain fluctuations can become a more important source of disorder. Lithography keeps static fabrication errors small, and residual detuning can be corrected by tuning. Resonator losses, however, can grow, from resistive losses on chip to ohmic damping in metals at optical frequencies. Compensating for these losses requires stronger gain, which can in turn increase gain fluctuations. The band gap then limits the resulting broadening of the resonance. Moreover, $\eta_\mathrm{c}$ and $R^*$ depend only on the ratio of the gain--loss strength to the couplings (Eqs.~\eqref{EqDressedY} and~\eqref{EqEPcShare}), so the mechanism can extend to frequencies at which the couplings are comparable to the loss. Plasmonic SSH chains milled into monocrystalline gold, for instance, reach couplings of about a tenth of the resonance frequency. Sub-nanometre control of the spacings allows the alternating couplings to open a band gap wide enough to separate the topological mid-gap modes from the bulk bands~\cite{schurr_plasmonic_2025}. Adding gain to one such chain, as in plasmonic nanolasers~\cite{azzam_ten_2020}, and coupling it to a lossy chain through a single bridge could therefore, in principle, realize the topological PT lattice at optical frequencies within a footprint of a few micrometres, although the noise of the optical gain medium would then add to the gain fluctuations. Likewise, photonic chips with gain--loss microrings and topological lattices with gain~\cite{hodaei_parity-timesymmetric_2014,parto_edge-mode_2018,bandres_topological_2018}, superconducting circuits with SSH arrays and parametric gain~\cite{kim_quantum_2021,aumentado_superconducting_2020}, and acoustic, optomechanical and magnonic systems with EPs~\cite{ding_emergence_2016,xu_topological_2016,zhang_observation_2017} offer similar routes. In each case, the broadening coefficient $\eta$ in Eq.~\eqref{EqGainShare} gives a measure of the resonance's sensitivity to local fluctuations before the device is built.

	\clearpage

	\section{Methods}
	\renewcommand{\figurename}{\textbf{Extended Data Fig.}}
	\setcounter{figure}{0}
	\renewcommand{\theHfigure}{ED\arabic{figure}}

	\subsection{Supporting measurements}

	\subsubsection{Noise analyses and stability of the ultrahigh-Q resonance}

	\begin{figure*}[t]
		\centering
		\includegraphics[width=1\textwidth]{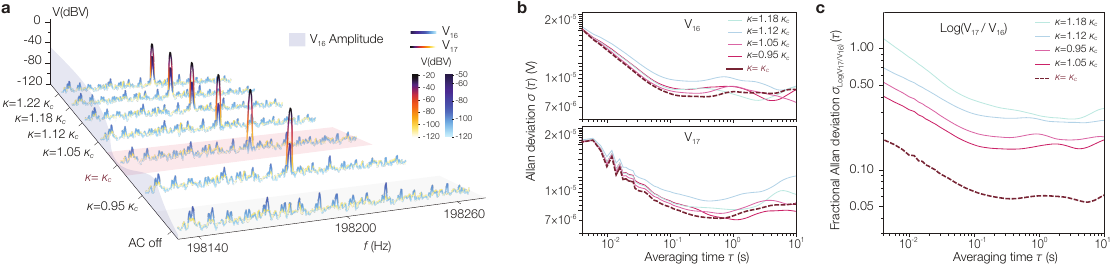}
		\caption{\textbf{Noise suppression and stability of the measured ultrahigh-Q resonance.}
			\textbf{a,} Voltage spectra of $V_{17}$ (injection, warm colours) and $V_{16}$ (extraction, cool colours) for six couplings from $0.95\,\kappa_c$ to $1.22\,\kappa_c$ ($\kappa_c=4.18$\,nF) and without AC excitation (AC off) show that, as $\kappa$ approaches $\kappa_c$ (pink shading), $V_{16}$ drops towards the noise floor at $\approx -120\,\mathrm{dBV}$. Because $Z \simeq R_\mathrm{shunt} V_{17}/V_{16}$, the fluctuations at the noise floor can substantially affect the impedance amplitude, which becomes crucial as $V_{16}$ vanishes at $\kappa_c$.
			\textbf{b,} The overlapping Allan deviation $\sigma(\tau)$ (Eq.~\eqref{EqMethodsAllanDev}) of the amplitudes of $V_{16}$ (top) and $V_{17}$ (bottom) for five couplings reveals that both signals show a white-noise decay and a stable plateau at $\sim 10^{-5}$--$10^{-6}$ for $\tau \gtrsim 0.1\,\mathrm{s}$, and $\kappa = \kappa_c$ (dashed curve) follows the same trend as the other couplings.
			\textbf{c,} The fractional Allan deviation of the measured impedance, that is, of $\log(V_{17}/V_{16})$, instead has a clear minimum at $\kappa_c$ (dashed curve), because the fluctuations of the two voltages are most strongly correlated there and cancel in their ratio, much like common-mode rejection.}
		\label{FigMethods_Noise}
	\end{figure*}

	In our system, the ultrahigh-Q resonance appears as a vanishing voltage at node~16 in the two-point impedance measurement, $Z_{17,16}\simeq R_\mathrm{shunt}V_{17}/V_{16}$ (Eq.~\eqref{EqMethodsZShunt}), where $V_{16}$ and $V_{17}$ are the measured voltages at the loss node~16 and the gain node~17. Extended Data Fig.~\ref{FigMethods_Noise}a shows the measured spectral response, obtained from an FFT-based spectrum analyser, for several values of the coupling capacitance, including the ultrahigh-Q condition at $\kappa=\kappa_c=4.18$\,nF. Without AC excitation (``AC off''), the measured spectra show a finite noise floor at approximately $-120\,\mathrm{dBV}$, which is amplified by the gain elements and appears as broadband fluctuations. This gain-amplified noise is negligible away from the high-Q regime, but it becomes relevant at $\kappa=\kappa_c$, where the amplitude of $V_{16}$ approaches the noise floor. In this limit, even small fluctuations of $V_{16}$ can cause large variations of the impedance through this ratio. The ultrahigh-Q regime is therefore, in principle, highly susceptible to noise unless the noise is suppressed (compare the 2-Node PT dimer in Supplementary Section~\ref{SuppMat2NodePT}).

	To characterize the noise and the long-term stability of the system, we analyse the overlapping Allan deviation $\sigma(\tau)$~\cite{allan_statistics_1966,riley_handbook_2008} of the measured voltage signals,
	\begin{equation}
		\sigma(\tau) = \sqrt{\frac{1}{2(M-2m+1)} \sum_{n=1}^{M-2m+1} \left( \bar{y}_{n+m} - \bar{y}_n \right)^2 },
		\label{EqMethodsAllanDev}
	\end{equation}
	where $M$ is the total number of samples and $\bar{y}_n$ denotes the $n$th overlapped average of the time series over an integration time $\tau = m/f_s$. Each signal was recorded continuously for five minutes (150,000 samples) at a sampling rate of $f_s=500\,\mathrm{Hz}$ under steady operating conditions. The analysis was applied separately to the individual voltage amplitudes, $V_{16}$ and $V_{17}$, and to their ratio $V_{17}/V_{16}$, which corresponds to the measured impedance.

	Extended Data Fig.~\ref{FigMethods_Noise}b shows the Allan deviation of the individual voltage signals for several coupling capacitances near $\kappa_c$. At short averaging times, both $V_{16}$ and $V_{17}$ show the decay characteristic of white noise, followed by a gradual saturation beyond $\tau \sim 10^{-1}\,\mathrm{s}$. For all couplings, the Allan deviation stays at the level of $\sim 10^{-5}$--$10^{-6}$, which indicates stable long-term operation over the whole measurement window. The ultrahigh-Q condition at $\kappa=\kappa_c$ follows nearly the same trend as the neighbouring couplings, which shows that the enhanced parametric sensitivity does not add long-term fluctuations to the individual voltage channels.

	Saliently, the Allan deviation of $\log(V_{17}/V_{16})$, the fractional Allan deviation of the measured impedance, shows a quantitatively important feature (Extended Data Fig.~\ref{FigMethods_Noise}c). Unlike the individual signals, the ratio has a pronounced minimum precisely at $\kappa = \kappa_c$, and its Allan deviation increases systematically as $\kappa$ is tuned away from $\kappa_c$ in either direction. The measured impedance therefore fluctuates least at the coupling where its response to $\kappa$ is steepest. The Allan deviations of $V_{16}$ and $V_{17}$ hardly change with $\kappa$ (Extended Data Fig.~\ref{FigMethods_Noise}b), whereas $V_{16}$ itself is smallest at $\kappa_c$ (Extended Data Fig.~\ref{FigMethods_Noise}a). On its own, the fractional fluctuation of $V_{16}$ would therefore be largest at $\kappa_c$. The minimum of the ratio thus shows that the fluctuations of the two voltages are most strongly correlated at $\kappa_c$, so that they largely cancel in their ratio.

	Finally, we address the relation between the near-zero eigenvalue condition and the lasing threshold. In gain--loss systems, EPs may coincide with or lie near the lasing threshold, where gain saturation drives the system into a nonlinear, self-oscillating regime~\cite{hashemi_linear_2022,ji_tracking_2023,alizadeh_lasing_2025,kullig_exceptional_2025,el-ganainy_exceptional_2014}. In our circuit, the negative-resistance elements operate far below gain saturation throughout all measurements, and saturation sets in only at amplitudes substantially larger than those required to reach the near-zero eigenvalue condition. Their weak amplitude dependence shifts the resonance slightly (see `Dependence of the resonance on the drive amplitude' below) but does not distort the waveforms, which remain sinusoidal (Supplementary Section~\ref{SuppMatTemporalWaveforms}). Consistently, no self-oscillation is observed without external driving, and the eigenvalue trajectories pass smoothly through the near-zero regime. All measurements are therefore performed below the oscillation threshold.

	\subsubsection{Measured admittance spectra}

	\begin{figure*}[t]
		\centering
		\includegraphics[width=0.9\linewidth]{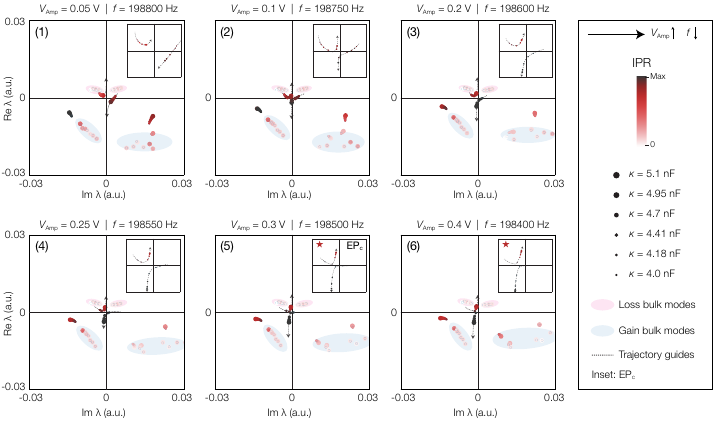}
		\caption{\textbf{Experimental admittance traces of the central exceptional point (EP$_\mathrm{c}$).} The admittance eigenvalues $\lambda$ are extracted from voltage-transfer measurements (Methods) for six couplings $\kappa$ (marker size, 4.0 to 5.1\,nF) and coloured by their inverse participation ratio (IPR). From (1) to (6), $V_\text{Amp}$ increases from 0.05 to 0.4\,V while $f$ decreases from 198.80 to 198.40\,kHz (panel titles), so that their opposite shifts along the imaginary axis nearly cancel. The bulk modes of the gain and loss rings stay at negative and positive real admittance (blue and pink ellipses), whereas the localized coupling modes (high IPR) undergo the imaginary-to-real transition of EP$_\mathrm{c}$ near the origin (dotted guides), with a residual avoided crossing (insets). The small net shift brings one coupling-mode trajectory to zero admittance in (5) and (6) (star).}
		\label{FigMethods_Spectra}
	\end{figure*}

	To probe the eigenvalue spectra associated with the ultrahigh-Q impedance resonance, we reconstruct an effective Laplacian-like operator from voltage-transfer measurements. In principle, the circuit Laplacian at a fixed excitation frequency $\omega=2\pi f$ can be obtained from independently measured voltage and current response matrices via $J(\omega)=I(\omega)V^{-1}(\omega)$~\cite{hofmann_chiral_2019,helbig_generalized_2020}. This approach is reliable for passive or weakly sensitive circuits, but it becomes problematic in circuits with gain, where even weak loading by the measurement renormalizes the effective onsite admittances and suppresses the most sensitive gain modes near vanishing admittance. As discussed in Supplementary Section~\ref{SuppMatVoltageTransfer}, shunt-based current sensing is therefore incompatible with Laplacian reconstruction in the present lattice.

	To avoid these limitations, we adopt a voltage-transfer measurement scheme implemented with network analysers. A sinusoidal voltage is injected directly into a selected node without a series shunt resistor, while the complex voltage responses at all nodes are recorded through high-impedance measurement channels. For each excitation node $i$, we measure the dimensionless complex voltage transfer ratios
	\begin{equation}
		T_{ji}(\omega)=\frac{V_j^{(i)}(\omega)}{V_i^{(i)}(\omega)},
		\label{EqMethodsVoltageTransferMatrix}
	\end{equation}
	which together form the $32\times32$ voltage-transfer matrix $\tilde{V}(\omega)$. This matrix encodes the linear voltage response of the grounded network up to an overall scale set by the excitation and measurement coupling. To analyse the spectral structure, we apply a bilinear transformation inspired by the impedance--scattering relation~\cite{gregory_q-factor_2022} and define an effective impedance matrix
	\begin{equation}
		Z_{\mathrm{eff}} = Z_0 (\mathds{I} + \tilde{V})(\mathds{I} - \tilde{V})^{-1},
		\label{EqMethodsEffectiveImpedanceMatrix}
	\end{equation}
	from which we define an effective Laplacian-like operator
	\begin{equation}
		J_{\mathrm{eff}} = Z_{\mathrm{eff}}^{-1}.
		\label{EqMethodsEffectiveLaplacianMatrix}
	\end{equation}
	Here, $Z_0 = 50\,\Omega$ is a reference impedance and $\mathds{I}$ denotes the identity matrix. Since $J_{\mathrm{eff}}$ is constructed solely from voltage-transfer measurements, it should be interpreted as an effective operator rather than the exact circuit Laplacian. Accordingly, the extracted eigenvalues are reported in arbitrary units.

	Experimentally, we use two synchronized Analog Discovery Pro ADP3450 devices, each providing four analog input channels with a sampling rate of 125\,MS/s per channel, which gives eight synchronized channels in total. The built-in waveform generators excite the circuit, and the network-analyser interface provides phase-resolved voltage acquisition. The excitation frequency is swept from 198.3\,kHz to 199.3\,kHz in 21 discrete steps. At each frequency point, a settling time of 50\,ms is applied before data acquisition, after which each signal is sampled over at least 128 periods to extract the voltage amplitude and phase accurately.

	To minimize instrumental asymmetries, all channels are connected with identical 50\,$\Omega$, 50\,cm coaxial cables, which remain fixed throughout the measurements. For each value of the coupling capacitance $\kappa$, a full $32\times32$ set of voltage-transfer measurements is performed. This procedure is repeated for six gate voltages of the variable coupling capacitor, corresponding to capacitance values of $(4.00,\,4.18,\,4.41,\,4.70,\,4.95,\,5.10)\,\mathrm{nF}$. The admittance eigenvalues $\lambda$ are then obtained by solving the eigenvalue problem $\det(J_{\mathrm{eff}}-\mathds{I}\lambda)=0$.

	The measured admittance eigenvalues allow us to trace the spectral origin of the ultrahigh-Q resonance through the evolution of the eigenvalues near zero admittance (Extended Data Fig.~\ref{FigMethods_Spectra})~\cite{dordevic_non-hermitian_2026}. The spectra are shown for different excitation amplitudes $V_\text{Amp}$, excitation frequencies $f$ and six discrete coupling capacitances $\kappa$. The marker size denotes $\kappa$, increasing from $\kappa = 4.0\,\mathrm{nF}$ (smallest dots) to $\kappa = 5.1\,\mathrm{nF}$ (largest dots), and the colour represents the inverse participation ratio (IPR), defined as $\text{IPR} = \sum_i|\psi_i|^4 / (\sum_i|\psi_i|^2)^2$, where $\psi$ denotes an eigenvector of the extracted Laplacian. Larger IPR values indicate stronger spatial localization.

	As noted above, the measured spectra should be interpreted as effective admittance pseudo-spectra, obtained in a measurement configuration that differs from the one used for the impedance measurements (Supplementary Section~\ref{SuppMatGrounding}).

	The spectra depend on both the drive frequency and the drive amplitude. A larger $V_\text{Amp}$ lowers the resonance frequency of the nodes (see `Dependence of the resonance on the drive amplitude' below) and thereby shifts the spectrum towards positive imaginary admittance, whereas a lower drive frequency shifts it towards negative imaginary admittance (Supplementary Section~\ref{SuppMatFreqShift}). We therefore increase $V_\text{Amp}$ while we reduce $f$, from $V_\text{Amp}=0.05$\,V at $f=198.80$\,kHz in (1) to $V_\text{Amp}=0.4$\,V at $f=198.40$\,kHz in (6) (panel titles and arrows in Extended Data Fig.~\ref{FigMethods_Spectra}), so that the two shifts nearly cancel and the spectra change only slightly. The remaining net change is a small shift of the whole spectrum towards negative imaginary admittance, set by the drive frequency. While the idealized model predicts a broad evolution of the coupling modes over a $1\,\mathrm{nF}$ change of $\kappa$, the intentional gain--loss imbalance, which can be read off from the unequal real-axis offsets of the gain and loss bulk modes, reduces their mobility in the experiment. The available tuning range is nevertheless sufficient to steer the modes into the near-zero regime (Supplementary Section~\ref{SuppMatMeasuredSpectra}).

	In all six panels, the central EP ($\mathrm{EP_c}$) lies close to the spectral origin. The bulk modes of the lossy SSH lattice (pink ellipses) cluster at positive real admittance, and those of the gain SSH lattice (blue ellipses) at negative real admittance. The coupling modes near the origin have much larger IPR values, which reflects their localization at the coupling interface, and their evolution with $\kappa$ is traced by the dotted lines. As $\kappa$ increases, the coupling modes undergo the characteristic imaginary-to-real (Im-to-Re) transition associated with EP$_\mathrm{c}$. The insets show magnified views of the near-zero region, where the residual avoided crossing is visible.

	From (1) to (6), the net shift towards negative imaginary admittance brings one branch progressively closer to the origin, most clearly in (5) and (6) (star), where the eigenvalue approaches the near-zero regime. The impedance reconstructed from the same measured spectra shows this approach to the origin as an enhancement of the impedance, for the topological PT circuit (Supplementary Section~\ref{SuppMatImpedanceFromSpectra}) and for the 2-Node PT circuit (Supplementary Section~\ref{SuppMat2NodePT}). Through combined tuning of $\kappa$, $f$ and $V_\text{Amp}$, the localized modes can thus be steered towards zero admittance.

	\subsubsection{Dependence of the resonance on the drive amplitude}

	\begin{figure}
		\centering
		\includegraphics[width=\linewidth]{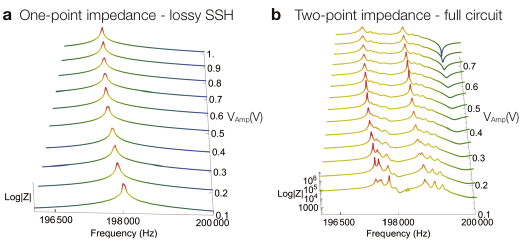}
		\caption{\textbf{Dependence of the resonance on the drive amplitude.}
			The excitation amplitude shifts the resonance towards lower frequencies and therefore provides an additional tuning parameter alongside $\kappa$.
			\textbf{a,} The one-point impedance measured at one bulk node of the detached lossy SSH circuit shows that the resonance peak shifts towards lower frequencies as the excitation amplitude $V_\text{Amp}$ increases.
			\textbf{b,} The two-point impedance measured between the coupling nodes of the full gain--loss SSH circuit as a function of $V_\text{Amp}$, with all other parameters fixed, shows the same shift of the impedance peaks, which helps to reach the near-zero-admittance regime. Here, $\kappa = 4.23\,\text{nF}$.
			}
		\label{FigMethods_VoltageAmp}
	\end{figure}

	The resonance of our circuit depends on the applied excitation amplitude $V_\text{Amp}$. While the nominal values of the passive components ($c_1$, $c_2$, $c_g$ and $R$) remain largely stable, weak second-order nonlinearities, such as the amplitude dependence of the incremental permeability of the composite inductor cores and the amplitude-dependent response of the negative-impedance converters, shift the collective resonance~\cite{li_observation_2025,li_nonlinear_2026}.

	To characterize this shift, we measured the one-point impedance of a node in the isolated lossy SSH circuit. As shown in Extended Data Fig.~\ref{FigMethods_VoltageAmp}a, the impedance resonance, initially centred at $\approx 198\,$kHz, shifts systematically towards lower frequencies as $V_\text{Amp}$ increases from $0.1\,$V to $1\,$V.

	A similar shift appears in the two-point impedance spectra of the full gain--loss circuit (Extended Data Fig.~\ref{FigMethods_VoltageAmp}b). With $\kappa$ held constant, increasing the voltage amplitude shifts the resonance peaks towards lower frequencies. This behaviour identifies $V_\text{Amp}$ as a useful tuning parameter alongside $\kappa$. Adjusting the excitation amplitude therefore provides an additional degree of freedom for steering the relevant eigenvalue towards zero admittance and accessing the ultrahigh-Q regime.

	\subsection{Circuit implementation and measurement setup}

	\subsubsection{Circuit design and fabrication}

	The experimental realization of high-Q resonances in a topological PT-symmetric circuit requires three considerations, namely (i) minimizing component tolerances and parasitic elements, since gain components also amplify noise, (ii) ensuring circuit stability and signal integrity in the presence of active elements, and (iii) achieving the desired gain and loss strengths in the presence of parasitics.

	With these considerations, we designed three separate 1.6\,mm-thick FR-4 printed circuit boards (PCBs), namely a gain SSH circuit, a coupling module and a lossy SSH circuit. For the gain SSH board, we used a four-layer PCB stack-up with two 2\,oz inner copper layers dedicated to the positive and negative DC supply, respectively. Power planes instead of traces for the DC supply ensure low-impedance current return paths and minimize crosstalk between signal lines. The power pins of the operational amplifiers (op-amps) were connected to the power layers through tented vias with a maximum drill diameter of 0.3\,mm to reduce parasitic inductance. Both outer layers were covered with 2\,oz copper ground planes. To improve signal integrity and reduce trace impedance, we avoided sharp angles in all routing and used a minimum trace width of 0.6\,mm where possible, and 1\,mm for power traces in the coupling module. The lossy SSH and coupling-module PCBs follow the same design principles as the gain SSH board but were implemented as two-layer boards. For the SSH circuits, we adopted a circular ring layout for a uniform structure. To access node voltages, we used vertical through-hole PCB-mount SMA coaxial RF connectors on all boards. The three PCBs were interconnected using jack edge-mount SMA coaxial RF connectors.

	\begin{figure}
		\centering
		\includegraphics[width=\linewidth]{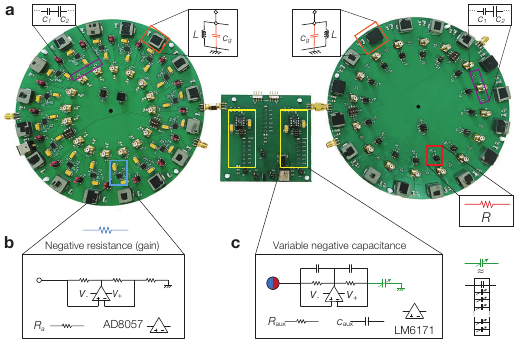}
		\caption{\textbf{Topological PT circuit board.}
			The topological PT lattice of Eq.~\eqref{EqMainTopoPT} is realized on printed circuit boards with standard surface-mount components.
			\textbf{a,} The circuit consists of two SSH rings of 8 unit cells each, coupled between node 16 of the lossy ring (B sublattice) and node 17 of the gain ring (A sublattice), and the insets show the elements of the nodes. The nominal values are $c_1=1.005$\,nF (intracell), $c_2=2.212$\,nF (intercell), $c_g=333$\,nF and $L=1.915\,\mu$H.
			\textbf{b,} Each gain node uses a negative-impedance converter with an AD8057 operational amplifier and $R=R_a=180\,\Omega$.
			\textbf{c,} The variable negative capacitance, which cancels the extra admittance of $\kappa$ at the coupling nodes, connects a bank of eleven variable capacitors and a fixed capacitor between the inverting input of an LM6171 operational amplifier and ground, with $R_\mathrm{aux}=191\,\Omega$ and $c_\mathrm{aux}=33$\,pF.}
		\label{FigMethods_Schematic}
	\end{figure}

	The detailed circuit schematic is shown in Extended Data Fig.~\ref{FigMethods_Schematic}a. The two SSH circuits share an identical bulk structure and differ only in the onsite gain and loss elements. All components were assembled by surface-mount soldering. For high stability and precision, all capacitors have C0G (NP0) ceramic dielectrics. The intra-cell and inter-cell coupling capacitors are $c_1 = 1.005$\,nF (Kemet C0603C102F2GECTU) and $c_2 = 2.212$\,nF (Kemet C0603C222F3GECTU), respectively. In addition, each node is connected to ground through a grounding capacitor $c_g = 333$\,nF (Murata GRM31C5C1E334GE01L), which also serves to stabilize the operational amplifiers (Supplementary Section~\ref{SuppMatNICStability}).

	Each node is also grounded through an inductor with nominal inductance $L = 1.915\,\mu\text{H}$ (Vishay Dale IHLP5050FDER2R2M0P). Although these inductors are magnetically shielded, they were placed uniformly along the circular layout and spaced adequately to further suppress variations due to mutual coupling. To balance gain and loss and minimize resistive losses, particular care was taken to select inductors with a low DC resistance (2.9\,m$\Omega$ per unit).

	The manufacturer-specified tolerances are 1\% for $c_1$ and $c_2$ and 2\% for $c_g$, and the inductors have a nominal tolerance of 20\%. To reduce this variation, each element was characterized individually with a BK Precision 891 bench LCR meter before assembly, which allowed us to preselect units and reduce the effective tolerance to approximately 2\%. The component values given in this work are therefore the measured values of the selected units rather than the datasheet values. To fine-tune the resonant frequency of each node and further reduce residual variations, we then applied small magnets and ferrite sheets to the inductors after assembly. This reduced the spread of the resonant frequencies of the LC nodes to 0.05\%. Full details of this tuning procedure are given in Supplementary Section~\ref{SuppMatNodeTuning}.

	\subsubsection{Calibration of gain and loss strength}

	Gain requires active components, whereas loss is implemented with regular resistors. The circular layout and the symmetric circuit design help keep the effective loss uniform across the nodes. To estimate the parasitic contribution of traces, contacts and cabling, we assembled the relevant components on a separate single-node test board and measured the residual resistance with the Kelvin four-wire method. The measured parasitic resistance was approximately $12\,\text{m}\Omega$ per node. Although this value is several orders of magnitude smaller than the nominal gain and loss resistors, its effect is magnified by the low reactance of the grounding capacitors and inductors at the operating frequency.

	In particular, the grounding capacitor $c_g=333\,\text{nF}$ and the inductor $L=1.915\,\mu\text{H}$ both have reactances of only $\approx 2.4\,\Omega$ near the operating frequency of $198\,\text{kHz}$. In such low-impedance branches, even milliohm-scale series resistances introduce a non-negligible dissipative admittance. A series resistance $r$ in a branch of reactance $X$ adds the shunt conductance $r/X^2$, so the parasitic resistance of about $12\,\text{m}\Omega$ per node corresponds to a dissipative admittance of about 2\,mS, consistent with the difference $(1/180-1/270)\approx1.9$\,mS between the negative and positive resistances used below. The gain--loss balance can therefore not be set from the nominal resistor values alone, since the effective admittances satisfy $|-1/R_\text{gain}+Y_\text{parasitic}|\neq |1/R_\text{loss}+Y_\text{parasitic}|$. To compensate for the additional dissipative admittance, the passive loss was implemented with $270\,\Omega$ resistors, whereas the negative-resistance converters used $180\,\Omega$ resistors (Panasonic Electronic Components ERA-3AEB181V). This choice brings the effective gain and loss close to balance once the component ESRs and parasitic series resistances are included. The circuit was intentionally kept slightly gain-favoured, which helps the relevant eigenvalue branch pass the residual avoided crossing near the exceptional point and approach the zero-admittance axis (Supplementary Sections~\ref{SuppMatSpectraOuterEPs} and~\ref{SuppMat2NodePT}).

	Overall, once the component ESRs, the parasitic series resistances and the loading of the measurement are included, the $270\,\Omega$ loss resistors and the $180\,\Omega$ negative resistances give a circuit that is close to balance with a slight excess of gain. Consistently, the ideal resistance of $270\,\Omega$ used in the analytical and numerical models corresponds to an effective resistance close to $180\,\Omega$ in LTspice simulations that include these parasitics. The small gain bias appears in the measured admittance spectra as a shift of the coupling-mode eigenvalues towards negative real admittance (Supplementary Section~\ref{SuppMatSpectraOuterEPs}).

	\subsubsection{Negative resistance and negative capacitance}

	The gain SSH circuit and the coupling module contain operational amplifiers configured as negative-impedance converters with current inversion (INICs), which implement negative resistance (gain) and negative capacitance~\cite{hofmann_chiral_2019,helbig_generalized_2020}. In the gain SSH circuit, we used AD8057ARZ (Analog Devices Inc.) 8-SOIC op-amps, selected for their low input capacitance, high slew rate and stable performance in high-frequency INIC configurations. To obtain phase inversion with unity gain, the two feedback resistors ($R_a$ in Extended Data Fig.~\ref{FigMethods_Schematic}b) of each INIC were matched to the negative resistance value, $R_a = R$.

	To ensure stable operation and suppress noise in all INICs, each op-amp was equipped with a pair of decoupling capacitors (22\,$\mu$F tantalum and 0.1\,nF ceramic) between the power supply pins and ground. This combination filters both low- and high-frequency voltage fluctuations on the supply rails. A regulated $\pm6$\,VDC supply was provided by the dual-output power supply of the Analog Discovery Pro ADP5250 (Digilent Inc.).

	In the coupling module, two additional INICs between the coupling nodes and ground realize a variable negative capacitance (Extended Data Fig.~\ref{FigMethods_Schematic}c). This design compensates for the additional capacitive load introduced by the coupling and keeps the onsite admittance uniform across both SSH circuits. For these INICs, we used LM6171BIN (Texas Instruments) 8-DIP op-amps, which offer a high output current and a fast settling time, making them suitable for capacitive inversion.

	\subsubsection{Variable inter-circuit coupling}

	A variable inter-circuit coupling of approximately 1.1\,nF was implemented with eleven surface-mount variable capacitors (Murata LXRW19V201-058), each with a tunable range of 100--200\,pF, connected in parallel. The capacitance of each capacitor varies with the applied control voltage between 0 and 5\,V, which allows precise tuning of the coupling between the two SSH circuits. Together, the eleven capacitors give 1.1 to 2.2\,nF. They were connected in parallel with a fixed ceramic capacitor of 2.9\,nF, so that the coupling has a fixed part of 4\,nF, drawn as one capacitor in Fig.~\ref{figMain_Experiment}b, and is tuned continuously from 4.0 to 5.1\,nF. All variable capacitors were biased from the same DC supply line to ensure an identical control voltage.

	\subsubsection{Two-point impedance measurement}

\begin{figure}
		\centering
		\includegraphics[width=0.9\linewidth]{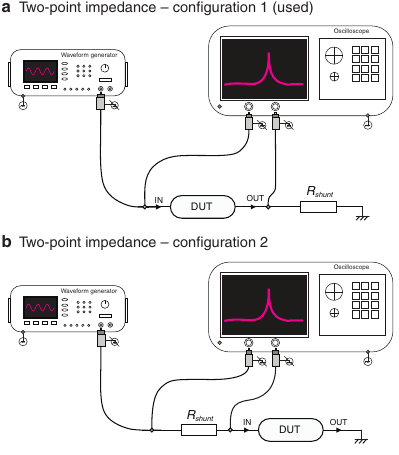}
		\caption{\textbf{Two-point impedance measurement configurations.}
			\textbf{a,} In the configuration used in this work, the shunt resistor $R_\mathrm{shunt}$ lies between the device under test (DUT) and ground, so it does not load the injection node.
			\textbf{b,} In the common configuration, $R_\mathrm{shunt}$ lies between the source and the DUT, where it loads the injection node and can detune the sensitive modes of the ultrahigh-Q resonance.}
		\label{FigMethods_ImpConfig}
	\end{figure}

	The two-point impedance is defined as $Z_{ab} = (V_a - V_b)/I$, where $V_a$ and $V_b$ are the voltages at nodes $a$ and $b$, and $I$ is the current flowing through the device under test (DUT). The current is usually measured through a shunt resistor $R_\mathrm{shunt}$ connected either between the source and the circuit or between the circuit and ground. Although the former configuration is more common, we adopted the latter (Extended Data Fig.~\ref{FigMethods_ImpConfig}a), because it avoids additional loading by the measurement equipment, which could otherwise perturb the node to which the probe is attached. In the configuration of Extended Data Fig.~\ref{FigMethods_ImpConfig}b, where the shunt resistor is placed between the source and the circuit, the shunt can load the injection node. Since the ultrahigh-Q resonance is highly sensitive to such parasitic influences, which detune the EP hybridization, we consistently used the configuration of Extended Data Fig.~\ref{FigMethods_ImpConfig}a. In this setup, the output voltage $V_b$ across the shunt resistor relates to the impedance as $Z = R_\mathrm{shunt}(V_a/V_b - 1)$. Near resonance, where $V_b \to 0$ and $Z \gg R_\mathrm{shunt}$, the impedance is approximately
	\begin{equation}
		Z \approx R_\mathrm{shunt} \frac{V_a}{V_b},
		\label{EqMethodsZShunt}
	\end{equation}
	which gives a sharp impedance peak as $V_b$ vanishes.

	The impedance was measured between the coupling nodes, nodes 16 and 17 of our circuit. We used Amphenol 50\,$\Omega$ coaxial RF cables for both the signal injection and the voltage measurement channels, without grounding their outer conductors. The measurements were performed with the impedance analyser function of the Analog Discovery Pro ADP5250 for the spectra of Fig.~\ref{figMain_Experiment}c and of the ADP3450, which provides a sampling rate of 125\,MS/s per channel, for the sweeps of Fig.~\ref{figMain_Experiment}d,e (both Digilent Inc.). A sinusoidal signal of constant amplitude (denoted $V_\text{Amp}$ in the main text) was applied at either node 16 or node 17 using the built-in arbitrary waveform generator of the same instrument. The current was measured through a 1\,M$\Omega$ shunt resistor between the node where the current was extracted and the ground of the measurement device. Because the waveform generator is single-ended, the current can be obtained directly from the voltage at the current extraction node. Each frequency sweep shown in Fig.~\ref{figMain_Experiment}d,e was performed from 197\,kHz to 200\,kHz with 5001 frequency points, that is, a step of 0.6\,Hz. At the sub-hertz level, small noise and jitter of the circuit and of the instruments shift the peak by amounts comparable to its width, which makes the measurement of such narrow lines challenging.

	For the measured spectra of Fig.~\ref{figMain_Experiment}c, Q is obtained from Lorentzian fits of $|Z(f)|$. In the two sweeps of Fig.~\ref{figMain_Experiment}d,e, whose step is comparable to the linewidth, the sharpest peak gives $\mathrm{Q}_\mathrm{max}=1.24\times10^5$ in both, at $(\kappa,V_\mathrm{Amp})=(4.18\,\mathrm{nF},88\,\mathrm{mV})$.

	\subsection{Theoretical model}

	\subsubsection{Circuit Laplacian of an SSH ring}

	Each SSH ring consists of $N$ identical LC resonators with the onsite admittance $Y_0=i\omega c_\Sigma+1/(i\omega L)$, where $c_\Sigma=c_g+c_1+c_2$, joined by alternating capacitors $c_1$ and $c_2$. Its circuit Laplacian~\cite{lee_topolectrical_2018,sahin_topolectrical_2025} is
	\begin{equation}
		\begin{split}
			J_\mathrm{SSH}={}&Y_0\,\mathds{I}_N-i\omega\sum_{n=1}^{N/2}\Big(c_1\ket{n,A}\bra{n,B}\\
			&+c_2\ket{n+1,A}\bra{n,B}+\mathrm{h.c.}\Big),
		\end{split}
		\label{EqMethodsSSH}
	\end{equation}
	where $n$ labels the unit cells, $A$ and $B$ are the two sites of a cell, and the ring is closed through $\ket{N/2+1,A}=\ket{1,A}$. The bulk eigenvalues of an isolated ring are $Y_0\pm i\omega|c_1+c_2e^{ik}|$, with $k$ in the Brillouin zone, so the two bands lie at $\omega|c_2-c_1|\leq|\mathrm{Im}\,(\lambda-Y_0)|\leq\omega(c_1+c_2)$, symmetric about the gap centre. In Eq.~\eqref{EqMainTopoPT}, the coupling node of the lossy ring is $a=(N/2,B)$, node 16 of the circuit, and that of the gain ring is $b=(1,A)$, node 17. The coupling therefore fixes the unit cells of Eq.~\eqref{EqMethodsSSH}, such that each coupling node forms a unit cell with its neighbour across $c_1$, and with this choice, each ring is in the topological phase of the SSH model with winding number one for $c_2>c_1$ (Supplementary Section~\ref{SuppMatCleanLimit}). An isolated ring under periodic boundary conditions is neither trivial nor nontrivial in an absolute sense, since its winding number depends on which bond is taken as intracell. Once coupled, however, every gapped ring ($c_1\neq c_2$) binds one mid-gap mode at its coupling node, for either ordering of $c_1$ and $c_2$, whereas the gapless ring ($c_1=c_2$) binds none. We therefore treat the gapped rings as topologically nontrivial and the gapless ring as the trivial reference (Supplementary Section~\ref{SuppMatTopoClass}). The onsite terms $\pm1/R$ shift the bands of the lossy and gain rings to positive and negative real admittance.

	\subsubsection{Impedance between the two lattices}

	The one-point impedance of an isolated ring, the impedance between one of its nodes and ground, follows from the local Green's function of its Bloch Laplacian (Supplementary Section~\ref{SuppMatOnePointZ}). It is the inverse of the input admittance of the main text, $Z_\pm=1/\mathcal{Y}(Y_\pm)$, where $Y_\pm=Y_0\pm1/R$ is the onsite admittance of the lossy ($+$) and gain ($-$) rings, and the sign function of Supplementary Eq.~\eqref{EqSuppMatSignFunction} selects the physical branch of the square root in Eq.~\eqref{EqDressedY}. Treating the coupling $\kappa$ as a perturbation of the isolated rings~\cite{economou_greens_2006,cserti_perturbation_2002,cserti_general_2026} gives the impedance between the two coupling nodes (Supplementary Section~\ref{SuppMatTwoPointZ}),
	\begin{equation}
		Z_{ab}(\omega,\kappa)=\frac{\mathcal{Y}(Y_+)+\mathcal{Y}(Y_-)-2i\omega\kappa}{\mathcal{Y}(Y_+)\,\mathcal{Y}(Y_-)+\omega^2\kappa^2}.
		\label{EqMethodsFinalZ}
	\end{equation}
	Its poles satisfy $\mathcal{Y}(Y_+)\,\mathcal{Y}(Y_-)+\omega^2\kappa^2=0$, where $Y_0=i\omega c_\Sigma+1/(i\omega L)$ and the input admittance of Eq.~\eqref{EqDressedY} is $\mathcal{Y}(y)=\sqrt{[y^2+\omega^2(c_2-c_1)^2][y^2+\omega^2(c_1+c_2)^2]}/y$. Replacing $Y_\pm\rightarrow Y_\pm-\lambda$ and setting $Y_0=0$ at the resonance frequency gives the eigenvalue condition of the main text, Eq.~\eqref{EqDressedDimer}.

	\subsubsection{Exceptional points of the coupled rings}

	The coupling needed for a lossless mode, Eq.~\eqref{EqHump}, has an extremum at the gap centre. Expanding $\ln|\mathcal{Y}(1/R+\lambda)|$ to second order in $\mathrm{Im}\,\lambda$ shows that this extremum is a maximum where $d^2\ln\mathcal{Y}/dy^2>0$ at $y=1/R$ and a minimum otherwise, with the curvature $d^2\kappa/d(\mathrm{Im}\,\lambda)^2=-\kappa_{\mathrm{EP_c}}\,d^2\ln\mathcal{Y}/dy^2$ that enters Eq.~\eqref{EqEPcShare}. For the dimer, $\mathcal{Y}=y$, and the centre is always a minimum, which gives a single EP. For the SSH rings, the band gap changes the shape of the curve. Writing $\mathcal{Y}^2=y^2+2\omega^2(c_1^2+c_2^2)+\omega^4(c_2^2-c_1^2)^2/y^2$, the last term, which exists only for a nonzero gap, makes $\mathcal{Y}$ grow as the node admittance $y$ approaches zero. For weak gain and loss and near the gap centre, this term dominates, the ring responds as a stiff element, and $|\mathcal{Y}|$ falls as $|\mathrm{Im}\,\lambda|$ moves away from zero towards the bulk bands. The centre is therefore a maximum. Far from the gap, $\mathcal{Y}\simeq y$ grows again with $|\mathrm{Im}\,\lambda|$, so the curve must pass through a minimum on each side of the maximum. Each of these three extrema is a coupling at which two lossless modes merge, which gives EP$_\mathrm{c}$ at the maximum and EP$_\mathrm{o1}$ and EP$_\mathrm{o2}$ at the two minima. With our capacitors, the centre remains a maximum for $R>182\,\Omega$. The minimum of $\mathcal{Y}$ on the real axis lies at $1/R^*=\omega\sqrt{(c_2-c_1)(c_1+c_2)}$, where $\kappa_{\mathrm{EP_c}}=2c_2$. At EP$_\mathrm{c}$, the $2\times2$ matrix of the coupling nodes is $\mathcal{Y}(1/R)\begin{psmallmatrix}1&-i\\-i&-1\end{psmallmatrix}$, a Jordan block with the single, self-orthogonal eigenvector $(1,-i)^\mathsf{T}$, and the eigenvectors of the full 32-node Laplacian coalesce at all three EPs~\cite{fritzsche_unveiling_2026}. For the gapless lattice ($c_1=c_2$), the centre is a single minimum for $R<398\,\Omega$, a range that contains our operating point. The derivations and numerical checks are given in Supplementary Section~\ref{SuppMatThreeEPs}.

	\subsubsection{Q factor, spectral width and broadening coefficient}

	As in the main text (Eq.~\eqref{EqQfactor}), the Q factor is the spectral Q of the driven response, $\mathrm{Q}=f_0/\Delta f$, where $f_0$ is the resonance frequency and $\Delta f$ is the full width at half maximum (FWHM) of the impedance power spectrum $|Z_{ab}(\omega)|^2$. Near a resonance, the eigenvalue $\lambda$ closest to the origin dominates Eq.~\eqref{EqImpEig}, and $Z_{ab}(\omega)\simeq A/\lambda(\omega)+Z_\mathrm{bg}$, where $A=(\psi_a-\psi_b)^2/\sum_j\psi_j^2$ is the residue of this eigenvalue and $Z_\mathrm{bg}$ collects the other eigenvalues, which vary slowly on the scale of the linewidth. We expand $\lambda(\omega)=\lambda_0+\partial_\omega\lambda\,(\omega-\omega_0)$ around the frequency $\omega_0$ at which $\mathrm{Im}\,\lambda$ vanishes. When $\partial_\omega\lambda$ is purely imaginary, $|Z_{ab}|^2\simeq|A|^2/[(\mathrm{Re}\,\lambda_0)^2+|\partial_\omega\lambda|^2(\omega-\omega_0)^2]$, a Lorentzian with the FWHM $\Delta\omega=2|\mathrm{Re}\,\lambda_0|/|\partial_\omega\lambda|$, which gives
	\begin{equation}
		\mathrm{Q}=\frac{\omega_0|\partial_\omega\lambda|}{2|\mathrm{Re}\,\lambda_0|}.
		\label{EqMethodsQLinewidth}
	\end{equation}
	The residue $A$ sets the height of the peak but not its width. The slope follows from first-order perturbation theory, $\partial_\omega\lambda=\psi^\mathsf{T}(\partial_\omega J)\psi/\psi^\mathsf{T}\psi$. For the circuits considered here, $\partial_\omega J=i\left(c_\Sigma+1/\omega^2L\right)\mathds{I}-iC_\mathrm{c}$, where $C_\mathrm{c}$ contains the coupling capacitances $c_1$, $c_2$ and $\kappa$ between the nodes. At $\omega\simeq\omega_0$ the first term equals $2ic_\Sigma$, and the second contributes at the level of $(c_1+c_2)/c_\Sigma\approx1\%$. Hence $\partial_\omega\lambda\simeq2ic_\Sigma$, which gives Eq.~\eqref{EqQLinewidth}, $\mathrm{Q}\simeq\omega_0c_\Sigma/|\mathrm{Re}\,\lambda_0|$. For the 32-node lattice and for the dimer at their operating points, with a gain change of 0.3\% or 1\% at the coupling node, the first expression agrees with the FWHM of the full $|Z_{ab}|^2$ to 0.1\%, and Eq.~\eqref{EqQLinewidth} to within 2\%. For a resonator described by a Hamiltonian with the complex eigenfrequency $\omega_0-i\gamma$, the role of $\lambda$ is played by $\gamma+i(\omega-\omega_0)$, so that $|\partial_\omega\lambda|=1$ and $\mathrm{Q}=\omega_0/2\gamma$.

	Disorder that varies slowly compared with a frequency sweep makes $\lambda_0$ a random variable. Each realization then has a Lorentzian line with $\mathrm{Q}=\omega_0c_\Sigma/|\mathrm{Re}\,\lambda_0|$. When the disorder spreads $\mathrm{Re}\,\lambda_0$ over the spectral width $\Delta W$, the largest net loss in the cloud bounds the linewidth, and for a cloud centred on the origin this net loss is $\Delta W/2$, which gives Eq.~\eqref{EqWidthQ}. The spread of $\mathrm{Im}\,\lambda_0$ shifts the line by $\mathrm{Im}\,\delta\lambda/(2c_\Sigma)$ in angular frequency without changing its width, but it broadens lines that are averaged over time.

	For a symmetric $J$, a small change $\delta$ of the admittance of node $b$ shifts a nondegenerate eigenvalue by $\delta\lambda=(\psi_b^2/\psi^\mathsf{T}\psi)\,\delta$ to first order, where $\psi$ is the right eigenvector and $\psi^\mathsf{T}$ the corresponding left eigenvector. The complex factor $\psi_b^2/\psi^\mathsf{T}\psi$ is the weight of the resonant mode on node $b$. Its real part is the broadening coefficient $\eta$ of Eq.~\eqref{EqGainShare}, and its imaginary part shifts the resonance frequency. Near an EP, $\psi^\mathsf{T}\psi\rightarrow0$, and the weight grows with the Petermann factor~\cite{wang_petermann-factor_2020}. Its closed forms for the dimer and the topological PT lattice are derived in Supplementary Sections~\ref{SuppMatTwoNodeReduction} and~\ref{SuppMatGainFluctuations}.

	The measured Q is further capped by measurement-level imperfections, mainly noise and jitter at the sub-hertz level and, for wide sweeps, the frequency step. A leakage admittance $Y_\mathrm{leak}$ in parallel with the circuit describes this cap with a single parameter, since it adds $AY_\mathrm{leak}$ to the net loss of the resonant eigenvalue (Supplementary Section~\ref{SuppMatLeakageModel}).

	\subsection{Numerical simulations}

	\subsubsection{Q factors in the clean limit}

	The Q factors in Fig.~\ref{figMain_CleanLimit}g,h are extracted from $|Z(f,\kappa)|$ (Eq.~\eqref{EqMethodsFinalZ}) on a uniform frequency grid of step $\delta f$. At each $\kappa$, we take the highest of up to four peaks and count the grid steps on either side of it until $|Z|$ drops to $|Z|_\mathrm{max}/\sqrt{2}$, which gives the FWHM and Q (Eq.~\eqref{EqQfactor}). When the true linewidth is narrower than $\delta f$, Q saturates at the resolution limit $f_0/(2\delta f)$. For Fig.~\ref{figMain_CleanLimit}h, every gain element has the resistance $269\,\Omega$ and every loss element $270\,\Omega$, in both the lattice and the dimer (inset). In the clean limit, this uniform imbalance moves every eigenvalue on the imaginary axis along the real axis by half of $1/269-1/270$\,S (Supplementary Fig.~\ref{FigSuppMatImbalanced}), which gives the common plateau $\mathrm{Q}\simeq2\omega_0c_\Sigma/(1/269-1/270)\approx6\times10^4$. The couplings at which a pole of $Z$ lies exactly on the real frequency axis, $4.4773$, $4.4960$ and $4.7299$\,nF for the lattice and $\kappa_c=1/(\omega_0\sqrt{R R_\mathrm{g}})=2.9774$\,nF for the dimer ($R_\mathrm{g}=269\,\Omega$), are added to the grid of $\kappa$, and there Q reaches the resolution limit, as in Fig.~\ref{figMain_CleanLimit}g.

	\subsubsection{LTspice disorder simulations}

	The disorder simulations of Fig.~\ref{figMain_Simulations} and Supplementary Section~\ref{SuppMatDisorderSimulations} use the parameters of the experiment, which are the values of the assembled components, $c_1=1.005$\,nF, $c_2=2.212$\,nF, $c_g=333$\,nF, $L=1.915\,\mu$H, $R=270\,\Omega$ and $N=16$ nodes per ring. The operating points at which one eigenvalue crosses the origin in the clean limit are $\kappa=3.072$\,nF and $f=198579$\,Hz for the 2-Node PT dimer, $\kappa=4.688$\,nF and $f=198514$\,Hz for the topological PT lattice, $\kappa=3.607$\,nF and $f=198917$\,Hz for the gapless PT lattice, and $f=199286$\,Hz for the 1-Node LC resonator.

	Following the measured disorder profile (Supplementary Section~\ref{SuppMatDisorderProfile}), the negative resistances of all gain nodes are randomized with an amplitude that decays exponentially with the distance from the gain-side coupling node, node 17, along the ring (Supplementary Eq.~\eqref{EqSuppMatLocalizedDisorder}). The loss elements keep their nominal values, and in the dimer and the 1-Node LC resonator, the single gain element carries this disorder. The disorder strength is $\Theta=0.1\%$, $1\%$, $5\%$ and $10\%$ in Fig.~\ref{figMain_Simulations}a and $\Theta=1\%$ in Fig.~\ref{figMain_Simulations}b.

	For Fig.~\ref{figMain_Simulations}a, $\kappa$ varies uniformly by $\pm10\%$ in each realization, and for the 1-Node LC resonator, which has no coupling, $c_g$ varies by $\pm0.5\%$. Both are reactive elements that move the eigenvalues mainly along the imaginary axis, so we compare their admittances rather than their capacitances. A variation of $c_g$ by $\pm0.5\%$ shifts $\mathrm{Im}\,\lambda$ by up to $\pm0.005\,\omega c_g\approx\pm2.1$\,mS, which matches the spread of about 4\,mS along the imaginary axis that the variation of $\kappa$ by $\pm10\%$ gives near the origin for the dimer (from $-1.0$ to $+3.0$\,mS). For Fig.~\ref{figMain_Simulations}b, the reactive element varies by less, by $\pm1\%$ of $\kappa$ for the topological PT lattice, $\pm3.7\%$ of $\kappa$ for the 2-Node PT dimer and $\pm0.25\%$ of $c_g$ for the 1-Node LC resonator, that is, by about $\pm0.06$, $\pm0.14$ and $\pm1.0$\,mS. These values are chosen so that the eigenvalue nearest the origin moves over a similar range along its trajectory in the three models. The auxiliary $-\kappa$ elements vary slightly around $\kappa$ to model their imperfect cancellation of the coupling (Supplementary Section~\ref{SuppMatDisorderSettings}).

	Each disorder level of Fig.~\ref{figMain_Simulations}a overlays 400 realizations at the operating frequency. For the insets of Fig.~\ref{figMain_Simulations}a and for the 500 realizations of Fig.~\ref{figMain_Simulations}b, each realization is also evaluated at 50 drive frequencies drawn uniformly within $f(1\pm5\times10^{-6})$, a window of about 2\,Hz. The spectra of Fig.~\ref{figMain_Simulations}a are computed in LTspice with operational-amplifier models of the gain and $-\kappa$ elements, controlled from Mathematica, and the Laplacian of each realization is reconstructed from the node voltages for a current injected at each node in turn. The impedances of Fig.~\ref{figMain_Simulations}b are obtained by direct inversion of the circuit Laplacian. The ideal Laplacian reproduces the LTspice eigenvalues with a median deviation of $2\times10^{-5}$\,S for the lattice (Supplementary Section~\ref{SuppMatDisorderSettings}).

	The colour of each eigenvalue in Fig.~\ref{figMain_Simulations}a is its weight in the impedance between the coupling nodes, the term of Eq.~\eqref{EqImpEig} evaluated with unit-normalized eigenvectors, on a logarithmic scale (Supplementary Section~\ref{SuppMatDisorderSettings}).

	\subsection{Data and code availability}
	All simulation and experimental data supporting this study will be made available in a Zenodo repository upon publication.

	\bibliography{references}

	\section{Acknowledgements}
	This work was supported by the Ministry of Education (MOE) of Singapore under Tier-II Grants No. MOE-T2EP50121-0014 (NUS Grant No. A-8000086-01-00) and MOE-T2EP50224-0007 (NUS Grant No. A-8003505-00-00), the MOE Tier-I FRC Grant (NUS Grant No. A-8000195-01-00), and NUS Grant No. A-8004258-00-00. F.O. acknowledges support from the Tokyo International University Personal Research Fund and the Tokyo International University Special Grant-in-Aid for Research Work. S.K.O. and R.E.-G. acknowledge support from the Air Force Office of Scientific Research (AFOSR) through the Multidisciplinary University Research Initiative (MURI) Award on Programmable Systems with Non-Hermitian Quantum Dynamics (Grant No.~FA9550-21-1-0202). S.K.O. acknowledges support from the Office of Naval Research (ONR) through Grant No.~N00014-26-1-2235-P00001.

	\subsection{Author contributions}
	H.S., C.H.L., and M.B.A.J. conceived the idea. C.H.L. and M.B.A.J. supervised the project. H.S. carried out the numerical and analytical analyses, performed circuit simulations, designed and conducted the experiments, analysed the data, and wrote the manuscript and Supplementary Information. F.O. performed a parallel experiment. S.Y. and M.S.H.R. contributed equally to the experimental work. Z.B.S. contributed extensively to the analytical and numerical analysis. S.K.O. analysed the experimental data and provided supervision to the project. J.F.K. co-supervised the project. S.M.R., R.E.-G., H.Y., and R.T. contributed to discussions. All authors reviewed and approved the final manuscript.

	\subsection{Competing interests}
	The authors declare no competing interests.

	\onecolumngrid
	\newpage
	\clearpage
	\begin{center}
		\textbf{\LARGE Supplementary Information}
	\end{center}

	\vspace{1em}

	\noindent\rule{\textwidth}{0.5pt}

	\tableofcontents

	\renewcommand{\theequation}{S\arabic{equation}}
	\setcounter{equation}{0}
	\renewcommand{\figurename}{\textbf{Fig.}}
	\renewcommand{\thefigure}{S\arabic{figure}}
	\renewcommand{\theHfigure}{S\arabic{figure}}
	\renewcommand{\theHtable}{S\arabic{table}}
	\renewcommand{\theHequation}{S\arabic{equation}}
	\setcounter{figure}{0}
	\setcounter{table}{0}
	\renewcommand{\thetable}{S\arabic{table}}

	\setcounter{figure}{0}

	\setcounter{secnumdepth}{3}

	\setcounter{section}{0}
	\setcounter{subsection}{0}
	\setcounter{subsubsection}{0}

	\renewcommand{\thesection}{S\arabic{section}}
	\renewcommand{\thesubsection}{S\arabic{section}.\arabic{subsection}}
	\renewcommand{\thesubsubsection}{S\arabic{section}.\arabic{subsection}.\arabic{subsubsection}}

	\makeatletter
	\renewcommand{\p@subsection}{}
	\renewcommand{\p@subsubsection}{}
	\makeatother

	\addtocontents{toc}{\protect\SIshowTOC}
	This Supplementary Information is organized in three parts. The experimental results come first (Supplementary Sections~\ref{SuppMat1NodeLC} to~\ref{SuppMatImplementation}), followed by the numerical simulations (Supplementary Sections~\ref{SuppMatDisorderSimulations} to~\ref{SuppMatReactivePerturbations}) and the analytical theory (Supplementary Sections~\ref{SuppMatImpedanceTheory} and~\ref{SuppMatEPTheory}). Each section states which part of the main text or Methods it supports.

	\SIpart{Experimental results}

	\section{Q factor of the 1-Node LC resonator with and without gain}

	\label{SuppMat1NodeLC}

	\begin{figure}[h]
		\centering
		\includegraphics[width=\textwidth]{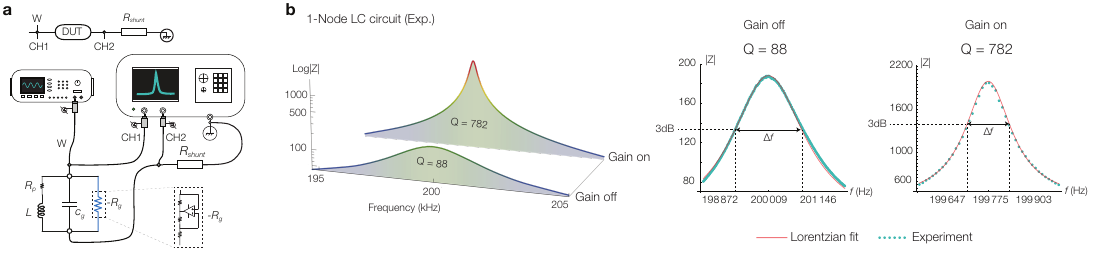}
		\caption{\textbf{Measured impedance of the 1-Node LC resonator with and without gain.} Gain sharpens the resonance of a single LC resonator, but its Q remains orders of magnitude below that of the topological PT lattice.
			\textbf{a,} The schematic shows the experimental setup used for the one-point impedance measurement performed with the ADP5250 device. The wave generator (W) provided a sinusoidal excitation with an amplitude of 1.684\,V, and the voltages were measured with Channel~1 (CH1) and Channel~2 (CH2). The current was monitored across a shunt resistor $R_\text{shunt}=1\,\mathrm{M}\Omega$, which corresponds to a floating one-point impedance measurement. The circuit consists of an inductor $L$ with a series resistance $R_p$, a shunt capacitor $c_g$ and a negative-impedance converter (NIC) connected to ground.
			\textbf{b,} In the measured impedance spectra with the gain activated (`Gain on') and deactivated (`Gain off'), the resonance becomes much sharper when gain is introduced, and the quality factor, obtained with Eq.~\eqref{EqQfactor} of the main text, increases from $\mathrm{Q}=88$ (lossy) to $\mathrm{Q}=782$ (gain). Despite this enhancement, the 1-Node LC circuit stays far below the Q $=3.2\times10^5$ of the topological PT lattice, which shows that the ultrahigh-Q resonance of the lattice does not arise from gain alone. The nominal component values used in the experiment are $(c_g, L, R_g) = (333\,\text{nF},\,1.915\,\mu\text{H},\,180\,\Omega)$.
		}
		\label{FigSuppMat1NodeLC}
	\end{figure}

	We experimentally tested a 1-Node LC tank circuit both with and without gain. To keep the structure and components identical to those of the SSH PT lattice, we assembled the same type of components on an otherwise empty board identical to that of the gain SSH circuit. Only a single node was soldered, so that the same parasitic losses and wiring effects are present. All measurements were carried out with the same setup and equipment as the main experiments.

	Figure~\ref{FigSuppMat1NodeLC}a shows the configuration used for the one-point impedance measurement. The ground of the circuit was connected through a shunt resistor, corresponding to a floating one-point impedance measurement. Without gain, the circuit is intrinsically lossy, which we represent by a series resistance $R_p$ that accounts for component and wiring losses. The circuit Laplacian in the presence of gain is
	\begin{equation}
		J_\text{1-Node}(\omega) = -\frac{1}{R_g} + \frac{1}{i\omega L + R_p} + i\omega c_g ,
	\end{equation}
	where the NIC provides a grounding resistance of $-R_g$. When the operational amplifier is not powered, its three internal resistors act as passive elements and effectively give a positive resistance of approximately $3R_g$.

	The measured impedance spectra for both cases are shown in Fig.~\ref{FigSuppMat1NodeLC}b. In the `Gain on' case, the operational amplifiers were powered by DC sources, while in the `Gain off' case, the DC supply was disconnected. The resulting quality factors are $\mathrm{Q}=782$ with gain and $\mathrm{Q}=88$ without gain. These measurements were optimized over parameters such as the excitation amplitude, and the highest Q observed with gain was 782. Although this value is reasonable for an active single LC resonator, it remains far below the Q $=3.2\times10^5$ of the topological PT lattice (Fig.~\ref{figMain_Experiment}c). This comparison shows that the Q enhancement in our system originates not merely from the presence of gain but from the spectral structure of the coupled lattices, which lowers the broadening coefficient of the gain fluctuations that reaches the linewidth (Eq.~\eqref{EqEPcShare} of the main text).

	\section{Spectral broadening of the 2-Node PT dimer}

	\label{SuppMat2NodePT}

	\begin{figure}[h]
		\centering
		\includegraphics[width=0.95\textwidth]{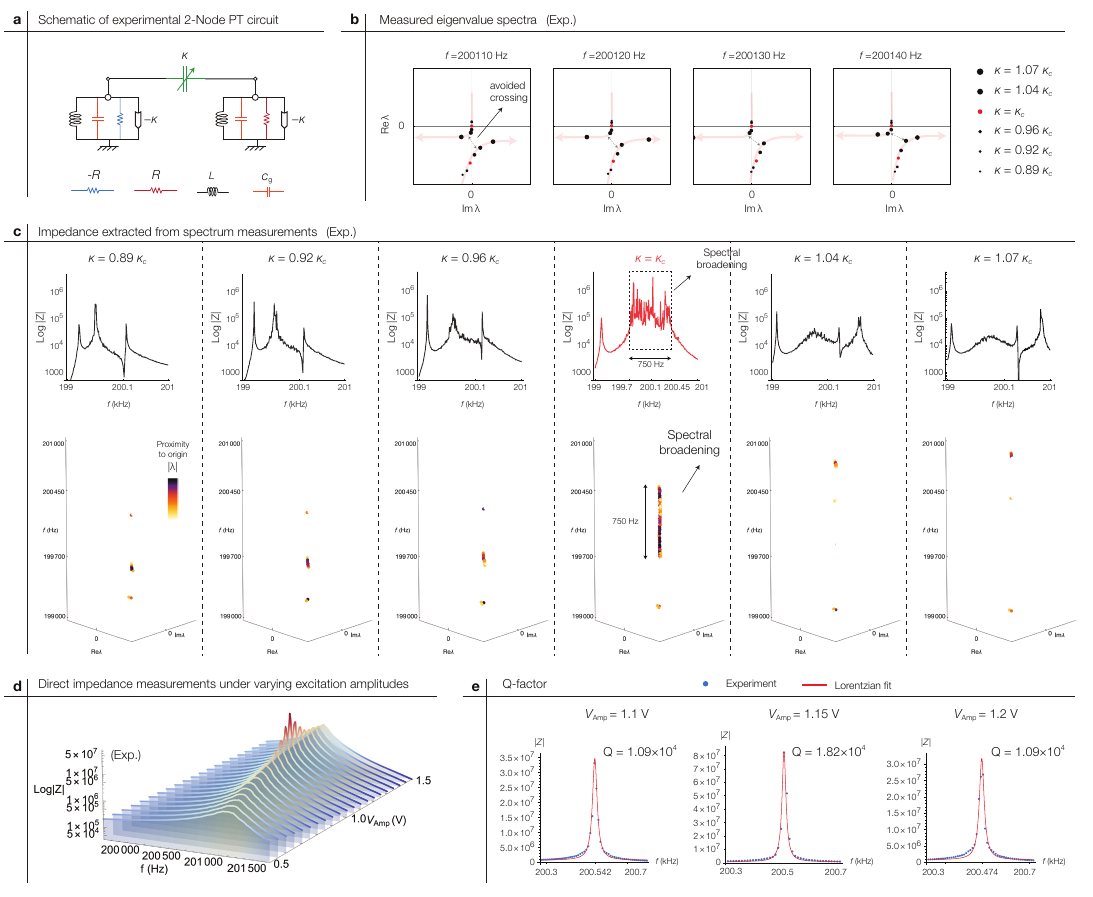}
		\caption{\textbf{Measured spectra, impedance and Q of the 2-Node PT dimer.} Near its EP, the eigenvalue of the dimer fluctuates strongly in frequency, which broadens the resonance and limits the Q.
			\textbf{a,} The schematic shows the experimental 2-Node PT circuit, which consists of two coupled LC resonators with gain and loss and the component values $(c_g,L,R,R_g)=(333\,\mathrm{nF},\,1.915\,\mu\mathrm{H},\,270\,\Omega,\,180\,\Omega)$. As in the topological PT circuit, the $180\,\Omega$ negative resistance nearly balances the $270\,\Omega$ loss once the parasitic losses of the nodes are included (Methods), and the coupling capacitance $\kappa$ is tuned through a voltage-controlled capacitor module.
			\textbf{b,} The measured admittance eigenvalue spectra are shown at four representative frequencies and a fixed excitation amplitude $V_\mathrm{Amp}=1.15\,\mathrm{V}$. The marker size denotes the coupling parameter $\kappa$, and the red marker highlights the critical coupling $\kappa=\kappa_c$, where one eigenvalue reaches its minimum distance from the spectral origin. The light-red curves guide the eye along the measured eigenvalue trajectories, and the dashed arrow indicates the avoided-crossing region near the exceptional-point transition.
			\textbf{c,} The upper panels show the impedance profiles reconstructed from the measured admittance spectra using Eq.~\eqref{EqMethodsEffectiveImpedanceMatrix} of Methods for representative coupling values, and the lower panels show the corresponding admittance spectra as a function of frequency. The colour indicates the proximity of an eigenvalue to the spectral origin, with darker colours corresponding to smaller $|\lambda|$. At $\kappa=\kappa_c$, the eigenvalue closest to the origin spreads over a frequency interval of approximately $750\,\mathrm{Hz}$ and forms a finite spectral cloud, which appears as a broadened impedance resonance that fluctuates between measurements. This interval is a range of fluctuation of the eigenvalue rather than the linewidth of a single resonance.
			\textbf{d,} Direct impedance measurements obtained with an impedance analyser at the critical coupling $\kappa=\kappa_c$ for different excitation amplitudes $V_\mathrm{Amp}$ show that the resonance strength is largest near $V_\mathrm{Amp}=1.15\,\mathrm{V}$.
			\textbf{e,} Lorentzian fits of representative impedance resonances for different excitation amplitudes give the largest Q factor, $\mathrm{Q}=1.82\times10^{4}$, at $V_\mathrm{Amp}=1.15\,\mathrm{V}$.
		}
		\label{FigSuppMat2NodePT}
	\end{figure}

	Here, we give the experimental characterization of the 2-Node PT circuit, including its measured admittance spectra, the corresponding impedance profiles and the highest observed Q factor. For a meaningful comparison with the topological PT circuit discussed in the main text, we use the same PCB platform and identical circuit elements wherever possible. Specifically, the circuit parameters are $c_g = 333\,\mathrm{nF}$, $L = 1.915\,\mu\mathrm{H}$, $R = 270\,\Omega$ and $R_g = 180\,\Omega$. The tunable coupling capacitance $\kappa$ is implemented with the same voltage-controlled module used in the topological PT circuit. For the dimer, the module uses a smaller fixed capacitor and six of its eleven voltage-controlled capacitors, which tune $\kappa$ from $2.9$ to $3.5\,\mathrm{nF}$, a range of $0.6\,\mathrm{nF}$, for a control voltage between $0$ and $5\,\mathrm{V}$. The critical coupling is $\kappa_c=3.26\,\mathrm{nF}$, and Figs.~\ref{FigSuppMat2NodePT}b,c show six couplings from $0.89\kappa_c$ to $1.07\kappa_c$.

	As in the topological PT implementation, the coupling module contains negative-capacitance INIC elements that compensate the reactive contribution of the coupling capacitor. The two-node prototype and the topological PT circuit therefore share a closely matched experimental configuration. The schematic of the circuit is shown in Fig.~\ref{FigSuppMat2NodePT}a. The two resonators are nominally identical, with gain introduced through a negative resistance $R_g$ on the left resonator and loss introduced through a positive resistance $R$ on the right resonator.

	Figure~\ref{FigSuppMat2NodePT}b shows the measured admittance eigenvalue spectra at a fixed excitation amplitude of $V_\mathrm{Amp}=1.15\,\mathrm{V}$ for four representative frequencies. For each frequency, the coupling capacitance is varied over the accessible range, and the corresponding eigenvalues are overlaid in the same complex-admittance plane. Marker size denotes the coupling parameter, and the red marker highlights the coupling value $\kappa=\kappa_c$ at which one eigenvalue reaches its minimum distance from the spectral origin.

	As $\kappa$ is varied, the eigenvalue pair follows the characteristic PT trajectory. The two branches approach one another, reach their minimum separation near the exceptional-point transition, and then separate again along a different direction in the complex plane. The dashed arrows indicate where the trajectories would intersect in the absence of imperfections. Instead, a finite avoided crossing remains. Consequently, even when the exceptional point is tuned close to the spectral origin, the eigenvalue trajectory does not generically pass through the origin itself. In the present experiment, a slight gain--loss imbalance shifts the exceptional point away from the exact origin, which allows one branch of the trajectory to pass closer to zero admittance than would otherwise be possible.

	To relate the measured spectra to the circuit response, we convert the measured admittance spectra into impedance using Eq.~\eqref{EqMethodsEffectiveImpedanceMatrix} of Methods. Unlike the topological PT circuit, where the full spectral reconstruction requires measurements of all $32$ nodes, the present system contains only two nodes and therefore permits much finer frequency sampling. Here, the frequency is swept from $199\,\mathrm{kHz}$ to $201\,\mathrm{kHz}$ in $201$ points. The resulting impedance profiles are shown in Fig.~\ref{FigSuppMat2NodePT}c for the same coupling values as in Fig.~\ref{FigSuppMat2NodePT}b.

	These impedance spectra provide a direct interpretation of the measured admittance eigenvalues, because both quantities are derived from the same dataset. For $\kappa<\kappa_c$, a pronounced central resonance is observed between approximately $199$ and $200.1\,\mathrm{kHz}$. As the system is tuned beyond $\kappa_c$, this resonance rapidly weakens and eventually disappears, consistent with the evolution of the admittance eigenvalues away from the spectral origin. The impedance enhancement therefore occurs only when an eigenvalue branch remains sufficiently close to zero admittance.

	More importantly, the measurements reveal pronounced fluctuations precisely at the critical coupling $\kappa=\kappa_c$. Although one eigenvalue reaches its smallest magnitude at this point, its frequency position fluctuates strongly over a range of approximately $750\,\mathrm{Hz}$. This behaviour appears directly in the impedance profile, where the central resonance becomes broadened and fragmented rather than forming a stable ultranarrow peak. The proximity of an eigenvalue to the spectral origin is thus not sufficient for a high-Q response. The eigenvalue must also remain spectrally localized.

	The same behaviour is visible in the lower three-dimensional spectral plots of Fig.~\ref{FigSuppMat2NodePT}c. Here, all measured admittance eigenvalues are shown as a function of frequency and coloured according to their distance from the spectral origin, with darker colours corresponding to smaller $|\lambda|$. The frequencies of these eigenvalues coincide with the resonance frequencies observed in the impedance spectra above. Away from $\kappa_c$, the eigenvalues remain relatively localized in frequency. At $\kappa=\kappa_c$, however, the eigenvalue closest to the origin spreads over approximately $750\,\mathrm{Hz}$ and forms a finite spectral cloud. This interval is not the linewidth of a single resonance. It is the frequency range over which the measured eigenvalue nearest the origin stays close to it, instead of crossing it at one frequency, which reveals strong fluctuations of the eigenvalue near the EP. The direct impedance measurement at the same coupling gives a Lorentzian line of about 11\,Hz ($\mathrm{Q}=1.82\times10^4$, Fig.~\ref{FigSuppMat2NodePT}e). This observation is a direct experimental visualization of the spectral broadening discussed in the main text, and it explains why the achievable Q factor remains limited despite the proximity of the eigenvalue to the origin. It agrees with Eq.~\eqref{EqDimerShare} of the main text, by which half of every gain fluctuation enters the linewidth of a dimer at any coupling, while the frequency jitter grows as the EP is approached.

	We next perform direct impedance measurements with an impedance analyser instead of reconstructing the impedance from the measured spectra. The coupling is fixed at $\kappa=\kappa_c$, and the excitation amplitude $V_\mathrm{Amp}$ is varied. As discussed in Methods, the effective circuit parameters depend weakly on the applied voltage because of the non-ideal response of the active circuit elements. Although this dependence is usually negligible in conventional resonant circuits, it becomes significant near the high-Q regime, because small parameter changes translate into appreciable shifts of the admittance spectrum. As shown in Fig.~\ref{FigSuppMat2NodePT}d, the resonance strength increases as $V_\mathrm{Amp}$ approaches $1.15\,\mathrm{V}$ and is largest at this value.

	The corresponding Lorentzian fits are shown in Fig.~\ref{FigSuppMat2NodePT}e. The largest measured Q factor is obtained at $V_\mathrm{Amp}=1.15\,\mathrm{V}$, giving $\mathrm{Q}=1.82\times10^{4}$. This is the highest Q factor observed in our 2-Node PT circuits. Further enhancement is prevented by the spectral broadening and frequency fluctuations evident in Fig.~\ref{FigSuppMat2NodePT}c. Although the eigenvalue can be tuned close to the spectral origin, the associated spectral cloud introduces an effective linewidth that limits the resonance. This limitation motivates the topological PT mechanism of the main text, in which disorder-induced spectral broadening is strongly suppressed and the measured Q reaches $3.2\times10^5$ (Fig.~\ref{figMain_Experiment}c). The suppression of spectral fluctuations and the resulting stability of the resonance are further supported by the impedance profiles reconstructed from the measured admittance spectra of the topological PT circuit (Supplementary Section~\ref{SuppMatImpedanceFromSpectra}) and by the Allan deviation analysis of Extended Data Fig.~\ref{FigMethods_Noise}.

	\section{Admittance spectra of the topological PT circuit}
	\label{SuppMatMeasuredSpectra}

	Here, we describe the experimental procedure for measuring the eigenvalue spectra and present the measured spectra at different frequencies and excitation amplitudes, which complement Extended Data Fig.~\ref{FigMethods_Spectra} and the method described in Methods.

	\subsection{Voltage-transfer measurement and loading by current sensing}
	\label{SuppMatVoltageTransfer}

	Extracting the circuit Laplacian requires the linear voltage response of the network to an external excitation at a fixed frequency $\omega = 2\pi f$. A conventional procedure is to excite individual nodes sequentially with a monochromatic signal and record the complex voltage responses (amplitude and phase) at all nodes~[S1,\,S2]. Repeating this process for each excitation node gives the voltage-response matrix $V(\omega)$ and the corresponding injected-current matrix $I(\omega)$, from which the circuit Laplacian follows as
	\begin{equation}
		J(\omega) = I(\omega)\,V^{-1}(\omega),
	\end{equation}
	provided that $V(\omega)$ is invertible and that measurement-induced perturbations are negligible. For passive or weakly sensitive circuits, this approach captures the intrinsic circuit behaviour accurately, since the loading by the measurement setup produces only small and approximately uniform spectral shifts that preserve the qualitative spectral structure. In parametrically sensitive systems, however, even weak loading modifies the effective onsite admittances and degrades the quantitative correspondence between the measured spectra and the intrinsic circuit Laplacian.

	In our experiment, measurement-induced loading arises from two sources, (i) probe-induced loading and (ii) shunt-induced loading. Probe-induced loading, discussed in Supplementary Section~\ref{SuppMatCapacitiveLoading}, originates from the finite input impedance of the coaxial cables and acquisition channels and locally renormalizes the node admittances. Shunt-induced loading is more severe and is associated with the finite output impedance of the waveform generator channels, nominally $50\,\Omega$.

	In the current-sensing attempts affected by loading (ii), we inserted a series shunt resistor $R_{\mathrm{shunt}}$ (10--100~$\Omega$) between the waveform-generator output and the excited node and inferred the injected current from the voltage drop across $R_{\mathrm{shunt}}$. In this configuration the excitation port is no longer an ideal voltage drive. The node is driven through an effective source impedance
	$Z_s(\omega)\simeq R_{\mathrm{out}}+R_{\mathrm{shunt}}$ (with $R_{\mathrm{out}}\approx 50~\Omega$), so that
	$V_{\mathrm{node}}(\omega)=V_s(\omega)/\!\left[1+Z_s(\omega)\,Y_{\mathrm{in}}(\omega)\right]$,
	where $Y_{\mathrm{in}}(\omega)$ is the input admittance of the lattice at the driven node (including the negative-resistance converter). For gain nodes, $Y_{\mathrm{in}}(\omega)$ is highly sensitive to external loading, and the added series impedance $Z_s$ renormalizes the effective onsite admittance and drives the negative-resistance converter out of its intended operating regime. Experimentally, for all tested values $R_{\mathrm{shunt}} = 10$, $20$, $50$ and $100~\Omega$, the gain signatures were suppressed already at the excitation stage, which shows that shunt-based current sensing is incompatible with Laplacian reconstruction in the present gain lattice.

	To avoid this limitation, we use the voltage-transfer scheme of Methods. A sinusoidal voltage is injected directly into one node, without a shunt resistor, and the complex voltages at all nodes are recorded through high-impedance channels of the ADP3450 network analysers. The transfer ratios $T_{ji}=V_j^{(i)}/V_i^{(i)}$ for all excitation nodes $i$ form the $32\times32$ matrix $\tilde V$. The effective impedance matrix $Z_\mathrm{eff}$ and the effective Laplacian $J_\mathrm{eff}=Z_\mathrm{eff}^{-1}$ then follow from the bilinear transformation of Eqs.~\eqref{EqMethodsEffectiveImpedanceMatrix} and~\eqref{EqMethodsEffectiveLaplacianMatrix}, which is modelled on the relation between impedance and scattering matrices~[S3]. Since $\tilde V$ is known only up to an overall scale, the eigenvalues of $J_\mathrm{eff}$ are given in arbitrary units (a.u.).

	\subsection{Grounding in spectral and impedance measurements}
	\label{SuppMatGrounding}

	Beyond the loading effects of the measurement probes, spectral and impedance measurements of multi-node circuits rely on different referencing (grounding). In the spectral measurements, the circuit ground is connected to the common ground of the instruments, so that all node voltages are measured against one fixed reference.

	In contrast, impedance measurements require the circuit to remain galvanically isolated (floating). This isolation ensures that the excitation current is confined to the path between the two measurement probes. If a ground reference is introduced, it creates an unintended return path, and leakage currents divert part of the injected current away from the node-to-node path. The measured response then no longer corresponds to the eigenmode expansion of the floating-circuit Laplacian used to define the two-point impedance.

	These different electrical configurations do not alter the physics of the circuit network or its spectral structure, so the qualitative structure of the spectral bands remains unchanged. They do, however, modify how the Laplacian is probed. The precise parametric correspondence between spectral and impedance measurements is therefore shifted, which prevents a direct numerical comparison between the two datasets, even though the underlying eigenvalue evolution is governed by the same circuit. The loading by the measurement channels adds a further frequency offset (Supplementary Section~\ref{SuppMatCapacitiveLoading}).

	\subsection{Measured spectra near the outer exceptional points}
	\label{SuppMatSpectraOuterEPs}

	\begin{figure}[h]
		\centering
		\includegraphics[width=\textwidth]{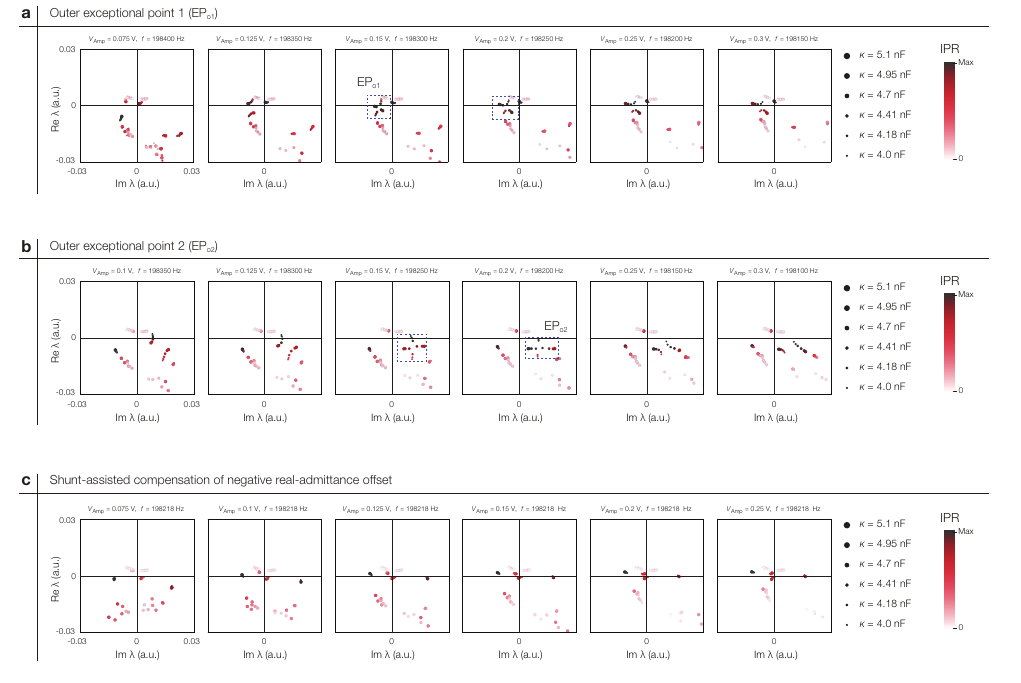}
		\caption{\textbf{Measured admittance spectra near the outer exceptional points.} The measured coupling modes coalesce at both outer EPs, as predicted by the theory. The colour scale indicates the inverse participation ratio (IPR), and the marker size encodes different values of $\kappa$.
			\textbf{a,} The measured complex eigenvalue spectra show the evolution of the coupling modes towards the outer exceptional point EP$_\mathrm{o1}$ as the injection amplitude $V_\text{Amp}$ is increased while the excitation frequency $f$ is reduced.
			\textbf{b,} The same measurement protocol, applied at frequencies lower by approximately $50\,\mathrm{Hz}$, reveals the other outer exceptional point, EP$_\mathrm{o2}$. In both panels, the bulk modes remain approximately stationary for different values of $\kappa$, while the coupling modes, identified by their larger IPRs, show a clear coalescence, highlighted by blue dashed rectangles.
			\textbf{c,} In a test measurement, a $400\,\Omega$ shunt resistor attached between the coupling nodes partially compensates the negative real-admittance offset and shifts the isolated coupling modes closer to the imaginary axis. This offset is associated with the parasitic resistance of the traces and contacts in series with the variable negative-capacitance converters between the coupling nodes and ground, which explains the offsets of EP$_\mathrm{o1,o2}$ towards the negative real axis. The shunt increases the mode localization but reduces the sensitivity to the coupling capacitance $\kappa$.}
		\label{FigSuppMatSpectraOuterEPs}
	\end{figure}

	Following the voltage-transfer measurement scheme of Supplementary Section~\ref{SuppMatVoltageTransfer}, we measured the effective circuit Laplacian (Eq.~\eqref{EqMethodsEffectiveLaplacianMatrix}) for several excitation frequencies $f$, injection amplitudes $V_\text{Amp}$ and six values of the coupling capacitance $\kappa$. The gain--loss imbalance that we introduce intentionally to pass the avoided crossing near zero admittance also reduces the sensitivity of the coupling modes to $\kappa$. Since the available tuning range of the variable capacitors, 1.1\,nF from 4.0 to 5.1\,nF, is therefore insufficient to capture the full evolution of the coupling-mode trajectories, we instead probe the spectra by jointly varying $f$ and $V_\text{Amp}$ while sweeping $\kappa$.

	In Figs.~\ref{FigSuppMatSpectraOuterEPs}a and b, we present the measured spectra for increasing $V_\text{Amp}$ together with decreasing $f$. As in Extended Data Fig.~\ref{FigMethods_Spectra}, a larger $V_\text{Amp}$ lowers the resonance frequency of the nodes, while a lower $f$ shifts the spectrum in the opposite direction along the imaginary axis (Methods). The bulk modes therefore remain approximately stationary, while the coupling modes show a clear evolution. For instance, in Fig.~\ref{FigSuppMatSpectraOuterEPs}a, two coupling modes, identified by their relatively large inverse participation ratios (IPRs), approach each other and coalesce at the outer exceptional point EP$_\mathrm{o1}$, highlighted by the blue dashed rectangle. At frequencies lower by approximately $50\,\mathrm{Hz}$, from $198350$ to $198100\,\mathrm{Hz}$ for nearly the same values of $V_\text{Amp}$, the other outer exceptional point, EP$_\mathrm{o2}$, is observed (Fig.~\ref{FigSuppMatSpectraOuterEPs}b).

	In the experiment, these EPs show avoided-crossing behaviour, which prevents the eigenvalues from reaching exactly zero admittance. Furthermore, the eigenvalue trajectories approaching the EPs have a negative real-admittance offset. This offset arises primarily from the gain-biased imbalance. In addition, the negative-capacitance converters used to implement the compensation $-\kappa$ introduce a residual negative resistance due to unavoidable trace and parasitic resistances. The combined effect is a systematic negative shift of the coupling-mode eigenvalues.

	This offset can be partially compensated by shunting the two coupling nodes with a positive resistance. In Fig.~\ref{FigSuppMatSpectraOuterEPs}c, we show the measured spectra obtained with a $400\,\Omega$ resistor between the coupling nodes. The additional shunt suppresses the negative offset, brings the isolated coupling modes closer to the imaginary axis and enhances their IPR contrast. However, it also increases mode localization and thereby reduces the sensitivity of the coupling modes to $\kappa$. For this reason, the shunt resistor is used only as a diagnostic test and is removed for all other measurements.

	\subsection{Impedance reconstructed from the measured spectra}
	\label{SuppMatImpedanceFromSpectra}

	\begin{figure}[h]
		\centering
		\includegraphics[width=\textwidth]{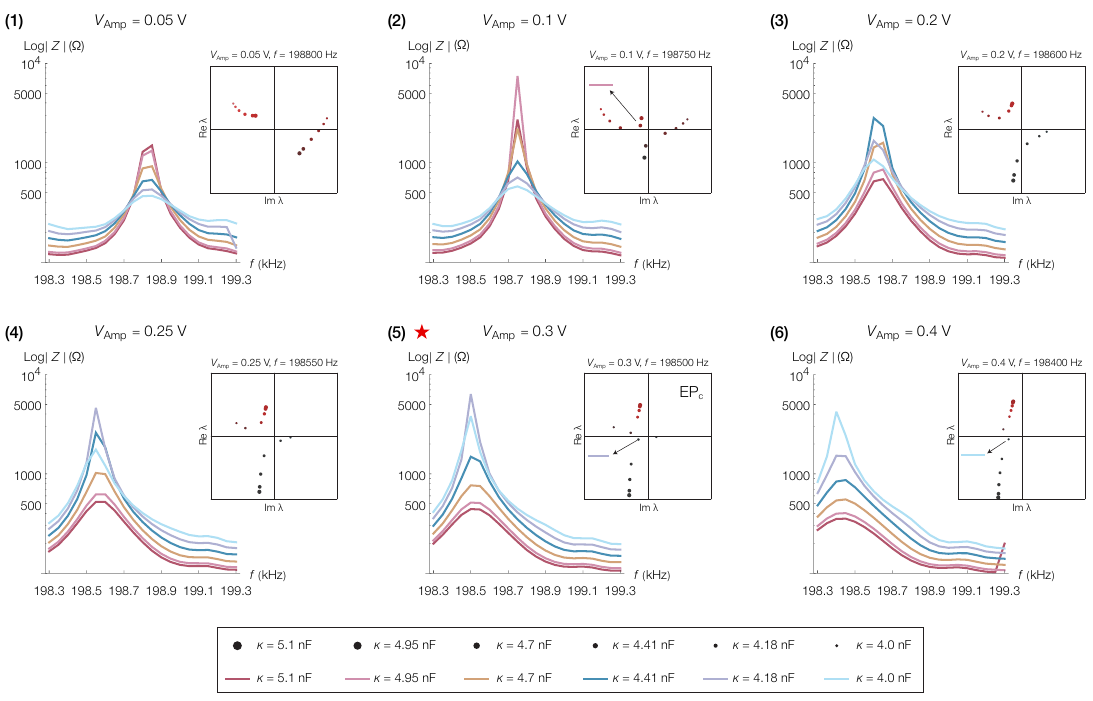}
		\caption{\textbf{Impedance reconstructed from the measured spectra of Extended Data Fig.~\ref{FigMethods_Spectra} (Experiment).} The reconstructed impedance peaks are stable, with no jitter or significant broadening, unlike those of the PT dimer shown in Fig.~\ref{FigSuppMat2NodePT}. The impedance magnitude between the coupling nodes 16 and 17, $|Z_{16,17}|$, is obtained from the effective impedance matrix $Z_{\mathrm{eff}}$ reconstructed from the voltage-transfer measurements of panels (1) to (6) of Extended Data Fig.~\ref{FigMethods_Spectra}, for excitation amplitudes from $V_\text{Amp}=0.05$ to 0.4\,V. The insets display the corresponding effective admittance eigenvalue spectra near the spectral origin at the frequencies indicated above each panel, and the marker size denotes the six coupling capacitances $\kappa=4.0$, 4.18, 4.41, 4.7, 4.95 and 5.1\,nF. In panels (2), (5) and (6), the largest impedance peak occurs at the coupling for which an isolated eigenvalue comes closest to zero admittance (arrows in the insets), consistent with the proximity of an eigenvalue trajectory to the spectral origin. The star marks panel (5), where this coupling is $\kappa=4.18$\,nF, the coupling of the sharpest measured peak in Fig.~\ref{figMain_Experiment}d,e.
		}
		\label{FigSuppMatImpedanceFromSpectra}
	\end{figure}

	Here, we present the impedance profiles reconstructed from the measured effective admittance spectra of Extended Data Fig.~\ref{FigMethods_Spectra}. Because of the acquisition time and the need to repeat the measurements for all excitation nodes, the frequency sweep was performed over 21 points from 198.3\,kHz to 199.3\,kHz.

	The effective impedance matrix $Z_{\mathrm{eff}}$ is obtained from the measured voltage-transfer matrix $\tilde{V}(\omega)$ through Eq.~\eqref{EqMethodsEffectiveImpedanceMatrix}. The two-point impedance between the coupling nodes 16 and 17 is then evaluated from the effective impedance matrix as
	\begin{equation}
		|Z_{16,17}(\omega)|=
		\left|Z^{\mathrm{eff}}_{16,16}(\omega)+Z^{\mathrm{eff}}_{17,17}(\omega)
		-Z^{\mathrm{eff}}_{16,17}(\omega)-Z^{\mathrm{eff}}_{17,16}(\omega)\right|.
	\end{equation}
	The resulting profiles are shown in Fig.~\ref{FigSuppMatImpedanceFromSpectra}. In each panel, the insets show the corresponding effective admittance eigenvalue spectra for six coupling capacitances $\kappa$, with the marker size labelling the different values of $\kappa$. Panels (1) to (3) of Fig.~\ref{FigSuppMatImpedanceFromSpectra} present the results for $V_\text{Amp}=0.05$, 0.1 and 0.2\,V. The largest impedance peak appears in panel (2), for $V_\text{Amp}=0.1$\,V near $f=198750$\,Hz at $\kappa=4.95$\,nF, where one eigenvalue comes closest to zero admittance.

	Panels (4) to (6) show the higher-amplitude regime with $V_\text{Amp}=0.25$, 0.3 and 0.4\,V. In this case, the strongest response occurs in panel (5), for $V_\text{Amp}=0.3$\,V near $f=198500$\,Hz at $\kappa=4.18$\,nF, again where an eigenvalue comes closest to zero admittance.

	Importantly, the impedance peaks remain smooth and well defined near the exceptional points, despite the coarse frequency sampling of only 21 points. This stable response contrasts with the 2-Node PT dimer, where the impedance near the EP is much more distorted and noisy (Supplementary Section~\ref{SuppMat2NodePT}). The reconstructed impedance thus links the measured eigenvalue trajectories of Extended Data Fig.~\ref{FigMethods_Spectra} with the impedance resonances of Fig.~\ref{figMain_Experiment}, although the two measurements use different grounding (Supplementary Section~\ref{SuppMatGrounding}).

	\section{Temporal waveforms and spatial profile of the fluctuations}
	\label{SuppMatFluctuations}

	\subsection{Temporal waveforms without self-oscillation}
	\label{SuppMatTemporalWaveforms}

	\begin{figure}[h!]
		\centering
		\includegraphics[width=\textwidth]{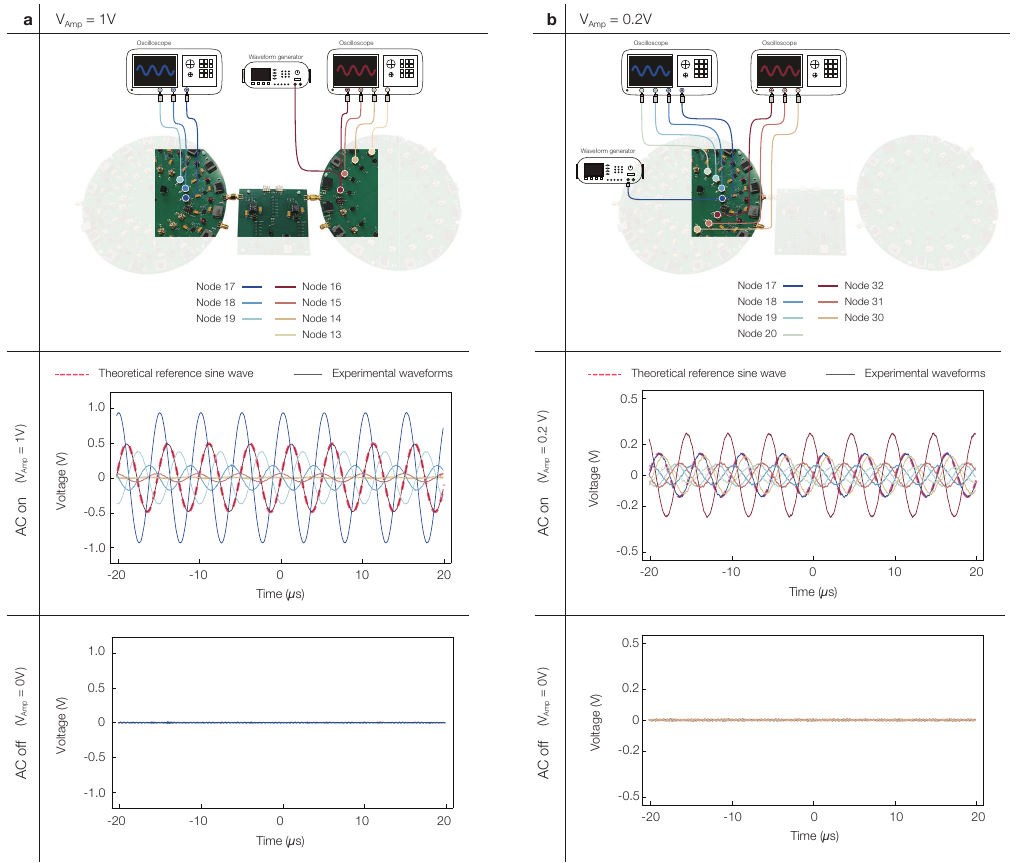}
		\caption{\textbf{Measured temporal waveforms.} The node voltages remain sinusoidal under excitation and show no spontaneous oscillation once the excitation is removed.
			\textbf{a,} Oscilloscope traces were recorded across the gain--loss coupled circuit for a sinusoidal excitation of $V_{\mathrm{Amp}}=1\,\mathrm{V}$ applied at node 16 in the lossy SSH circuit. The second row shows the waveforms measured at nodes 13--19 together with the theoretical sinusoidal reference (red dashed line), which indicates the integrity of the node oscillations. The third row shows the corresponding traces when the AC excitation is removed while the active elements remain powered, and this ``AC off'' profile shows no oscillation even though the op-amps are powered, which highlights that the circuit operates below the oscillation threshold.
			\textbf{b,} In the equivalent measurements for a $V_{\mathrm{Amp}}=0.2\,\mathrm{V}$ excitation applied at node 17 in the gain SSH circuit, the neighbouring node N32 has a voltage amplitude larger than the injected signal, consistent with the gain. In both cases, the absence of spontaneous oscillations in the ``AC off'' measurements confirms the stable operation of the active circuit.}
		\label{FigSuppMatTemporalWaveforms}
	\end{figure}

	Figure~\ref{FigSuppMatTemporalWaveforms} presents representative time-domain measurements used to assess the signal integrity and dynamical stability of the circuit. Since the system contains active negative-resistance elements, an important concern is the possible emergence of self-sustained oscillations or other instabilities that could distort the impedance measurements. Such effects would be particularly harmful in the ultrahigh-Q regime, where even small perturbations may substantially modify the measured resonance profile.

	The measurements were acquired with a Digilent ADP3450 oscilloscope at a sampling rate of $100\,\mathrm{MHz}$ with $32{,}768$ samples per acquisition, corresponding to a record of approximately $328\,\mu\mathrm{s}$. The excitation frequency was fixed at $198.4\,\mathrm{kHz}$, the operating frequency of the ultrahigh-Q resonance. The oscilloscope channels were AC-coupled. The dashed red curves in Fig.~\ref{FigSuppMatTemporalWaveforms} are ideal sinusoidal reference signals at the same excitation frequency.

	Figure~\ref{FigSuppMatTemporalWaveforms}a shows measurements obtained by injecting a sinusoidal signal with amplitude $V_{\mathrm{Amp}}=1\,\mathrm{V}$ at node N16, at the coupling interface on the lossy SSH side. The oscilloscope traces were recorded simultaneously at nodes N13--N19, which span both the gain and the loss circuits. As shown in the second row, all measured waveforms remain highly sinusoidal and closely follow the reference signal. No waveform clipping, harmonic distortion or other signature of strong nonlinear behaviour is observed. The circuit therefore operates in the linear-response regime during the impedance measurements.

	The dynamical stability of the active circuit is further verified by the ``AC off'' measurements in the third row. In these measurements, the external excitation is removed while the op-amps remain powered. If the gain exceeded the stability threshold of the circuit, persistent self-sustained oscillations would appear even without an external signal. Instead, the measured voltages remain close to the noise floor and no periodic oscillations are observed. These measurements confirm that the active circuit remains dynamically stable over the operating range investigated in this work.

	Figure~\ref{FigSuppMatTemporalWaveforms}b presents an equivalent measurement with a smaller excitation amplitude of $V_{\mathrm{Amp}}=0.2\,\mathrm{V}$ applied directly at node N17 in the gain SSH circuit. The measured waveforms again remain highly sinusoidal and agree well with the reference signal. The voltage amplitude at node N32 exceeds the injected amplitude at N17, which is consistent with the action of the negative-resistance elements of the gain SSH circuit.
	Despite this amplification, the waveforms remain well preserved and free of visible distortion, which indicates stable operation in the linear regime.

	The ``AC off'' measurements in the third row again show no self-sustained oscillations. Small residual fluctuations remain visible near the noise floor, because the active elements amplify both intrinsic electronic noise and small circuit imperfections. Although such fluctuations are negligible for most resonant systems, they become relevant in the ultrahigh-Q regime, because even millivolt-scale perturbations may influence the linewidth and peak amplitude of the resonance. In Supplementary Section~\ref{SuppMatDisorderProfile}, we use the same temporal records to quantify these fluctuations and to construct the experimentally motivated disorder model of the numerical simulations.

	\subsection{Spatial profile of the measured fluctuations}
	\label{SuppMatDisorderProfile}

	The disorder model of the simulations (Supplementary Section~\ref{SuppMatDisorderSettings}) is motivated by direct measurements of temporal fluctuations in the experimental circuit. Rather than assuming an arbitrary spatial distribution of disorder, we extract the local fluctuation strength from measured waveforms and use it to identify the dominant sources of instability. The strongest fluctuations originate at the coupling interface between the gain and loss circuits, where the tunable coupling capacitors and the negative-resistance converters are located. These active components are the dominant source of instability, because any amplitude or phase fluctuation is amplified by the gain circuitry.

	Starting from the time-domain voltage traces measured at a given node (the same data as in Supplementary Section~\ref{SuppMatTemporalWaveforms}), we divide the signal into multiple temporal windows and fit each window with the reference waveform
	\begin{equation}
		V_{\mathrm{fit}}(t) =
		A\sin(2\pi f t+\phi)+c,
	\end{equation}
	where $A$, $\phi$ and $c$ denote the local amplitude, phase and DC offset, respectively, and $f=198.4\,\mathrm{kHz}$ is the drive frequency. This procedure converts the continuous waveform into a sequence of local amplitude and phase states, $\{A_i,\phi_i\}$, which characterizes the temporal evolution of the circuit. The amplitude and phase fluctuations are quantified through the variances
	\begin{equation}
		\sigma_A^2=\frac{1}{M-1}\sum_{i=1}^{M}\left(A_i-\bar A\right)^2,\qquad \sigma_{\phi}^2=\frac{1}{M-1}\sum_{i=1}^{M}\left(\phi_i-\bar\phi\right)^2,
	\end{equation}
	where $M$ is the number of windows and $\bar A$ and $\bar\phi$ are the mean values. We then define the local disorder power density as
	\begin{equation}
		P_{\mathrm{disorder}} = \sigma_A^2+\bar A^2\sigma_{\phi}^2.
	\end{equation}
	This quantity captures both amplitude modulation and phase jitter and therefore directly measures the local instability experienced by each node.

\begin{figure}[h]
		\centering
		\includegraphics[width=\textwidth]{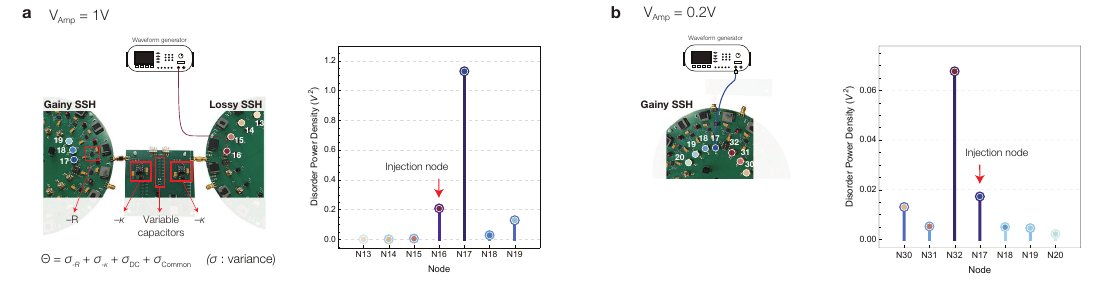}
		\caption{\textbf{Measured disorder profile of the topological PT circuit.} In both measurements, the dominant disorder is concentrated near the gain--loss coupling region, which contains the negative-resistance converter and the tunable coupling elements, and this motivates the disorder model of the numerical simulations.
			\textbf{a,} For a $1\,\mathrm{V}$ excitation applied at node N16 in the lossy SSH ring, the extracted disorder power density peaks at the gain-side coupling node N17 and decreases away from the coupling interface.
			\textbf{b,} For a $0.2\,\mathrm{V}$ excitation applied at node N17 in the gain SSH ring, the largest fluctuation occurs at node N32, while N17 shows the second-largest fluctuation because of the stabilizing effect of the stiff source attached at the injection node. The disorder power density is obtained from the temporal amplitude and phase fluctuations of the measured waveforms.}
		\label{FigSuppMatDisorderProfile}
	\end{figure}

	Figure~\ref{FigSuppMatDisorderProfile}a shows the extracted disorder profile when a $1\,\mathrm{V}$ signal is injected at node N16 in the lossy SSH ring. The disorder power density is strongly concentrated near the coupling interface, with the largest value at node N17 and the second-largest value at the neighbouring coupling node N16. Away from the interface, the disorder decreases rapidly towards the bulk. This observation indicates that the dominant source of instability is the coupling region containing the negative-resistance converter and the tunable coupling network.

	Figure~\ref{FigSuppMatDisorderProfile}b shows a second measurement in which the excitation is applied directly to node N17 in the gain SSH ring. In this case, the largest disorder power density is observed at node N32, and node N17 has the second-largest value. At first sight, this appears inconsistent with the previous measurement. However, the wave generator is connected directly to N17 and therefore acts as a stiff source, which partially suppresses local waveform fluctuations at the driven node. The measured disorder is consequently redistributed towards the neighbouring sites. Since the coupling between N17 and N32 is stronger than that between N17 and N18, the largest fluctuation appears at N32. Despite this source-induced redistribution, the disorder remains concentrated around the coupling interface.

	The two measurements therefore lead to the same physical conclusion. The dominant disorder originates at the coupling boundary between the gain and loss circuits and decays away from the interface. This observation directly motivates the disorder profile used in the numerical simulations, in which the gain disorder decays from node 17 over a few nodes (Eq.~\eqref{EqSuppMatLocalizedDisorder} with $\alpha=10$, Supplementary Section~\ref{SuppMatDisorderSettings}). More generally, it suggests that realistic disorder in the experimental system is spatially localized rather than uniformly distributed over the lattice.

	\section{Circuit implementation and measurement details}
	\label{SuppMatImplementation}

	\subsection{Experimental setup}
	\label{SuppMatExperimentalSetup}

	\begin{figure}[t!]
		\centering
		\includegraphics[width=0.9\textwidth]{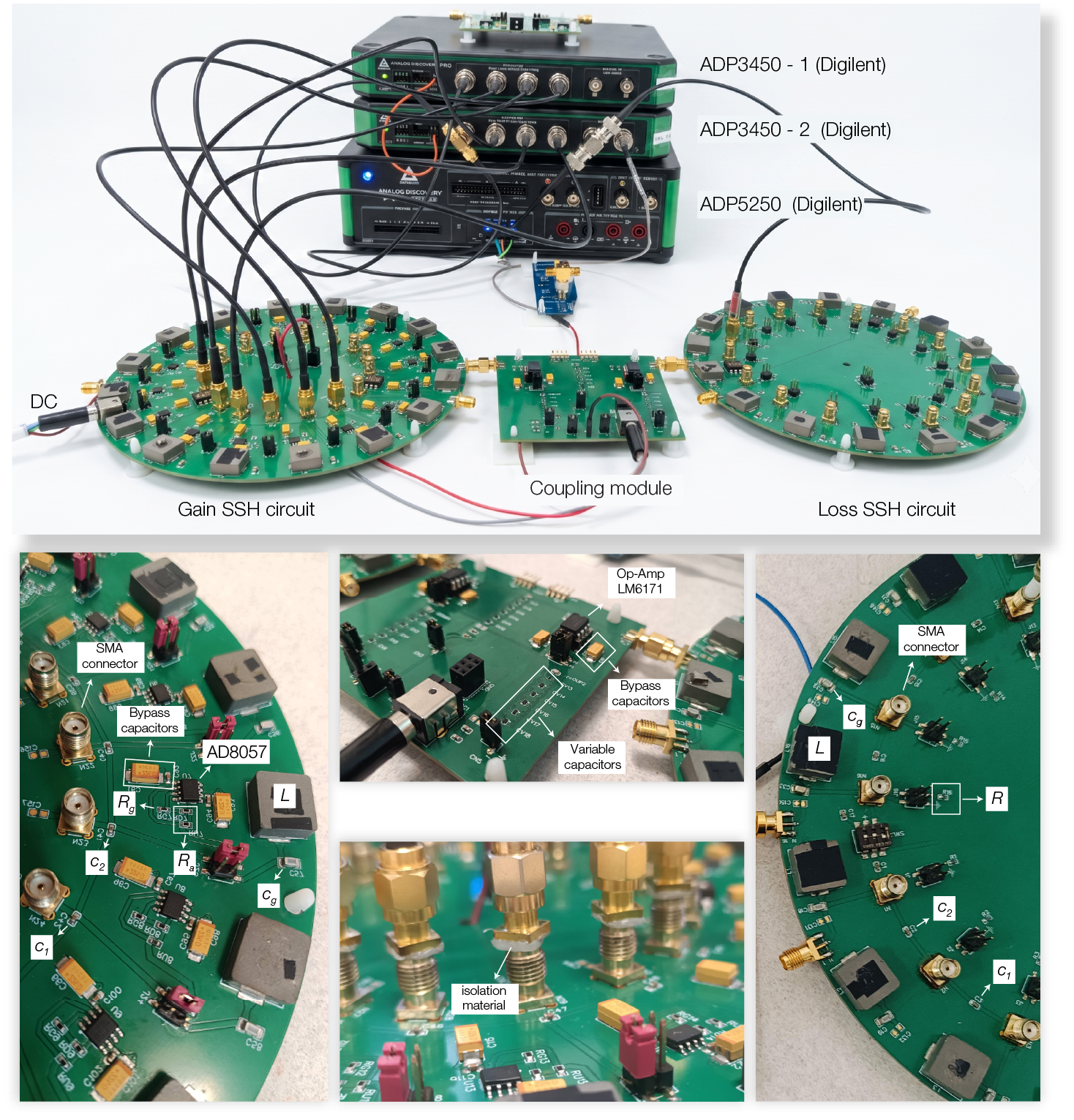}
		\caption{\textbf{Photographs of the topological PT circuit and measurement setup.} The platform consists of gain and loss SSH circuits connected through a varactor-based coupling module, and it is characterized with two ADP3450 units and one ADP5250 unit, which are all-in-one benchtop test and measurement stations connected to a PC. The lower panels show representative circuit elements, including the AD8057-based negative-impedance converter, the LM6171 control circuitry, inductors, capacitors, resistors, SMA connectors, bypass capacitors and isolation structures, which isolate the ground shields of the coaxial cables to prevent loading through the return path.}
		\label{FigSuppMatExperimentalSetup}
	\end{figure}

	Figure~\ref{FigSuppMatExperimentalSetup} presents the experimental implementation of the topological PT circuit. The platform consists of a 16-node gain SSH circuit, a 16-node loss SSH circuit and an intermediate coupling module mounted on separate printed circuit boards. The gain SSH circuit implements the negative-resistance network with AD8057-based impedance converters, while the loss SSH circuit contains the corresponding dissipative elements. The coupling module contains the varactor-based tunable coupling network together with the LM6171 control circuitry used to adjust the inter-lattice coupling $\kappa$.

	Electrical characterization was performed with two ADP3450 units and one ADP5250 impedance analyser. The ADP3450 instruments were used for waveform acquisition, spectral measurements and network characterization, while the ADP5250 was used for the high-resolution impedance measurements of Fig.~\ref{figMain_Experiment}c and provided the DC bias voltages for the active and tunable circuit elements. The impedance sweeps of Fig.~\ref{figMain_Experiment}d,e were taken with the ADP3450. The lower panels of Fig.~\ref{FigSuppMatExperimentalSetup} show representative circuit elements, including the alternating capacitors $c_1$ and $c_2$, grounding capacitors $c_g$, grounding inductors $L$, gain and loss resistors $R_g$ and $R$, negative-impedance converter resistors $R_a$, SMA measurement ports, and the isolation structures used to decouple the outer shield of the coaxial cables from the circuit ground during selected measurements. Bypass capacitors were placed at all op-amp supply pins to suppress power-supply noise. Ferrite sheets attached to the inductors were used to reduce the effective inductance tolerance, as discussed in Supplementary Section~\ref{SuppMatNodeTuning}. Jumper headers, DIP switches and auxiliary SMA connectors were added to facilitate reconfiguration and experimental access during circuit characterization.

	\subsection{Tuning of the node resonances with ferrite sheets and magnets}
	\label{SuppMatNodeTuning}

	\begin{figure}[h]
		\centering
		\includegraphics[width=\textwidth]{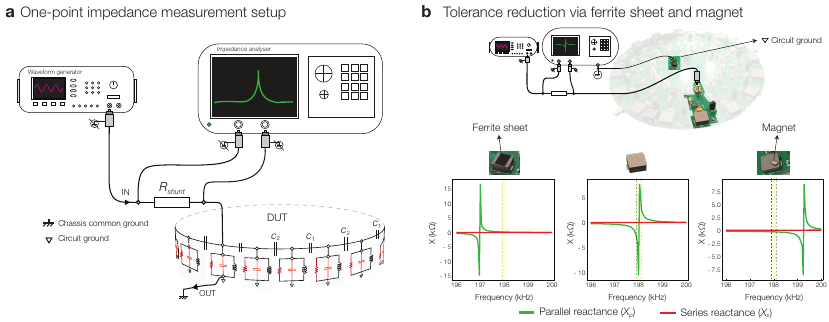}
		\caption{\textbf{Fine-tuning of the node resonance frequencies with ferrite sheets and magnets (Experiment).} Through iterative tuning with suitably sized ferrite sheets and magnets, all node resonances were aligned to the target frequency with an overall effective static tolerance below 0.05\%.
			\textbf{a,} The one-point impedance measurement setup is used to identify and tune the local resonance of each node in an isolated SSH circuit board, where the impedance response is measured across a shunt resistor while the excitation frequency is swept.
			\textbf{b,} The resonance mismatch is reduced through local magnetic control. A suitable piece of adhesive ferrite sheet placed on top of the inductor increases the local magnetic permeability, enhances the flux linkage and shifts the resonance to lower frequencies. Conversely, a small magnet partially cancels the self-field of the inductor, reduces the effective inductance and shifts the resonance upward. The resonance frequency is identified at the intersection of the parallel ($X_p$, green) and series ($X_s$, red) reactances.}
		\label{FigSuppMatNodeTuning}
	\end{figure}

	As the simulations of Supplementary Section~\ref{SuppMatLatticeSize} show, reducing the component tolerance substantially improves the stability of the ultrahigh-Q resonance, because it relies on a precise interaction between the modes of the lattice. Although each component was preselected before assembly, which reduced the overall tolerance to approximately 2\%, this precision is still insufficient for the gain SSH circuit. The effective tolerance of the gain lattice is amplified by the active elements, so that even small deviations appear much larger in the overall response. This effect is visible in the measured eigenvalue spectra of our circuit (Extended Data Fig.~\ref{FigMethods_Spectra}), where the eigenvalues with negative real parts, which belong to the gain SSH circuit, are more scattered than those of the lossy SSH circuit on the positive side of the real axis.

	Because the component values at each node determine its local resonant frequency, uniform resonance across all nodes is essential. To minimize the frequency mismatch, we developed a fine-tuning approach that allows the resonant frequency of each node to be adjusted to a target value, which compensates for the residual tolerances of the components. To implement this method, we first performed one-point impedance measurements, as illustrated in Fig.~\ref{FigSuppMatNodeTuning}a. In these measurements, an AC voltage was injected through a shunt resistor while the circuit ground was connected to the common ground reference.

	To minimize nonuniformities in the local resonances, we used ferrite sheets and small permanent magnets to fine-tune the resonant frequency of each node. The self-inductance of an inductor is $L = \Phi / I$, where $\Phi$ is the magnetic flux linked to the coil and $I$ is the current through it. The local magnetic environment therefore directly affects $\Phi$ and modifies the effective inductance. When a ferrite sheet is placed near the inductor, it increases the magnetic permeability of the surrounding region and increases $\Phi$, which gives a higher effective inductance and a downward shift of the resonant frequency. Conversely, when a small magnet is placed near the inductor, its field partially counteracts the self-field of the inductor and reduces the net flux linkage and thus the inductance, which shifts the resonance towards higher frequencies.

	As shown in Fig.~\ref{FigSuppMatNodeTuning}b, these effects are clearly observed in the changes of both the parallel ($X_p$) and series ($X_s$) reactances, where the resonance occurs at the frequency at which $X_p$ and $X_s$ intersect. During the tuning process, we continuously monitored the impedance response with the impedance analyser function while sweeping the frequency. By placing and adjusting small adhesive ferrite pieces on top of each inductor, or magnets where a correction towards higher frequencies was required, we aligned all local resonances to the desired reference frequency. This procedure reduced the residual frequency mismatch among the nodes to approximately 0.05\% across the SSH circuits.

	\subsection{Stability of the negative-impedance converters}
	\label{SuppMatNICStability}

\begin{figure}[h]
		\centering
		\includegraphics[width=0.85\textwidth]{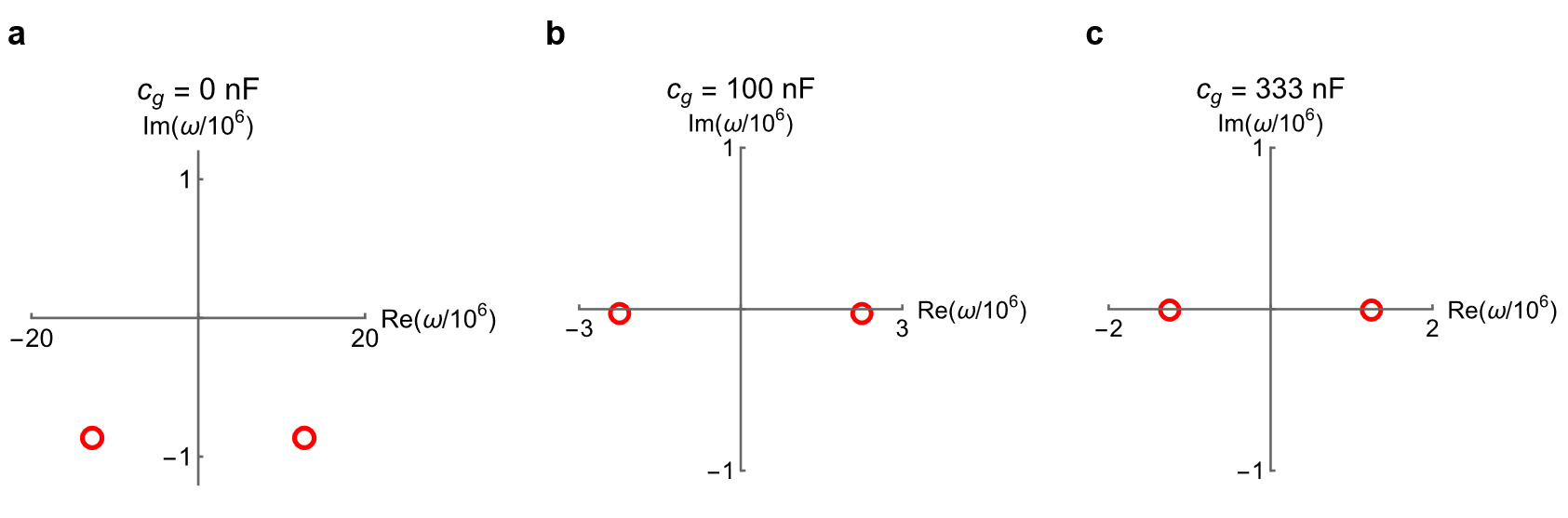}
		\caption{\textbf{Stabilizing gain nodes through a parallel common grounding admittance $c_g$ (Calculated).} Particular care must be taken to stabilize the operational amplifiers that realize the gain against self-oscillation, that is, an oscillation without an AC drive while the op-amps are powered.
			\textbf{a,} Without a grounding capacitor ($c_g = 0\,\text{nF}$), the roots have a large negative imaginary part, which corresponds to a fast-growing oscillation.
			\textbf{b,} For $c_g = 100\,\text{nF}$, the roots move closer to the real axis, but the growth remains noticeable.
			\textbf{c,} At $c_g = 333\,\text{nF}$, the value used in the experiment, the imaginary part is reduced to about $0.7\%$ of the real part. In the assembled circuit, the coupling to the lossy ring and the parasitic losses damp all modes, and no self-oscillation was observed (Fig.~\ref{FigSuppMatTemporalWaveforms}). The parameters are $c_1=1.005\,\text{nF}$, $c_2=2.212\,\text{nF}$, $L = 1.915\,\mu\text{H}$ and $R_g=180\,\Omega$.
		}
		\label{FigSuppMatNICStability}
	\end{figure}

	Adding a resistive load to each node is a common way to stabilize operational amplifiers, but it is not compatible with balanced gain and loss. A negative-impedance converter with current inversion (INIC) realizes a negative resistance through positive feedback~[S1], and its stability depends on the impedance seen at its port. For the INICs used here, a low port impedance at high frequencies, where the loop gain of the op-amp is still large, suppresses parasitic high-frequency oscillation. The large grounding capacitor $c_g = 333\,\text{nF}$ provides this low impedance ($|Z| \approx 2.4\,\Omega$ at $200$\,kHz and lower above), and it acts as a short circuit for high-frequency noise at every node. The capacitor also changes the natural dynamics of a gain node. Its admittance is
	\begin{equation}
		Y_{-}(\omega) = -\frac{1}{R_g} + \frac{1}{i \omega L} + i \omega c_\Sigma,
		\label{EqSuppMatStabilityPoles}
	\end{equation}
	with $c_\Sigma=c_1+c_2+c_g$, $c_1=1.005$\,nF, $c_2=2.212$\,nF, $L=1.915\,\mu$H and $R_g=180\,\Omega$, the negative resistance of the converters in our experiment (Methods). Its natural frequencies follow from $Y_-(\omega)=0$, that is, from $1-\omega^2Lc_\Sigma-i\omega L/R_g=0$, whose roots are $\omega = \pm\sqrt{1/(Lc_\Sigma)-1/(4R_g^2c_\Sigma^2)} - i/(2R_gc_\Sigma)$. With the time dependence $e^{i\omega t}$, the negative imaginary part describes a growth rate $1/(2R_gc_\Sigma)$ of an isolated gain node. Increasing $c_g$ reduces this rate relative to the oscillation frequency, $|\mathrm{Im}\,\omega|/\mathrm{Re}\,\omega \approx \sqrt{L/c_\Sigma}/(2R_g)$, from a strong instability at $c_g=0$ to about $0.7\%$ at $c_g = 333$\,nF (Fig.~\ref{FigSuppMatNICStability}). An isolated node with net negative conductance is, however, never strictly stable. In the assembled circuit, the gain nodes are coupled to the lossy ring and carry parasitic losses (inductor core and series resistances, Methods), and all natural modes of the full network are damped. The absence of any self-oscillation in the `AC off' measurements (Supplementary Section~\ref{SuppMatTemporalWaveforms}) confirms this.

	\subsection{Capacitive loading by the measurement channels}
	\label{SuppMatCapacitiveLoading}

	In our experimental setup, we identified a parasitic capacitive loading of about 82\,pF per channel due to the measurement equipment, originating mainly from the coaxial cables, the input channels and stray wiring. For the Laplacian matrix measurements, we used grounded Johnson/Cinch Connectivity Solutions RG174 50\,cm 50 Ohm coaxial cables (415-0199-MM500), whose datasheet specifies a capacitance of 105\,pF/m. Since the outer conductors of all coaxial cables were connected to the common ground during these measurements, this capacitance acts fully as a shunt load to ground. In addition, the datasheet of the ADP3450 specifies an input capacitance of 15\,pF for each analog channel, and we estimated a stray capacitance of $\sim$15\,pF from several manual measurements. Together, these parasitic loads amount to $52.5+15+15\approx82$\,pF per probed node. The return current of this capacitance flows through the grounded outer conductor of the same cable, so it is already included. The cable that injects the signal does not add a further load, because the voltage of the driven node is set by the waveform generator and the transfer ratios to the other nodes do not depend on it. To quantify the influence of this loading, we evaluate the unloaded and loaded resonant frequencies of the circuit,
	\begin{equation}
		f_{\text{unloaded}}=\frac{1}{2\pi\sqrt{L(c_1+c_2+c_g)}},\qquad
		f_{\text{loaded}}=\frac{1}{2\pi\sqrt{L(c_1+c_2+c_g+c_\mathrm{load})}},
	\end{equation}
	where $c_\mathrm{load}=82$\,pF per channel. For one node, this load lowers the resonance frequency by about 24\,Hz. In the network-analyser measurements, six coaxial channels were connected simultaneously, each to a distinct node, which gives a total load of $6\times82$\,pF\,$\approx\,490$\,pF. The shift of a mode that extends over several nodes depends on its weight on the loaded nodes, so adding the six loads gives an upper estimate of the shift. With our component values $c_1=1.005$\,nF, $c_2=2.212$\,nF, $c_g=333$\,nF and $L=1.915\,\mu$H, the corresponding frequency shift is
	\begin{equation}
		\Delta f = f_{\text{unloaded}}-f_{\text{loaded}} \lesssim 145~\text{Hz},
		\qquad
		\frac{\Delta f}{f_{\text{unloaded}}}\lesssim 7\times10^{-4}.
	\end{equation}
	This loading lowers the frequencies of the measured spectra relative to the impedance measurements, in which the channels are not grounded. Together with the different grounding (Supplementary Section~\ref{SuppMatGrounding}), it contributes to the frequency offset between the two datasets.

	\SIpart{Numerical simulations}

	\section{Disorder simulations of the four circuit models}
	\label{SuppMatDisorderSimulations}

	\subsection{Models, operating points and disorder settings}
	\label{SuppMatDisorderSettings}

Here, we give additional details of the disorder simulations presented in the main text. To understand how different resonant architectures respond to imperfections, we compare four representative systems, namely a conventional 1-Node LC resonator, a 2-Node PT dimer, a gapless PT lattice and the topological PT lattice (Figs.~\ref{FigSuppMatDisorderAnalyses}a--d). The gapless PT lattice serves as an additional reference. It has the same coupled-ring geometry as the topological case but uniform couplings ($c_1=c_2$).

	For a fair comparison, all four systems are first tuned to show comparable high-Q impedance resonances without disorder ($|Z| \sim 10^7\, \Omega$). Each disorder-free spectrum therefore contains at least one eigenvalue at the spectral origin, which gives the impedance peak shown in the adjacent panel. These eigenvalues are highlighted by the red markers in the middle column. In the impedance plots, the coloured markers indicate the frequencies at which individual eigenvalues approach the spectral origin. As discussed in Supplementary Section~\ref{SuppMatFreqShift}, varying the excitation frequency translates the entire spectrum along the imaginary axis. Different eigenvalue trajectories therefore reach the spectral origin at different frequencies and give the multiple impedance peaks visible in the black-background impedance spectra. For the topological PT lattice, the blue, green and yellow markers indicate additional resonances associated with the other EP trajectories.

	All simulations use the values of the assembled components, $c_1=1.005$\,nF, $c_2=2.212$\,nF, $c_g=333$\,nF, $L=1.915\,\mu$H and $R=270\,\Omega$, with $N=16$ nodes per ring in the lattice models. The operating couplings are $\kappa=4.688$\,nF for the topological PT lattice, $3.607$\,nF for the gapless PT lattice and $3.072$\,nF for the 2-Node PT dimer, at which one eigenvalue lies at the origin in the clean limit (Methods).

	The disorder profile is not chosen arbitrarily. As shown experimentally in Supplementary Section~\ref{SuppMatDisorderProfile}, the dominant fluctuations originate at the gain--loss coupling interface, where the tunable coupling capacitors and the active gain elements are located. Physically, these components are the primary source of instability, because fluctuations of amplitude and phase are amplified by the negative-resistance circuitry. Motivated by these measurements, we model the disorder as a perturbation localized near the coupling interface rather than distributing it uniformly over the lattice. Specifically, the local negative resistance is randomized according to
	\begin{equation}
		\frac{1}{R_n(x)}=\frac{1}{R}\left[1+\xi_x\,\Theta_{R_n}\exp\left(-\alpha\,\frac{d(x,x_c)}{N/2}\right)\right],
		\label{EqSuppMatLocalizedDisorder}
	\end{equation}
	where $\xi_x\in[-1,1]$ is drawn independently and uniformly for each node $x$ of the gain ring, $\Theta_{R_n}$ denotes the disorder strength, $x_c$ is the gain-side coupling node (node 17), $d(x,x_c)$ is the number of bonds between $x$ and $x_c$ along the ring, and $\alpha$ controls the spatial localization of the disorder. Large values of $\alpha$ concentrate the disorder near the coupling node, whereas $\alpha=0$ reduces Eq.~\eqref{EqSuppMatLocalizedDisorder} to a spatially uniform disorder distribution. For all simulations, we used $\alpha=10$, comparable to that of the measured profile (Supplementary Section~\ref{SuppMatDisorderProfile}). The loss elements keep their nominal values. The ideal Laplacian with these settings reproduces the eigenvalues of LTspice simulations with operational-amplifier models (AD8057 for the gain and LT1363 for the $-\kappa$ elements (equivalent to LM6171 based on their specifics)), with a median deviation of $2\times10^{-5}$\,S for the lattice and $7\times10^{-5}$\,S for the dimer.
	Similarly, the coupling capacitance is randomized according to
	\begin{equation}
		\kappa \rightarrow \kappa \left( 1+\Theta_{\kappa}\xi \right).
	\end{equation}
	For the 1-Node LC resonator, which contains no coupling interface, disorder is instead introduced directly into the onsite capacitor ($\pm0.5\%$ of $c_g$). Like $\kappa$, $c_g$ is a reactive element, and this variation spreads the eigenvalue along the imaginary axis by about as much as the variation of $\kappa$ does in the dimer (Methods). The exponentially localized profile in Eq.~\eqref{EqSuppMatLocalizedDisorder} reproduces the experimentally observed concentration of disorder near the coupling boundary and therefore represents the dominant instability sources of the physical circuit realistically.

	For the impedance statistics of Fig.~\ref{figMain_Simulations}b, the gain disorder is $\Theta_{R_n}=1\%$, and each of the 500 realizations is evaluated at 50 drive frequencies drawn uniformly within $f(1\pm5\times10^{-6})$, a window of about 2\,Hz. The reactive element varies by less than in Fig.~\ref{figMain_Simulations}a, namely $\Theta_\kappa=1\%$ for the topological PT lattice and $3.7\%$ for the 2-Node PT dimer, and $c_g$ varies by $0.25\%$ for the 1-Node LC resonator. In admittance, these are about $\pm0.06$, $\pm0.14$ and $\pm1.0$\,mS. They are chosen so that the eigenvalue nearest the origin moves over a similar range along its trajectory in the three models. In the circuit, the auxiliary $-\kappa$ elements are negative-impedance converters, which cancel the coupling capacitance at the two coupling nodes only approximately. To include this, the $-\kappa$ elements at nodes 16 and 17 vary independently by a relative amount of $5\times10^{-5}$ around the realized $\kappa$. The spectra of Fig.~\ref{figMain_Simulations}a and Fig.~\ref{FigSuppMatDisorderAnalyses} are computed in LTspice with operational-amplifier models, and the impedances of Fig.~\ref{figMain_Simulations}b by direct inversion of the circuit Laplacian (Methods).

	In Fig.~\ref{FigSuppMatDisorderAnalyses} and Fig.~\ref{figMain_Simulations}a of the main text, the coupling-capacitance tolerance is fixed at $\Theta_{\kappa}=10\%$, and the negative-resistance tolerance $\Theta_{R_n}$ is increased from $0.1\%$ to $10\%$ to isolate its influence. For each tolerance window, 400 independent disorder realizations are generated and their admittance eigenvalue spectra are overlaid. For every realization, a new set of randomized circuit parameters is generated and the complete eigenvalue spectrum is calculated. The coloured points in the right columns are the accumulated eigenvalues of all realizations, and the colour scale indicates the weight of each eigenvalue in the impedance between the coupling nodes, $w_n=(\hat\psi_{n,a}-\hat\psi_{n,b})^2/(\lambda_n+\epsilon)$, the term of Eq.~\eqref{EqImpEig} for this eigenvalue evaluated with the unit-normalized right eigenvector $\hat\psi_n$ and $\epsilon=10^{-12}$\,S. Regions with warmer colours therefore correspond to eigenvalues that contribute more strongly to the impedance response, whereas diffuse eigenvalue clouds indicate stronger disorder-induced spectral broadening.

	\subsection{Simulated spectra under disorder}
	\label{SuppMatDisorderSpectra}

	\begin{figure}[h]
		\centering
		\includegraphics[width=\textwidth]{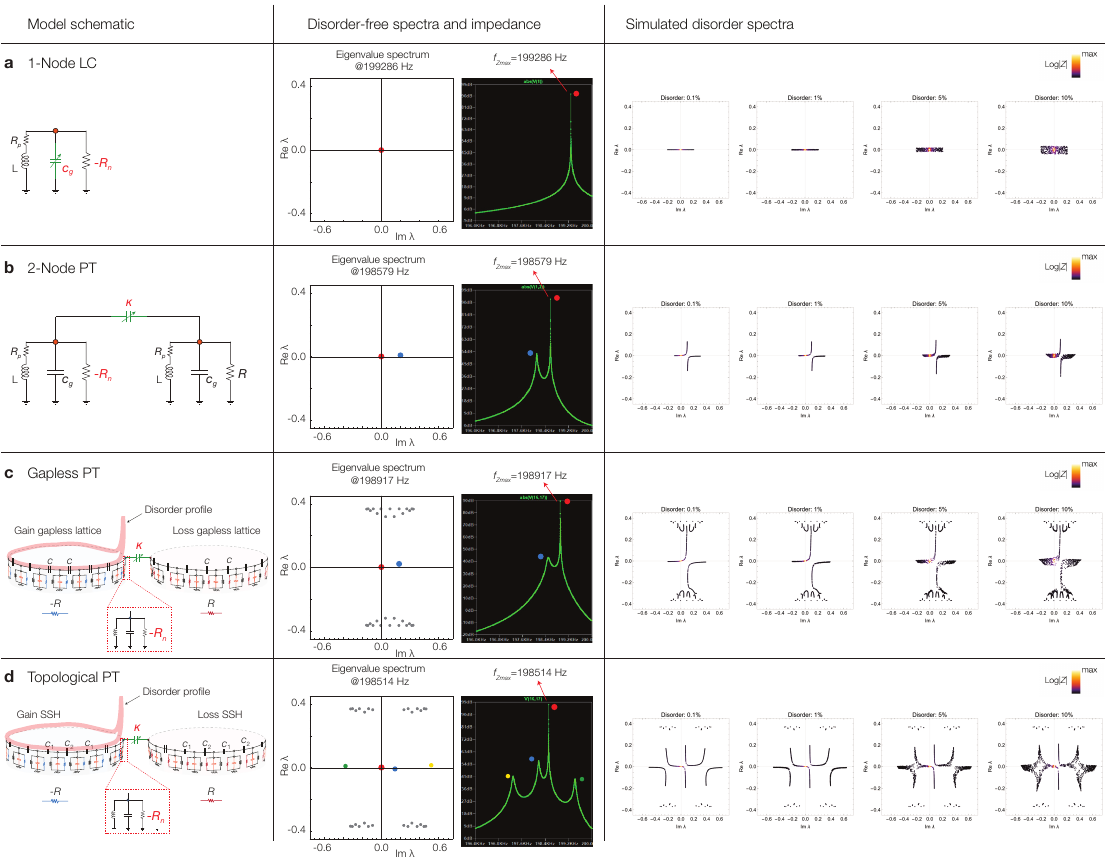}
		\caption{\textbf{Simulated spectra of the four circuit models under disorder and simulation procedure.} Disorder broadens the eigenvalue clouds in all four models, but the topological PT lattice shows much less spreading near the spectral origin, particularly along the central exceptional-point trajectory responsible for the ultrahigh-Q resonance. The rows show the
			\textbf{a,} 1-Node LC resonator,
			\textbf{b,} conventional 2-Node PT dimer,
			\textbf{c,} gapless PT lattice and
			\textbf{d,} topological PT lattice. For each model, the left column shows the circuit schematic and the disorder profile considered, and the middle column shows the disorder-free eigenvalue spectrum together with the corresponding impedance resonance. All models are first tuned to a comparable clean-limit impedance peak of $\sim 10^7\,\Omega$, which arises from one eigenvalue at the origin (red) at the drive frequency indicated above each admittance spectrum. The right columns show the eigenvalue spectra under increasing disorder. The coupling capacitance is varied randomly within a $\Theta_{\kappa}=10\%$ window to examine the dynamical disorder over a wide range of the trajectories, and the negative-resistance disorder is increased from $\Theta_{R_n}=0.1\%$ to $10\%$ with the experimentally determined profile (Supplementary Section~\ref{SuppMatDisorderSettings}), where $R_n$ is the negative resistance at node $n$. For the 1-Node LC resonator, $c_g$ varies by $\pm0.5\%$ instead of $\kappa$, which gives a similar spread along the imaginary axis in units of mS. The colours denote the weight of each eigenvalue in the impedance between the coupling nodes (through $1/|\lambda|$ for the single node of the 1-Node LC resonator), accumulated over 400 disorder realizations.}
		\label{FigSuppMatDisorderAnalyses}
	\end{figure}

	Figure~\ref{FigSuppMatDisorderAnalyses} summarizes the disorder simulations. The left column shows the circuit schematics, and the red labels indicate the components to which disorder is applied. The middle column presents the disorder-free eigenvalue spectra together with the corresponding impedance responses. The frequency indicated above each spectrum is the frequency of the maximum impedance resonance, namely $f_{Z_{\mathrm{max}}}=199286\,\mathrm{Hz}$, $198579\,\mathrm{Hz}$, $198917\,\mathrm{Hz}$ and $198514\,\mathrm{Hz}$ for the 1-Node LC, 2-Node PT, gapless PT lattice and topological PT lattice, respectively. These frequencies are then used as the injection frequencies to evaluate the disorder-induced spectra shown in the right columns.

	The four systems respond in qualitatively different ways. In the 1-Node LC resonator (Fig.~\ref{FigSuppMatDisorderAnalyses}a), the eigenvalues broaden into a nearly rectangular cloud around the spectral origin. Since no spectral mechanism holds the resonance close to zero, the resonance quickly becomes sensitive to parameter fluctuations. The 2-Node PT dimer (Fig.~\ref{FigSuppMatDisorderAnalyses}b) shows the characteristic EP trajectory in the clean limit. Under disorder, however, the EP unfolds into an avoided crossing and the eigenvalue cloud spreads broadly around the EP region. As $\Theta_{R_n}$ increases, the eigenvalues deviate more and more from their clean-limit trajectories, and the spectral broadening near the origin increases substantially.

	The gapless PT lattice in Fig.~\ref{FigSuppMatDisorderAnalyses}c embeds the same PT gain--loss structure in a lattice without a band gap. Its clean-limit spectrum is qualitatively similar to that of the isolated PT dimer, and its disorder response follows the same trend. The eigenvalues scatter increasingly around the EP trajectory as $\Theta_{R_n}$ increases, which shows that embedding in a lattice alone does not suppress disorder-induced spectral broadening.

	The topological PT lattice behaves markedly differently. As shown in Fig.~\ref{FigSuppMatDisorderAnalyses}d, its spectrum contains three exceptional-point trajectories, labelled EP$_\mathrm{c}$, EP$_\mathrm{o1}$ and EP$_\mathrm{o2}$. While the outer trajectories (EP$_\mathrm{o1}$ and EP$_\mathrm{o2}$) broaden similarly to those of the gapless lattice, the central trajectory EP$_\mathrm{c}$, which is responsible for the ultrahigh-Q resonance, remains much more confined. Even at the largest disorder strengths, the eigenvalues approaching the spectral origin spread much less than in the other three models. Since the resonance linewidth is governed by the width of the eigenvalue cloud around the spectral origin, this reduced broadening directly corresponds to a smaller spectral width $\Delta W$ and, through Eq.~\eqref{EqWidthQ} of the main text, a higher attainable Q factor.

	These differences follow the broadening coefficient $\eta$ of a gain fluctuation that reaches the linewidth (Eq.~\eqref{EqGainShare} of the main text). At the operating points used here, the full Laplacians give $\eta=1$ for the single resonator, $\eta=0.50$ for the dimer at any coupling, $\eta=1.04$ for the gapless lattice and $\eta=-0.28$ for the topological lattice (Supplementary Section~\ref{SuppMatGainFluctuations}). The gapless lattice has a single EP at our gain--loss strength (Supplementary Section~\ref{SuppMatThreeEPs}), and its broadening coefficient is larger than that of the dimer, which explains why lattice embedding alone does not suppress the broadening. With $\kappa$ fixed and a gain disorder of 1\% with the profile of Eq.~\eqref{EqSuppMatLocalizedDisorder}, the standard deviation of $\mathrm{Re}\,\lambda$ near the origin is 0.72 times that of the dimer for the topological PT lattice and 2.1 times for the gapless PT lattice, so that the order of the clouds follows $\eta$. With disorder at node 17 alone, the ratio for the topological PT lattice is 0.54, close to $|\eta_\mathrm{c}|/\eta_\mathrm{dimer}$. The difference comes from node 32, the neighbour of node 17 across $c_2$, on which the resonant mode has a real weight of $+0.80$ (Supplementary Section~\ref{SuppMatGainFluctuations}). Weaker gain and loss, $R>398\,\Omega$, lie outside the regime considered here (Methods).

	These simulations show that the ultrahigh-Q resonance of the topological PT lattice does not originate from exceptional-point physics alone. It arises from the combined action of the exceptional-point trajectories and the topological band gap of the SSH lattice, which directs the response of EP$_\mathrm{c}$ to gain fluctuations into the resonance frequency and strongly suppresses disorder-induced broadening near the spectral origin.

	\section{Topological classification and clean-limit evolution of the coupled rings}
	\label{SuppMatCleanLimit}

	\begin{figure}[h]
		\centering
		\includegraphics[width=\textwidth]{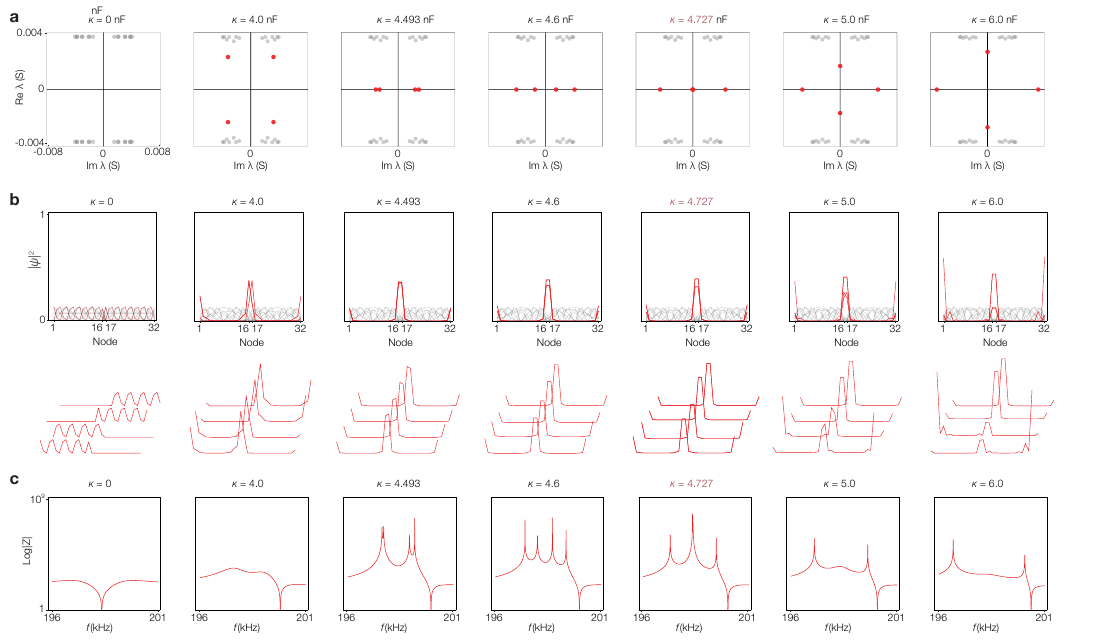}
		\caption{\textbf{Evolution of eigenvalues, eigenmodes and impedance spectra of the topological PT lattice (Calculated).} With increasing coupling, the interface modes evolve from defect-like states into the boundary modes of the topological SSH rings.
			\textbf{a,} The complex eigenvalue spectra of the circuit Laplacian are shown for varying $\kappa$. At $\kappa=0$, the two SSH circuits are uncoupled and show mirror-symmetric bulk bands about the imaginary axis. As $\kappa$ increases, four interface-related eigenvalues separate from the bulk bands, and around $\kappa=4.48\,\text{nF}$ they reach the imaginary axis, which marks the onset of undamped hybridized modes. At $\kappa=4.727\,\text{nF}$, the two eigenvalues closest to the origin coalesce there (EP$_\mathrm{c}$). Beyond this point, this pair becomes purely real, with one eigenvalue on the gain side and one on the loss side, and moves through the bulk gap.
			\textbf{b,} The spatial amplitude profiles $|\psi|^2$ of all eigenmodes (upper row) and of the four interface eigenmodes (bottom row) show that all modes are extended for small $\kappa$, whereas localized modes appear at the interface nodes 16 and 17 near the transition. Beyond $\kappa=4.727\,\text{nF}$, the real-valued eigenmodes are exponentially localized on alternating sites, the signature of the SSH interface modes, while the imaginary-valued ones remain confined at the interface.
			\textbf{c,} The theoretical impedance spectra $\log|Z|$ for representative values of $\kappa$ are featureless for weak coupling, consistent with the absence of localized modes. As $\kappa$ approaches the transition, sharp impedance resonances appear at the frequencies of the undamped (purely imaginary) eigenvalues, whose clean-limit Q is limited only by the frequency resolution (Fig.~\ref{figMain_CleanLimit}g). Beyond EP$_\mathrm{c}$ at $\kappa=4.727\,\text{nF}$, two main impedance peaks remain, which come from the pair of eigenvalues that stays purely imaginary. The parameters are $c_1 = 1.005\,\text{nF}$, $c_2 = 2.212\,\text{nF}$, $c_g = 0.333\,\mu\text{F}$, $L = 1.915\,\mu\text{H}$, $R=270\,\Omega$, $f = 198347\,\text{Hz}$ and $N=16$ per ring.}
		\label{FigSuppMatCleanLimitSpectra}
	\end{figure}

	This section first sets out how we classify the coupled rings topologically and then follows the coupling modes from defect-like states to topological boundary states.

	\subsection{Winding number of a closed SSH ring}
	\label{SuppMatWindingRing}

	Unlike generic gapless one-dimensional lattices, the Su--Schrieffer--Heeger (SSH) lattice has a gapped bulk spectrum except at the transition, which allows a bulk topological invariant~[S4]. For a chiral-symmetric two-band model with off-diagonal block $h(k)$, the winding number is
	\begin{equation}
		\mathcal{W}=\frac{1}{2\pi i}\int_{-\pi}^{\pi} dk \frac{d}{dk} \log h(k),
		\label{EqSuppMatWindingNumber}
	\end{equation}
	where for the capacitive SSH lattice we may take $h(k) = c_1 + c_2 e^{i k}$ (a global factor such as $-i\omega$ does not affect $\mathcal{W}$). A nonzero winding ($|c_2|>|c_1|$) corresponds to a loop of $h(k)$ around the origin of the complex plane and gives boundary-localized states under open boundary conditions (OBC).

	For a closed ring under periodic boundary conditions (PBC), translational symmetry holds, and $\mathcal{W}$ depends only on which bond is taken as intracell. Taking the other bond exchanges $c_1$ and $c_2$ in $h(k)$ and changes $\mathcal{W}$ from 1 to 0, and for an isolated ring both choices are equally valid~[S4]. An isolated ring is therefore neither trivial nor nontrivial in an absolute sense, and the only property of its bulk that does not depend on this choice is the band gap $\omega|c_2-c_1|$, which closes only at $c_1=c_2$, where $\mathcal{W}$ changes. The familiar rule that $c_2>c_1$ is nontrivial and $c_2<c_1$ is trivial assumes that a boundary fixes the unit cell, as the two ends of an open chain do. It cannot be carried over one to one to a ring under PBC, and an open chain with $c_2<c_1$ is a physically different system, with two ends and different boundaries, rather than a trivial counterpart of our rings.

	\subsection{Interface mode at the coupling node}
	\label{SuppMatInterfaceMode}

	The coupling, however, removes this freedom, since it breaks the translational symmetry and provides a natural reference point that fixes the unit cell at the coupling node, in the same way as the end of an open chain or the domain wall of an SSH heterostructure does~[S4]. With the unit cells of Eq.~\eqref{EqMethodsSSH} in Methods, where each coupling node forms a unit cell with its neighbour across $c_1$, each ring has $\mathcal{W}=1$ for $c_2>c_1$. Exchanging $c_1$ and $c_2$ mirrors each ring about its coupling node, so that the mirrored ring has the same spectrum and, read from the coupling node in the opposite direction, the same unit cells and $\mathcal{W}=1$. It therefore describes the same nontrivial system. The physical signature of this phase is the pair of mid-gap interface modes that decay from the coupling nodes on alternating sites (Fig.~\ref{FigSuppMatCleanLimitSpectra}b).

	The same conclusion follows without any choice of unit cell. Seen from its coupling node, each ring continues across $c_1$ on one side and across $c_2$ on the other, and for strong coupling, the remaining $N-1$ nodes act as an open SSH chain whose two ends terminate on different bonds. The end that terminates on the weaker bond has $\mathcal{W}=1$ and the other end has $\mathcal{W}=0$, so the winding numbers on the two sides of the coupling node differ by one for any $c_1\neq c_2$. As at an SSH domain wall, this difference binds one mid-gap mode to the coupling node~[S4,\,S11]. Exchanging $c_1$ and $c_2$ only moves the mode to the other side of the node, and for $c_1=c_2$ the gap closes and the mode spreads over the whole ring. The full 32-node Laplacian confirms this. Beyond EP$_\mathrm{c}$ ($\kappa=6$\,nF), each mode of the real pair carries 58\% of its weight on a neighbour of a coupling node, across $c_2$ for $c_2>c_1$ (node 1 or 32) and across $c_1$ after the exchange (node 15 or 18), with identical spectra. In the gapless lattice, the same pair spreads evenly over eight nodes of each ring and has no weight on the coupling nodes.

	\subsection{Gapped and gapless rings as nontrivial and trivial phases}
	\label{SuppMatTopoClass}

	The topological distinction that is well defined for our rings is therefore between the gapped rings ($c_1\neq c_2$), which bind one interface mode at their coupling node and which we regard as topologically nontrivial, and the gapless ring ($c_1=c_2$), which binds none and serves as the trivial reference. For this reason, we compare gapped and gapless lattices throughout this work (Figs.~\ref{figMain_CleanLimit}d,e and~\ref{figMain_SpectralBroadening}d of the main text and Supplementary Sections~\ref{SuppMatDisorderSpectra} and~\ref{SuppMatThreeEPs}), rather than rings with $c_2>c_1$ and $c_2<c_1$, which are the same system. A single invariant of the whole coupled structure is not required for this classification. The coupled rings form a non-Hermitian heterostructure without translational symmetry, for which the textbook bulk invariant is not defined. Nonlinear lattices are in a similar position, since no Bloch winding number exists for them, and their topology is nevertheless identified through the gap of the underlying lattice and the protected modes it supports~[S12,\,S13].

	\subsection{\texorpdfstring{Role of the band gap in the stabilization of EP$_\mathrm{c}$}{Role of the band gap in the stabilization of EPc}}
	\label{SuppMatGapRole}

	The gapped phase is also what the stabilization of EP$_\mathrm{c}$ needs. The gap enters the input admittance of each ring through the gap edge $Y_\mathrm{gap}=\omega|c_2-c_1|$ (Eq.~\eqref{EqSuppMatDressedY}). It places EP$_\mathrm{c}$ at the centre of a spectral region free of bulk modes, where the interface modes described above form it, and together with the band edge, it sets $\eta_\mathrm{c}$ and $R^*$ (Supplementary Section~\ref{SuppMatGainFluctuations}). All of these depend only on $|c_2-c_1|$, as the mirror argument requires. The three EPs exist for any nonzero gap when the gain and loss are weak enough (Supplementary Section~\ref{SuppMatThreeEPs}), and $\eta_\mathrm{c}$ has a zero only with a gap, whereas the gapless lattice has a single EP at our gain--loss strength and $|\eta|\approx1.1$ (Table~\ref{TabSuppMatEta}). The topology of the rings therefore enters the mechanism through the gapped phase and its interface modes, not through the value of $\mathcal{W}$ for a particular choice of unit cell.

	\subsection{Evolution of the interface modes with the coupling}
	\label{SuppMatInterfaceEvolution}

	In the pristine configuration with two uncoupled SSH circuits, the rings are periodic. Each circuit also carries a uniform onsite admittance shift $\pm1/R$ (corresponding to net loss or gain), which rigidly shifts its bulk bands without changing the topology. As shown for $\kappa = 0$ in Fig.~\ref{FigSuppMatCleanLimitSpectra}a, the SSH bands with opposite onsite shifts are mirror-symmetric about the imaginary axis in the complex admittance plane. Because both lattices are periodic, their eigenmodes are extended bulk states confined to their respective lattices (Fig.~\ref{FigSuppMatCleanLimitSpectra}b, $\kappa=0$). Even when the two lattices are coupled up to $\kappa \approx 4\,\text{nF}$, the impedance spectrum remains featureless without a pronounced peak (Fig.~\ref{FigSuppMatCleanLimitSpectra}c), which indicates the absence of localized or resonant modes.

	As the inter-lattice coupling $\kappa$ increases, four interface-related complex eigenvalues of the circuit Laplacian gradually move away from the bulk bands. Their real parts decrease while their imaginary parts dominate, and the eigenvalues approach the imaginary axis, which corresponds to undamped, purely reactive modes of the circuit. This regime begins around $\kappa = 4.48\,\text{nF}$, where two of the eigenvalues move outward along the imaginary axis while the other pair approaches the origin, forming hybridized interface modes between the two lattices.

	At the critical coupling $\kappa = 4.727\,\text{nF}$, the two eigenvalues closest to the origin coalesce there, forming EP$_\mathrm{c}$. Beyond this point, this pair becomes purely real, with one eigenvalue on the gain side and one on the loss side. In the admittance picture, purely imaginary eigenvalues correspond to lossless (reactive) modes, whereas real eigenvalues correspond to modes with net loss or net gain. EP$_\mathrm{c}$ therefore marks a PT transition of the interface pair, from the PT-symmetric (unbroken) regime, in which each mode is shared equally by gain and loss, to the broken regime. As $\kappa$ increases further, the real pair moves through the band gap, and the corresponding eigenmodes become exponentially localized on alternating sites, as expected for SSH-type boundary modes.

	This evolution is clearly reflected in the spatial eigenmode profiles in Fig.~\ref{FigSuppMatCleanLimitSpectra}b. Just above the outer EPs ($\kappa = 4.493\,\text{nF}$ in Fig.~\ref{FigSuppMatCleanLimitSpectra}b), the hybridized modes are confined to the interface nodes (16 and 17), while all other modes remain extended. Beyond EP$_\mathrm{c}$, the pair of real eigenmodes shows exponential localization over alternating nodes, a signature of SSH-type topological states, whereas the other pair, which corresponds to the imaginary eigenvalues, stays confined as defect-like modes at the interface.

	The corresponding impedance spectra in Fig.~\ref{FigSuppMatCleanLimitSpectra}c show the signatures of this transition. Sharp impedance resonances appear at the frequencies associated with the undamped (purely imaginary) eigenvalues, with a clean-limit Q that is limited only by the frequency step. For $\kappa$ between $4.48$ and $4.727\,\text{nF}$, four distinct resonance peaks appear. They reduce to two prominent peaks beyond EP$_\mathrm{c}$, because one eigenvalue pair becomes purely real (net loss or gain) and no longer gives a sharp peak. This window is the PT-symmetric regime of interest, in which the impedance peaks directly reveal the long-lived, nearly lossless modes that give rise to ultrahigh-Q resonances.

	\section{Shifts of the spectrum by the drive frequency and by gain--loss imbalance}
	\label{SuppMatSpectrumShifts}

	\subsection{Shift along the imaginary axis by the drive frequency}
	\label{SuppMatFreqShift}

	\begin{figure}[h]
		\centering
		\includegraphics[width=\textwidth]{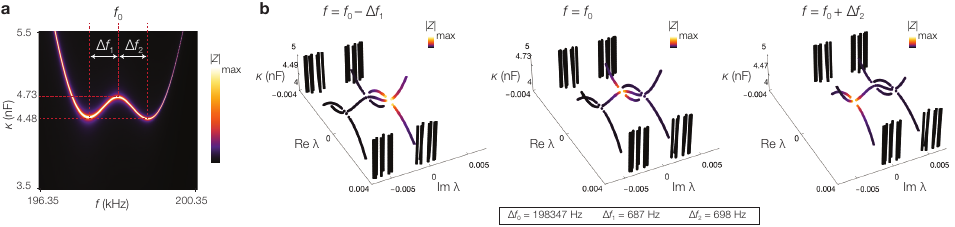}
		\caption{\textbf{The drive frequency shifts the spectrum along the imaginary axis (Calculated).} Since the drive frequency shifts the entire spectrum along the imaginary axis, each lossless eigenvalue reaches the spectral origin at its own pair of $\kappa$ and $f$ and gives a distinct impedance peak.
			\textbf{a,} The colour map of the impedance magnitude $|Z|$ as a function of the coupling capacitance $\kappa$ and the frequency $f$ shows bright branches at the $\kappa$--$f$ pairs with high-impedance resonances. Within the interval $\kappa = 4.48$--$4.73\,\text{nF}$, four distinct impedance peaks appear in a frequency window around $f_0 = 198347\,\text{Hz}$. The dashed horizontal lines mark the representative couplings $\kappa = 4.48$ and $4.73\,\text{nF}$ used in panel~\textbf{b}, and the vertical dashed lines mark the asymmetric detunings $\Delta f_1$ and $\Delta f_2$.
			\textbf{b,} The complex admittance spectra are shown as a function of $\kappa$ at three representative drive frequencies, $f=f_0-\Delta f_1$, $f=f_0$ and $f=f_0+\Delta f_2$, with each eigenvalue colour-coded by its contribution to the two-point impedance. The frequency detuning shifts the entire admittance bands solely along the imaginary axis, with a negative imaginary offset for a drive frequency below the resonance frequency, no offset at the resonance frequency and a positive imaginary offset above it. Each frequency therefore brings a different eigenvalue on the imaginary axis to the spectral origin at a different $\kappa$--$f$ pair. The parameters are $c_1 = 1.005\,\text{nF}$, $c_2 = 2.212\,\text{nF}$, $c_g = 0.333\,\mu\text{F}$, $L = 1.915\,\mu\text{H}$, $R=270\,\Omega$ and $N=16$ per ring.}
		\label{FigSuppMatFreqShift}
	\end{figure}

	Away from the resonance frequency, the onsite admittance $Y_0(\omega)=i\omega c_\Sigma+1/(i\omega L)\simeq2ic_\Sigma(\omega-\omega_0)$ adds the same imaginary amount to every node and therefore shifts the whole spectrum along the imaginary axis. The coupling admittances $i\omega c_{1,2}$ and $i\omega\kappa$ change only by a relative amount of order $(\omega-\omega_0)/\omega_0$, below $4\times10^{-3}$ here. Figure~\ref{FigSuppMatFreqShift}a shows the impedance magnitude as a function of frequency and coupling capacitance $\kappa$. The bright branches highlight $\kappa$--$f$ pairs with strong impedance resonances. Within the range $\kappa \in [4.47,\,4.73]\,\text{nF}$, four distinct impedance peaks appear in a frequency window between $f_0 - \Delta f_1$ and $f_0 + \Delta f_2$, where $f_0 = 198347\,\text{Hz}$. The dashed horizontal lines mark the couplings of the outer EPs and of EP$_\mathrm{c}$, $\kappa = 4.48$ and $4.73\,\text{nF}$, and the dashed vertical lines mark the three drive frequencies of Fig.~\ref{FigSuppMatFreqShift}b, each of which places one EP at the origin. In this regime, the relevant eigenvalues remain purely imaginary, corresponding to undamped modes that shift along the imaginary axis as the driving frequency varies.

	In Fig.~\ref{FigSuppMatFreqShift}b, we show the corresponding admittance spectra in the complex plane as a function of $\kappa$ for the three representative frequencies $f=f_0-\Delta f_1$, $f=f_0$ and $f=f_0+\Delta f_2$. At $\kappa = 4.47\,\text{nF}$, $4.49\,\text{nF}$ and $4.73\,\text{nF}$, the exceptional points reach zero admittance, each at its own driving frequency. Each eigenvalue is colour-coded according to its weight in the two-point impedance between the coupled nodes. As the driving frequency changes, the admittance bands shift vertically along the imaginary axis, so that different eigenvalues cross zero and give distinct impedance resonances.

	At the circuit resonant frequency $f_0$, the central pole reaches zero and produces the strongest impedance peak. When the frequency is detuned by $\Delta f_1$ or $\Delta f_2$, the spectra move along the imaginary axis and the outer EPs reach zero admittance at $f = f_0 - \Delta f_1$ and $f = f_0 + \Delta f_2$, respectively. Because the circuit admittance contains both capacitive terms proportional to $i\omega c$ and inductive terms proportional to $1/(i\omega L)$, the admittance eigenvalues are not symmetric about the resonance frequency $f_0$. Equal detunings above and below $f_0$ therefore lead to unequal spectral shifts, and we find $\Delta f_1 = 687\,\text{Hz}$ and $\Delta f_2 = 698\,\text{Hz}$. Since the coupling admittance $i\omega\kappa$ also changes slightly with frequency, the outer EPs reach zero admittance at slightly different couplings ($4.49$ and $4.47$\,nF) at these two frequencies, whereas at $f_0$ they form at the same coupling, $\kappa\approx4.48$\,nF.

	\subsection{Shift along the real axis by gain--loss imbalance}
	\label{SuppMatImbalanceShift}

	\begin{figure*}
		\centering
		\includegraphics[width=\textwidth]{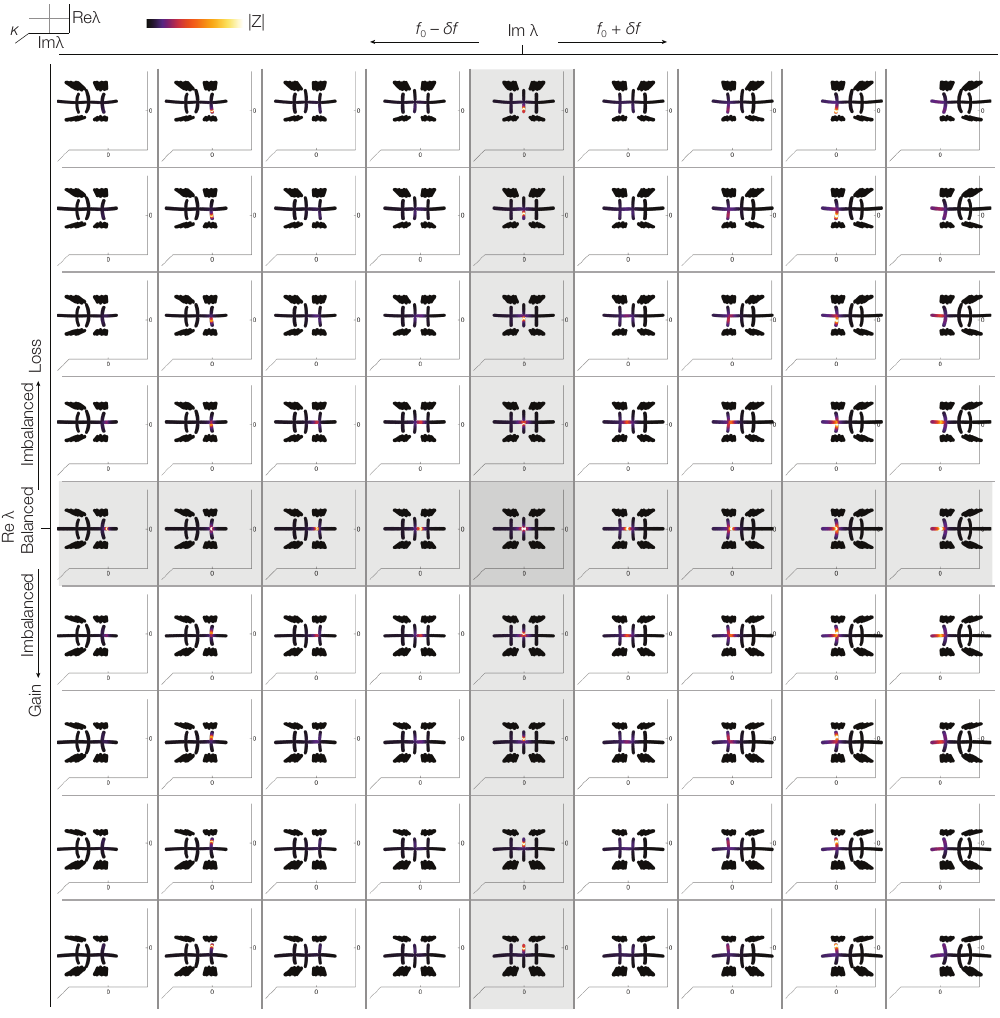}
		\caption{\textbf{Admittance spectra under frequency detuning and gain--loss imbalance (Calculated).} While the drive frequency shifts the entire spectrum along the imaginary axis, the gain--loss imbalance shifts it along the real axis, so that tuning both can give the spectrum any offset in the complex plane. The complex admittance eigenvalues are plotted as a function of the coupling $\kappa$ for various combinations of frequency (horizontal axis, from $f_0-\delta f$ to $f_0+\delta f$, where $f_0$ denotes the natural resonance frequency of the circuit) and gain--loss imbalance (vertical axis, from loss-dominant at the top to gain-dominant at the bottom), and the colour encodes the contribution of each mode to the two-point impedance. In the perfectly balanced case (centre row), the modes aligned along the imaginary axis give rise to ultrahigh-Q impedance resonances through the same mechanism as in Fig.~\ref{FigSuppMatFreqShift}. In the imbalanced cases, the entire spectrum shifts along the real axis towards negative or positive values, depending on whether the imbalance favours the gain or the loss. Even in these imbalanced regimes, an eigenvalue crosses the origin of the complex plane for specific $\kappa$--$f$ combinations and produces a high-$|Z|$ resonance. The parameters are $c_1 = 1.005\,\text{nF}$, $c_2 = 2.212\,\text{nF}$, $c_g = 333\,\text{nF}$, $L = 1.915\,\mu\text{H}$, $R=270\,\Omega$ and $N=16$ per ring.}
		\label{FigSuppMatImbalanced}
	\end{figure*}

	Here, we evaluate the impedance resonances under imbalanced conditions, that is, when the gain and loss strengths deviate from perfect balance. While detuning the excitation frequency by $\delta f$ from the natural resonant frequency $f_0$ shifts the admittance bands along the imaginary axis, an imbalance between gain and loss shifts them along the real axis. Figure~\ref{FigSuppMatImbalanced} presents a grid of admittance spectra as a function of the coupling $\kappa$, with the horizontal direction corresponding to frequency detuning, from $f_0-\delta f$ to $f_0+\delta f$, and the vertical direction to the gain--loss imbalance, from loss-dominant (top) to gain-dominant (bottom).

	Each spectrum shows the complex admittance eigenvalues colour-coded by their contribution to the two-point impedance between the coupling nodes. A strong impedance resonance arises when an eigenvalue approaches the origin of the complex plane. The modes lying between the bulk bands dominate this response, while the bulk modes remain largely undeformed for all $\kappa$. For a specific combination of $\kappa$ and $f$, a pronounced impedance peak appears whenever one of these intermediate modes passes through zero. The proximity of the eigenvalue to zero determines the Q factor of the resonance, and the closer it is, the higher the Q.

	At perfect gain--loss balance (centre row), at least two modes lie on the imaginary axis. They are undamped states that give rise to ultrahigh-Q impedance resonances. As the imbalance increases (moving vertically away from the centre), the bands shift along the real axis, which indicates growing damping or amplification. The impedance peaks then broaden and the Q factor decreases. However, because the coupling strength $\kappa$ moves the modes continuously along their trajectories, a mode can still cross the imaginary axis at certain $\kappa$--$f$ values even under imbalance. This behaviour, illustrated by isolated high-$|Z|$ regions in the off-centre panels of Fig.~\ref{FigSuppMatImbalanced}, shows that ultrahigh-Q resonances can still be reached in imperfect systems by tuning frequency and coupling together. For a uniform imbalance $\delta$ of all gain elements, every eigenvalue on the imaginary axis moves along the real axis by $\delta/2$ in every PT model, because the gain nodes carry half of the real weight of each such mode (Supplementary Section~\ref{SuppMatGainFluctuations}). This gives the common Q plateau of Fig.~\ref{figMain_CleanLimit}h (Methods).

	\section{Response of a lattice-embedded resonator to static reactive perturbations}
	\label{SuppMatReactivePerturbations}

	\subsection{Isolated and lattice-embedded resonators under static perturbations}
	\label{SuppMatIsolatedEmbedded}

	\begin{figure}[h]
		\centering
		\includegraphics[width=\textwidth]{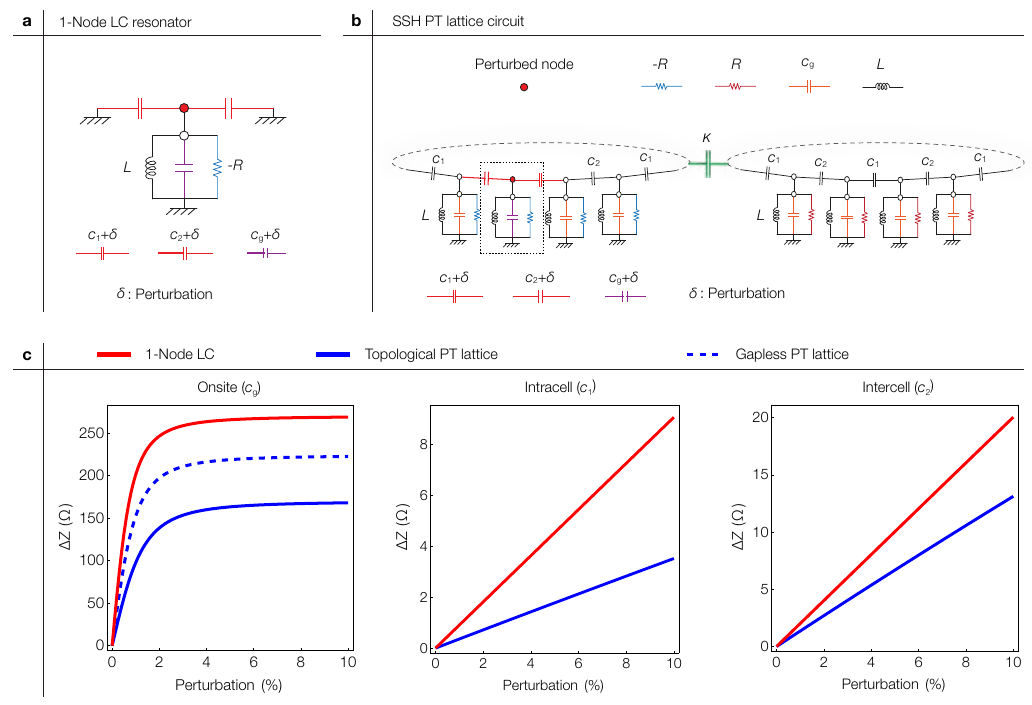}
		\caption{\textbf{Robustness of an isolated and a lattice-embedded resonator under static perturbations (Calculated).} The same resonator deviates more under static perturbations, such as imperfections in the component values, when it is isolated, whereas its resonance deviates less when it is embedded in a lattice, and least when the lattice is dimerized as in the topological SSH lattices.
			\textbf{a,} The schematic of the 1-Node LC circuit represents an isolated resonator, where perturbations are applied to the capacitors $c_1$, $c_2$ and $c_g$ as $c_i\rightarrow c_i(1+\delta)$.
			\textbf{b,} In the PT lattice, an identical LC resonator is embedded in the lattice bulk, and the same perturbation scheme is applied to the capacitors of the perturbed node (dashed rectangle).
			\textbf{c,} The deviation of the impedance amplitude $\Delta Z$ as a function of the perturbation strength for onsite ($c_g$), intracell ($c_1$) and intercell ($c_2$) variations is largest for the 1-Node LC (red), moderate for the gapless PT lattice (dashed line) and smallest for the topological PT lattice (blue). The parameters are $f_0 = 198347\,\text{Hz}$, $c_1 = 1.005\,\text{nF}$, $c_2 = 2.212\,\text{nF}$, $c_g = 0.333\,\mu\text{F}$, $L = 1.915\,\mu\text{H}$, $\kappa = 4\,\text{nF}$ and $R=270\,\Omega$.}
		\label{FigSuppMatReactivePerturbation}
	\end{figure}

	Although conventional exceptional-point (EP) systems show a square-root sensitivity, they are prone to instability because their eigenbasis becomes defective at the EP, which amplifies the effect of noise and imperfections~[S5,\,S6]. This sensitivity is particularly pronounced in macroscale devices, where isolation from ambient fluctuations is limited. In contrast, our model embeds the EP in a topological lattice. In this subsection, we show that embedding a resonator in a lattice already reduces the effect of local perturbations, because the response of a node is shared with the surrounding network. The perturbations considered here are static changes of the capacitors of one bulk node, which mainly shift the resonance frequency. They differ from the gain fluctuations at the coupling nodes that set the linewidth (main text), for which embedding in a gapless lattice alone does not help (Supplementary Section~\ref{SuppMatDisorderSpectra}).

	To examine the robustness of the resonance against perturbations such as component variations, we compare the effect of a perturbation applied to a single node in a lattice with that in an isolated LC circuit. The 1-Node LC circuit of Fig.~\ref{FigSuppMatReactivePerturbation}a can be regarded as an isolated resonator, whereas in the PT lattice of Fig.~\ref{FigSuppMatReactivePerturbation}b, the same resonator is embedded in an extended coupled network. The resonant frequency of both configurations is
	\begin{equation}
		f_0 = \frac{1}{2\pi\sqrt{L(c_1 + c_2 + c_g)}},
	\end{equation}
	which gives $f_0 = 198347\,\text{Hz}$ for the experimental parameters $c_1 = 1.005\,\text{nF}$, $c_2 = 2.212\,\text{nF}$, $c_g = 0.333\,\mu\text{F}$ and $L = 1.915\,\mu\text{H}$ (with $\kappa = 4\,\text{nF}$ and $R=270\,\Omega$ in the lattice). Since the perturbed node lies in the bulk of a ring, the value of $\kappa$ has no noticeable effect, and the same results are obtained at $\kappa_{\mathrm{EP_c}}$. We first calculate the one-point impedance of both circuits at this resonant frequency, denoted $Z_0^\text{1-Node}$ and $Z_0^\text{SSH-PT}$. We then apply perturbations to three different capacitors of one bulk node, (i) $c_1\rightarrow c_1(1+\delta)$ (intracell coupling capacitor), (ii) $c_2\rightarrow c_2(1+\delta)$ (intercell coupling capacitor) and (iii) $c_g\rightarrow c_g(1+\delta)$ (grounding capacitor), where $\delta$ varies from 0 to 0.1, corresponding to an increase of the capacitor value by up to 10\%.

	For each perturbation level $\delta$, we evaluate the nodal impedance at the fixed unperturbed resonance $f_0$ and define $\Delta Z(\delta)=\lvert Z(\delta;f_0)-Z_0\rvert$. Figure~\ref{FigSuppMatReactivePerturbation}c compares $\Delta Z$ for the 1-Node LC, the topological PT lattice and the gapless PT lattice. In the left panel, where the onsite capacitor $c_g$ is perturbed, the deviation is largest for the 1-Node LC and smallest for the topological PT lattice. In particular, the initial slope at $\delta\to 0^+$ is markedly higher for the 1-Node LC than for the topological PT lattice, which indicates that embedding the resonator in the lattice strongly reduces its sensitivity to small onsite variations. Conversely, the larger deviation of the isolated resonator shows its higher responsivity to onsite perturbations, which is an advantage when such a perturbation is the quantity to be sensed. The same responsivity, however, also amplifies unwanted disorder, such as drifts of the components and fluctuations of the active elements, which a sensor must then separate from the signal. Within this comparison, the isolated resonator may therefore suit sensing when the perturbation to be detected dominates over the disorder, although other lattice designs or operating conditions may give a higher responsivity.

	We next compare the deviations of the impedance when the intra- and intercell coupling capacitances are perturbed in the 1-Node LC and the topological PT lattice. Since intra- and intercell couplings are not distinguished in the gapless lattice, we restrict this comparison to the topological PT lattice and the 1-Node LC. As shown in the middle and right panels of Fig.~\ref{FigSuppMatReactivePerturbation}c, the deviation of the resonance amplitude is consistently smaller when the resonator is embedded in the lattice. The same resonator shows larger deviations when isolated, which indicates the stabilizing role of the lattice. This can be understood from the impedance of a node in a lattice. The node is connected in parallel to its grounding elements, with the admittance $Y_g=\pm1/R+1/(i\omega L)+i\omega c_g$, and to the two parts of the ring on either side. For a bulk node of an SSH ring, these parts start with the bonds $c_1$ and $c_2$, whose impedances are $Z_{c_1}=1/(i\omega c_1)$ and $Z_{c_2}=1/(i\omega c_2)$. Each part is again a chain of bonds and grounded nodes, so that, for a long ring, the impedance of the node is the continued fraction
	\begin{equation}
		Z_\mathrm{node}=
		\cfrac{1}{
			Y_g
			+\cfrac{1}{
				Z_{c_1}
				+\cfrac{1}{
					Y_g
					+\cfrac{1}{
						Z_{c_2}
						+\cfrac{1}{
							Y_g+\cdots}}}}
			+\cfrac{1}{
				Z_{c_2}
				+\cfrac{1}{
					Y_g
					+\cfrac{1}{
						Z_{c_1}
						+\cfrac{1}{
							Y_g+\cdots}}}}}.
		\label{EqSuppMatLadder}
	\end{equation}
	The first term in the denominator is the node itself, and each further level of the fraction adds the next bond and the next node of the ring on either side, so that the impedance of the node depends on its neighbours, on their neighbours, and so on. A local change at the perturbed node (e.g., $c\rightarrow c(1+\delta)$) enters only the first level of Eq.~\eqref{EqSuppMatLadder}, whereas the deeper levels, which describe the rest of the ring, remain unchanged. Because each level contributes through reciprocal sums, the effect of the local change is distributed over and progressively attenuated by the surrounding network. This collective dependence on the neighbouring sections reduces the weight of local deviations relative to the isolated oscillator and underpins the observed robustness.

	We perturb the capacitors here, but perturbations of other components, such as the inductors, give the same qualitative outcome. In the next subsection, we quantify statistically how the lattice size affects the stability of the near-zero eigenvalue.

	\subsection{Dependence on the lattice size}
	\label{SuppMatLatticeSize}

	\begin{figure}[h]
		\centering
		\includegraphics[width=\textwidth]{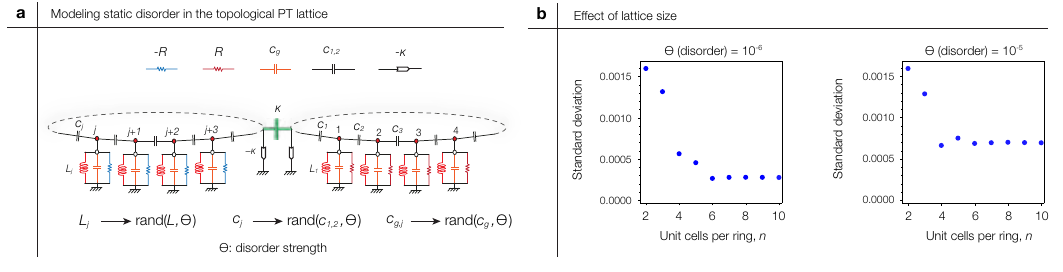}
		\caption{\textbf{Simulated spread of the near-zero eigenvalue for different lattice sizes of the topological PT lattice.} Larger lattices show an enhanced spectral stability, and the spread of the near-zero eigenvalue saturates once the rings exceed about five unit cells.
			\textbf{a,} In the statistical model of the disordered topological PT lattice, random deviations are introduced through $L_{j} \rightarrow \mathrm{rand}(L,\Theta)$, $c_{j} \rightarrow \mathrm{rand}(c_{1,2},\Theta)$ and $c_{g,j} \rightarrow \mathrm{rand}(c_g,\Theta)$ for each node $j$, where $\Theta$ denotes the disorder strength.
			\textbf{b,} The standard deviation of the smallest eigenvalue is shown as a function of the lattice size, given by the number of unit cells per ring $n$, for two disorder strengths. The nominal component values used in the simulations are $(\kappa_{\mathrm{EP_c}},c_1,c_2,c_g,L,R)=(4.727\,\text{nF},1.005\,\text{nF},2.212\,\text{nF},333\,\text{nF},1.915\,\mu\text{H},270\,\Omega)$.}
		\label{FigSuppMatLatticeSize}
	\end{figure}

	We performed a statistical analysis to quantify the effect of the lattice size on the stability of each SSH PT lattice. Each element shown in Fig.~\ref{FigSuppMatLatticeSize}a was perturbed with a random deviation. For the inductors, for example, $\mathrm{rand}(L,\Theta)$ draws the inductance of each node from a normal distribution centred at $L$ with the standard deviation $\Theta L$. Here, we consider static circuit disorder.

	In the clean limit, two of the four interface eigenvalues reach the origin at $\kappa_{\mathrm{EP_c}}$, Eq.~\eqref{EqMainKappaEPc} of the main text (Supplementary Section~\ref{SuppMatEPCouplings}).
	As $\kappa$ approaches $\kappa_{\mathrm{EP_c}}$, two eigenvalues approach the spectral origin in the clean system, corresponding to the EP responsible for the ultrahigh-Q impedance resonance. To quantify how this near-zero mode is affected by disorder for different lattice sizes, we performed a statistical eigenvalue analysis of the disordered SSH PT lattice.

	For each lattice size, given by the number of unit cells per ring $n$ (a ring has $N=2n$ nodes), random Gaussian perturbations of strength $\Theta$ were applied independently to all circuit parameters, including the inductances $L_{j}$, the coupling capacitances $c_{1,j}$ and $c_{2,j}$, the shunt capacitances $c_{g,j}$, the gain and loss resistances $R_{j}$, and the inter-ring coupling capacitor $\kappa$. For every disorder realization, the full admittance matrix was constructed and diagonalized at the nominal resonance frequency. The calculation was repeated 1000 times for each value of $n$.

	Because disorder can cause nearby eigenvalues to interchange order, the near-zero mode was identified with an eigenvector-overlap tracking procedure. Starting from the smallest disorder strength, the eigenvalue closest to the spectral origin was selected, and for larger disorder strengths the corresponding mode was followed by maximizing the overlap with the eigenvector obtained at the preceding disorder value. This procedure ensures that the same physical eigenmode is tracked throughout the statistical ensemble.

	For each lattice size, we then calculated the standard deviation of the tracked near-zero eigenvalue over the 1000 disorder realizations. Since this mode determines the largest impedance response, its statistical spread directly quantifies the disorder-induced broadening of the resonance condition. As shown in Fig.~\ref{FigSuppMatLatticeSize}b, the standard deviation decreases rapidly as the lattice size increases and approaches a nearly constant value for $n\gtrsim5$. This behaviour is observed for both $\Theta=10^{-6}$ and $\Theta=10^{-5}$, which shows that larger topological PT lattices statistically suppress fluctuations of the near-zero exceptional-point mode. The spread thus decreases as the lattice grows and saturates once the rings are longer than the localization length of the interface modes. Our rings, with $n=8$, lie in this saturated regime.

	\SIpart{Analytical theory}

	\section{Analytical derivation of the impedance}
	\label{SuppMatImpedanceTheory}

	\subsection{Analytical impedance of the isolated SSH circuit}
	\label{SuppMatOnePointZ}

	We begin by analysing the impedance of each Su--Schrieffer--Heeger (SSH) circuit in the uncoupled case, where the inter-circuit coupling capacitance is zero, $\kappa = 0$. Our system consists of two independent periodic SSH circuits, one with uniform gain and the other with uniform loss. Each unit cell contains two sublattice nodes, labelled $A$ and $B$, which are connected through intra-cell ($c_1$) and inter-cell ($c_2$) capacitive couplings. In addition, each node is grounded through three elements in parallel, a grounding inductor $L$, a capacitor $c_g$, and either a resistor $R$ (in the lossy circuit) or a negative resistor $-R$ (in the gain circuit). The negative resistance is realized with negative-impedance converters with current inversion~[S1].

	Each isolated SSH circuit is described by the momentum-space Laplacian
	\begin{equation}
		J_\pm(k_x;\omega)=
		\begin{pmatrix}
			Y_\pm(\omega) & - i \omega c_1 - i \omega c_2 e^{-i k_x} \\
			-i \omega c_1 - i \omega c_2 e^{i k_x}  & Y_\pm(\omega)
		\end{pmatrix},
		\quad Y_\pm(\omega) = \pm \frac{1}{R} + \frac{1}{i \omega L} + i \omega(c_1 + c_2 + c_g)=Y_0(\omega)\pm\frac{1}{R},
		\label{EqSuppMatBlochLaplacian}
	\end{equation}
	where $\omega$ is the AC frequency and $k_x$ is the wavevector. The subscript $\pm$ denotes whether the circuit is lossy ($+$) or with gain ($-$), corresponding to the sign of the resistance term, and $Y_0$ is the onsite admittance of the main text. The diagonal entries ($Y_\pm(\omega)$) represent the total admittance to ground at each node, while the off-diagonal terms encode the capacitive coupling between sublattices with $k$-dependent phase factors. To evaluate the impedance at a given node $(\ell,\mu)$, where $\ell$ is the unit cell index and $\mu \in \{A, B\}$ is the sublattice label, we inject a current $I$ at that node and measure the resulting voltage $V$. The impedance is defined as
	\begin{equation}
		Z = \frac{V}{I}, \qquad V = IG, \qquad \Rightarrow \quad Z = G,
	\end{equation}
	where $G$ is the Green's function entry of the same node $(\ell,\mu)$. Since voltage and current are both defined at the same location, the impedance equals the diagonal component of the Green's function. Each SSH circuit is periodic, so translational symmetry is preserved, and the Green's function can be obtained from the inverse of the momentum-space Laplacian,
	\begin{equation}
		G_\pm(k_x;\omega) = J_\pm^{-1}(k_x;\omega).
	\end{equation}
	The real-space Green's function is then obtained by a discrete inverse Fourier transform,
	\begin{equation}
		G_\pm(\ell,\mu,\ell',\mu';\omega) = \frac{1}{n} \sum_{x=0}^{n-1} [G_\pm(k_x;\omega)]_{\mu,\mu'} e^{i k_x (\ell-\ell')}, \qquad k_x = \frac{2\pi x}{n},
	\end{equation}
	where $n=N/2$ is the total number of unit cells of a ring with $N$ nodes, and $x$ indexes the discrete momentum values from 0 to $n-1$. The one-point impedance at node $(\ell,\mu)$ is given by the diagonal entry $G_\pm(\ell,\mu,\ell,\mu;\omega)$, obtained by setting $\ell=\ell'$ and summing over the Brillouin zone,
	\begin{equation}
		Z_\pm(\omega) = G_\pm(\ell,\mu,\ell,\mu;\omega) = \frac{1}{n} \sum_{x=0}^{n-1} [G_\pm(k_x;\omega)]_{\mu,\mu}.
		\label{EqSuppMatZGreenFourier}
	\end{equation}
	Substituting the inverse of the $k$-space Laplacian gives the expression for the impedance at any node in terms of Bloch modes,
	\begin{equation}
		Z_\pm(\omega) = \frac{1}{n} \sum_{x=0}^{n-1} \frac{Y_\pm(\omega)}{\omega^2 (c_1^2 + c_2^2 + 2 c_1 c_2 \cos k_x ) + \left( Y_\pm(\omega) \right)^2},
		\label{EqSuppMatOnePointZSum}
	\end{equation}
	where $Y_\pm(\omega) = \pm \frac{1}{R} + \frac{1}{i \omega L} + i \omega(c_1 + c_2 + c_g)$ and $k_x = \frac{2\pi x}{n}$ with the wavevector index $x$ running from $0$ to $n-1$.

	To derive a closed-form expression from Eq.~\eqref{EqSuppMatOnePointZSum}, we consider the limit of a large number of unit cells $n$, in which the discrete sum can be approximated by an integral over the first Brillouin zone,
	\begin{equation}
		Z_{\pm}(\omega) \approx \frac{ Y_{\pm}(\omega) }{ 2\pi } \int_{-\pi}^{\pi} \frac{dk_x}{\omega^2(c_1^2 + c_2^2 + 2 c_1 c_2 \cos k_x) + (Y_{\pm}(\omega))^2}.
	\end{equation}
	This integral can be evaluated by contour integration, which gives
	\begin{equation}
		Z_{\pm}(\omega) = \frac{Y_{\pm}(\omega)}{\omega^2 \sqrt{\left(c_1^2 + c_2^2 + \frac{(Y_{\pm}(\omega))^2}{\omega^2}\right)^2 - (2 c_1 c_2)^2}}s_\pm,
	\end{equation}
	where
	\begin{equation}
		s_\pm = \mathrm{sgn}\left( \left|\frac{ Y_\pm^2 + (c_1^2 +c_2^2)\omega^2 + \sqrt{\left(Y_\pm^2 + (c_1^2+c_2^2)\omega^2\right)^2 - \left(2c_1c_2\omega^2\right)^2}}{c_1c_2\omega^2}\right| - 2\right).
		\label{EqSuppMatSignFunction}
	\end{equation}
	The sign function selects the root of the contour integral that lies inside the unit circle.

	Multiplying by $\omega^4$ inside the square root gives the final analytical form
	\begin{equation}
		Z_{\pm}(\omega) = \frac{Y_{\pm}(\omega)}{\sqrt{\left( Y_{\pm}(\omega)^2 + \omega^2(c_1^2 + c_2^2) \right)^2 - 4 \omega^4 c_1^2 c_2^2}}s_\pm.
		\label{EqSuppMatOnePointZ}
	\end{equation}
	This expression is the closed-form impedance of an SSH circuit with grounding onsite terms $Y_\pm(\omega) = \pm \frac{1}{R} + \frac{1}{i \omega L} + i \omega(c_1 + c_2 + c_g)$ and intra- and inter-cell capacitances $c_1$ and $c_2$. By modifying these grounding elements, the expression can be adapted to other SSH circuit configurations. With the gap edge $Y_\mathrm{gap}=\omega|c_2-c_1|$ and the band edge $Y_\mathrm{band}=\omega(c_1+c_2)$, the expression under the square root factorizes as $(Y_\pm^2+Y_\mathrm{gap}^2)(Y_\pm^2+Y_\mathrm{band}^2)$, so that $Z_\pm=1/\mathcal{Y}(Y_\pm)$, where $\mathcal{Y}$ is the input admittance of Eq.~\eqref{EqDressedY} of the main text. The sign function makes $\mathcal{Y}(y)\rightarrow y$ for $|y|\rightarrow\infty$ (Supplementary Section~\ref{SuppMatTwoNodeReduction}). We next use Eq.~\eqref{EqSuppMatOnePointZ} to obtain an analytical solution for the two-point impedance of our coupled circuit.

	\subsection{Analytical impedance of the coupled system}
	\label{SuppMatTwoPointZ}

	The two-point impedance between the coupling nodes of our periodic topological SSH circuits can be expressed in terms of the Green's function $G$. To obtain the Green's function of the circuit, we treat the coupling between the two SSH rings as a perturbation of the Laplacian of the two uncoupled rings~[S7,\,S8]. Specifically, the real-space circuit Laplacian is written as
	\begin{equation}
		J = J_0 + J_1,
	\end{equation}
	where $J_0$ is the real-space Laplacian of the two SSH rings and $J_1$ the coupling between them. We denote the Green's function of the uncoupled rings by $G_0$ and that of the full circuit by $G$. Denoting the index of the last site of the lossy ring by $a$ (node 16) and that of the first site of the gain ring by $b$ (node 17), and the indices of arbitrary sites on the two rings by $\rho$ and $\tau$, respectively, $J_1$ has the explicit form
	\begin{equation}
		J_1 = -i\omega \kappa \left(|b\rangle\langle a| + |a\rangle\langle b|\right),
		\label{EqSuppMatJ1}
	\end{equation}
	which has no diagonal elements, because the $-\kappa$ elements cancel the onsite terms $i\omega\kappa$ that the coupling capacitor adds at $a$ and $b$, while the Green's function of the uncoupled rings is
	\begin{equation}
		G_0 = \sum_{\rho, \rho'} |\rho\rangle G_{+;\rho,\rho'} \langle \rho'| + \sum_{\tau, \tau'} |\tau\rangle G_{-;\tau,\tau'} \langle \tau'|.
		\label{EqSuppMatG0}
	\end{equation}
	From $JG=\mathbb{1}$ and $J_0 G_0=\mathbb{1}$, where $\mathbb{1}$ is the identity matrix, we have
	\begin{equation}
		(J_0 + J_1) G = \mathbb{1},
		\label{EqSuppMatGreens}
	\end{equation}
	where $G$ is the coupled Green's function associated with $J$. Using $J_0 G_0 =\mathbb{1}$ and Eq.~\eqref{EqSuppMatGreens}, left-multiplying by $G_0$ gives $(1+G_0 J_1) G=G_0$, which rearranges to the Dyson equation
	\begin{equation}
		G = G_0 - G_0 J_1 G.
		\label{EqSuppMatDyson}
	\end{equation}
	By iteratively substituting the expression for $G$ on the right-hand side, $G$ can be obtained as a power series~[S7]
	\begin{equation}
		G = \sum_{n=0}^\infty (-G_0J_1)^n G_0.
		\label{EqSuppMatDysonSeries}
	\end{equation}

\begin{figure*}[h]
		\centering
		\includegraphics[width=\textwidth]{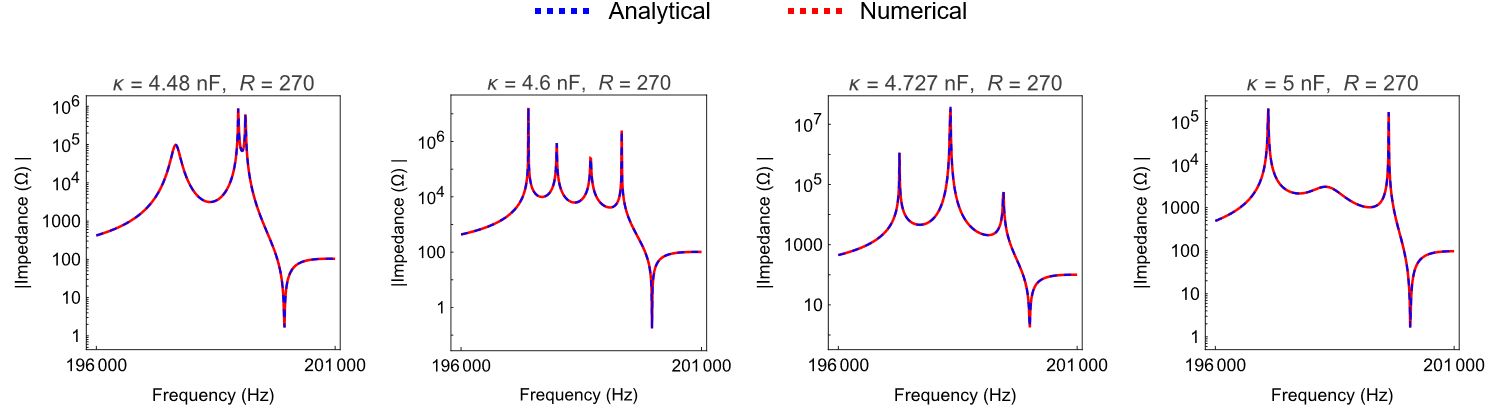}
		\caption{\textbf{Calculated two-point impedance, numerical and analytical.} The analytical two-point impedance (blue dashed curve, Eq.~\eqref{EqSuppMatFinalZ}) agrees closely with the numerical one (red dashed curve) for the values of $\kappa$ used in Fig.~\ref{figMain_CleanLimit}f. The parameters are $N=16$ nodes per ring (8 unit cells), $c_1=1.005$\,nF, $c_2=2.212$\,nF, $c_g=333$\,nF, $L=1.915\,\mu$H and $R=270\,\Omega$.}
		\label{FigSuppMatImpComparison}
	\end{figure*}

	Using the explicit forms of $J_1$ and $G_0$ in Eqs.~\eqref{EqSuppMatJ1} and~\eqref{EqSuppMatG0}, the factor $(-G_0J_1)$ in Eq.~\eqref{EqSuppMatDysonSeries} is
	\begin{equation}
		-G_0J_1 = i\omega \kappa \left( \sum_{\tau} |\tau\rangle G_{-;\tau,b} \langle a| + \sum_{\rho} |\rho\rangle G_{+;\rho,a} \langle b| \right).
	\end{equation}
	Because the final result for the impedance of the PT system involves only the $a$th and $b$th elements of $G$, we introduce the compact notation
	\begin{equation}
		(-G_0J_1) \simeq \begin{pmatrix} (-G_0J_1)_{a,a} & (-G_0J_1)_{a,b} \\ (-G_0J_1)_{b,a} & (-G_0J_1)_{b,b} \end{pmatrix} =  i\omega\kappa \begin{pmatrix} 0 & G_{+;a,a} \\ G_{-;b,b} & 0 \end{pmatrix},
	\end{equation}
	from which it follows that for even $n=2m$ and odd $n=2m+1$,
	\begin{align}
		(-G_0J_1)^{2m} &\simeq (-1)^m (\omega \kappa)^{2m} \begin{pmatrix}  (G_{+;a,a}G_{-;b,b} )^m  & 0 \\ 0 & (G_{+;a,a}G_{-;b,b} )^m  \end{pmatrix}, \label{EqSuppMatG0J1Even} \\
		(-G_0J_1)^{2m+1} &\simeq (-1)^m i (\omega\kappa)^{2m+1} \begin{pmatrix}  0 & G_{+;a,a}(G_{+;a,a}G_{-;b,b} )^m   \\ G_{-;b,b}(G_{+;a,a}G_{-;b,b} )^m & 0  \end{pmatrix}. \label{EqSuppMatG0J1Odd}
	\end{align}
	Substituting Eqs.~\eqref{EqSuppMatG0J1Even} and~\eqref{EqSuppMatG0J1Odd} into the Dyson series of Eq.~\eqref{EqSuppMatDysonSeries}, with $G_0$ restricted to the nodes $a$ and $b$ given by $\mathrm{diag}(G_{+;a,a},G_{-;b,b})$, therefore leads to
	\begin{align}
		\begin{pmatrix} G_{a,a} & G_{a,b} \\ G_{b,a} & G_{b,b}\end{pmatrix} &=  \begin{pmatrix}  G_{+;a,a} & i\omega\kappa G_{+;a,a}G_{-;b,b} \\ i\omega\kappa  G_{+;a,a}G_{-;b,b} & G_{-;b,b} \end{pmatrix} \left( \sum_{m=0}^\infty \left(-(\omega\kappa)^2 G_{+;a,a}G_{-;b,b} \right)^m\right) \nonumber \\
		&=\frac{1}{1+\omega^2\kappa^2G_{+;a,a}G_{-;b,b}}\begin{pmatrix}  G_{+;a,a} & i\omega\kappa G_{+;a,a}G_{-;b,b} \\ i\omega\kappa  G_{+;a,a}G_{-;b,b} & G_{-;b,b} \end{pmatrix}.
		\label{EqSuppMatGab}
	\end{align}
	Using $G_{\pm;\alpha,\alpha} = Z_\pm$ for any site $\alpha$, with $Z_\pm$ derived in the previous subsection, the matrix elements take the explicit forms
	\begin{align}
		G_{a,a}  &= \frac{1}{D}\left(s_+Y_{+}\sqrt{\left(Y_{-}^2 + (c_1^2 + c_2^2) \omega^2\right)^2-\left(2c_1c_2\omega^2\right)^2}\right), \\
		G_{b,b}  &= \frac{1}{D} \left(s_-Y_{-}\sqrt{\left(Y_{+}^2 + (c_1^2 + c_2^2) \omega^2\right)^2-\left(2c_1c_2\omega^2\right)^2}\right), \\
		G_{a,b} &= G_{b,a} = \frac{i\omega\kappa s_+s_-Y_+Y_-}{D},
	\end{align}
	where
	\begin{equation}
		D = \kappa^2\omega^2 s_+s_-Y_-Y_+ +
		\sqrt{\left(Y_-^2 + (c_1^2 + c_2^2) \omega^2\right)^2-\left(2c_1c_2\omega^2\right)^2}
		\sqrt{\left(Y_+^2 + (c_1^2 + c_2^2) \omega^2\right)^2-\left(2c_1c_2\omega^2\right)^2}.
	\end{equation}
	The two-point impedance between nodes $a$ and $b$, $Z_{ab}(\omega,\kappa)$, is hence
	\begin{align}
		Z_{ab} &= G_{a,a} + G_{b,b} - G_{a,b}-G_{b,a} \nonumber \\
		&= \frac{1}{D} \Bigg( -2 i\omega\kappa s_+ s_- Y_+ Y_- + s_+Y_+\sqrt{\left(Y_-^2 + (c_1^2 + c_2^2) \omega^2\right)^2-\left(2c_1c_2\omega^2\right)^2} \nonumber \\
		&\qquad +  s_-Y_- \sqrt{\left(Y_+^2 + (c_1^2 + c_2^2) \omega^2\right)^2-\left(2c_1c_2\omega^2\right)^2}\Bigg).
		\label{EqSuppMatFinalZ}
	\end{align}
	Dividing the numerator and the denominator by $s_+s_-Y_+Y_-$ and using $Z_\pm=1/\mathcal{Y}(Y_\pm)$ gives the compact form $Z_{ab}=[\mathcal{Y}(Y_+)+\mathcal{Y}(Y_-)-2i\omega\kappa]/[\mathcal{Y}(Y_+)\,\mathcal{Y}(Y_-)+\omega^2\kappa^2]$, which is Eq.~\eqref{EqMethodsFinalZ} of Methods. Its poles give the eigenvalue condition of Supplementary Section~\ref{SuppMatTwoNodeReduction}. Figure~\ref{FigSuppMatImpComparison} compares Eq.~\eqref{EqSuppMatFinalZ} with the impedance obtained by numerical inversion of the full Laplacian, and the two agree closely.

\begin{figure}[h]
		\centering
		\includegraphics[width=\textwidth]{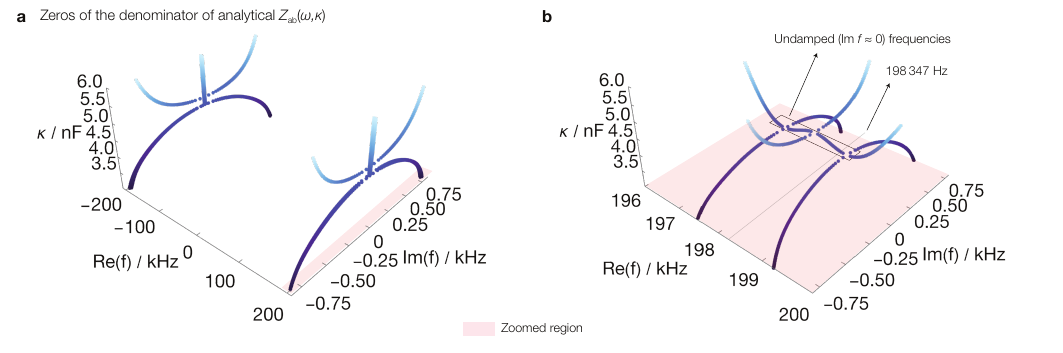}
		\caption{\textbf{Calculated poles of the two-point impedance in the complex frequency plane.} The poles of the two-point impedance reach the real frequency axis at the couplings where the circuit has undamped resonances.
			\textbf{a,} The zeros of the denominator $D$ of Eq.~\eqref{EqSuppMatFinalZ} are shown in the complex frequency plane, $\mathrm{Re}\,f$ and $\mathrm{Im}\,f$, for $\kappa$ from 3.5 to 6\,nF, where the colours only distinguish the values of $\kappa$.
			\textbf{b,} The magnified view near $f_0=198\,347$\,Hz (shaded region in \textbf{a}) shows where the poles reach the real frequency axis, $\mathrm{Im}\,f\approx0$. The parameters are $c_1 = 1.005\,\text{nF}$, $c_2 = 2.212\,\text{nF}$, $c_g = 0.333\,\mu\text{F}$, $L = 1.915\,\mu\text{H}$ and $R=270\,\Omega$.}
		\label{FigSuppMatComplexFrequency}
	\end{figure}

	Since Eq.~\eqref{EqSuppMatFinalZ} is an analytic function of $\omega$, it can also be evaluated at complex frequency. Its poles, the zeros of its denominator $D$, are then the complex natural frequencies of the coupled circuit. Figure~\ref{FigSuppMatComplexFrequency} follows these poles as $\kappa$ increases. Where a pole reaches the real frequency axis, the circuit has an undamped resonance, which appears as a sharp peak of the impedance at real frequency (Fig.~\ref{figMain_CleanLimit}c,f).

	\subsection{Leakage model of the measured linewidth}
	\label{SuppMatLeakageModel}

	In the ideal case, the theoretical peak impedance diverges when the denominator $\mathcal{D}$ of Eq.~\eqref{EqMethodsFinalZ} becomes exactly zero at $\omega=2\pi f_0$, which corresponds to a vanishing linewidth and an infinite Q factor. In practice, an infinite Q factor cannot be realized. Disorder in the device spreads the eigenvalues, and imperfections of the measurement, such as the finite frequency step of a sweep, instrument noise and residual leakage currents, bound the measurable response.

	The influence of these effects can be modelled phenomenologically by a leakage admittance $Y_{\mathrm{leak}}$ in parallel with the ideal circuit of impedance $Z_{ab}$ in Eq.~\eqref{EqMethodsFinalZ}, where $Y_{\mathrm{leak}}$ is taken to be real. (Any imaginary component can be absorbed into $Z_{ab}$, which is purely imaginary at resonance.) Near a resonance frequency $\omega_0 = 2\pi f_0$ at which $\mathcal{D}(\omega)$ has a simple zero, we write
	\begin{equation}
		Z_{ab}(\omega) \simeq \frac{i\mathcal{N}}{\mathcal{D}'_0\,(\omega-\omega_0)},\qquad \mathcal{D}'_0 \equiv \left.\frac{d\mathcal{D}}{d\omega}\right|_{\omega_0},
		\label{EqSuppMatZabPole}
	\end{equation}
	where $i\mathcal{N}$ is the numerator of $Z_{ab}$ at $\omega_0$ and $\mathcal{D}'_0$ is the slope of $\mathcal{D}$ at $\omega_0$. Both are constants at a given operating point. The effective impedance, $Z_{\mathrm{eff}} = (Y_{\mathrm{leak}} + 1/Z_{ab})^{-1}$, is then
	\begin{equation}
		Z_{\mathrm{eff}}(\omega) = \left( Y_{\mathrm{leak}} - i 	\frac{\mathcal{D}'_0(\omega-\omega_0)}{\mathcal{N}}\right)^{-1},
		\label{EqSuppMatZeffLeak}
	\end{equation}
	which corresponds to the impedance power spectrum
	\begin{equation}
	|Z_{\mathrm{eff}}(\omega)|^2 = \frac{\mathcal{N}^2}{\mathcal{N}^2 Y_{\mathrm{leak}}^2 + (\mathcal{D}'_0)^2(\omega-\omega_0)^2}.
	\label{EqSuppMatLorentzianLeak}
	\end{equation}
	This has a Lorentzian form with a FWHM of $2|\mathcal{N} Y_{\mathrm{leak}}/\mathcal{D}'_0|$, so that
	\begin{equation}
		\mathrm{Q} = \left|\frac{\mathcal{D}'_0\, \omega_0}{2\mathcal{N} Y_{\mathrm{leak}}}\right| .
		\label{EqSuppMatQLeak}
	\end{equation}
	For the lattice, the expression for $\mathcal{D}'_0$ is long and not very informative, so we evaluate it numerically. The Q factor diverges for $Y_{\mathrm{leak}}\rightarrow0$ and decreases as $Y_{\mathrm{leak}}$ grows. Since $Z_{ab}\simeq A/\lambda$ near the pole, a leakage admittance in parallel shifts the resonant eigenvalue to $\lambda+AY_\mathrm{leak}$ and adds to its net loss. $Y_\mathrm{leak}$ thus plays the role of $|\mathrm{Re}\,\lambda|$ in Eq.~\eqref{EqQLinewidth} for a loss added by the measurement rather than by the device, and it gives one fit parameter for all contributions to the measured linewidth, including the spectral width $\Delta W$ of Eq.~\eqref{EqWidthQ} (Fig.~\ref{figMain_SpectralBroadening}).

\begin{figure}[h!]
		\centering
		\includegraphics[width=0.5\linewidth]{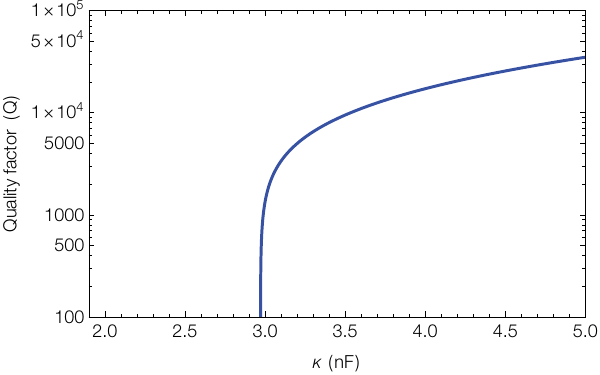}
		\caption{\textbf{Calculated Q factor of the 2-Node PT circuit with a leakage admittance.} The leakage admittance offers an alternative way to see what sets the upper limit of high-Q resonances. Q is obtained from Eq.~\eqref{EqSuppMatQ2NodePT} as a function of $\kappa$ for the standard parameter set with $R=270\ \mathrm{\Omega}$ and $Y_{\mathrm{leak}}=5\times10^{-5}\ \mathrm{S}$. The curve starts at the EP of the dimer, $\kappa_\mathrm{th}\approx2.97$\,nF, and increases sharply, but instead of diverging it approaches a smooth ceiling set by the leakage admittance.}
		\label{FigSuppMatLeakageQ}
	\end{figure}

	As an illustrative example, the ideal response of the 2-Node PT circuit of Eq.~\eqref{Eq2NodePTLaplacian}, with $Y_0=i\omega c_\Sigma+1/(i\omega L)$, is
	\begin{equation}
		Z_{ab}(\omega) = -\frac{2i\omega LR^2\left[\omega^2L(c_\Sigma-\kappa)-1\right]}{R^2 + \omega^2 L (L - 2 c_{\Sigma} R^2) + \omega^4 L^2 R^2 (c_{\Sigma}^2 - \kappa^2)},
		\label{EqSuppMatZ2NodePT}
	\end{equation}
	where $c_{\Sigma} = c_1 + c_2 + c_{g}$ is the total node capacitance. The denominator $\mathcal{D}(\omega)$ of this expression is a biquadratic polynomial in $\omega$,
	\begin{equation}
		\mathcal{D}(\omega)=\mathcal{A}(\kappa)\,\omega^4+\mathcal{B}\,\omega^2+\mathcal{C},\qquad \mathcal{A}(\kappa)=L^2R^2(c_{\Sigma}^2-\kappa^2),\qquad \mathcal{B}=L(L-2c_{\Sigma}R^2),\qquad \mathcal{C}=R^2.
		\label{EqSuppMatDimerDenominator}
	\end{equation}
	Setting $\mathcal{D}(\omega) = 0$ gives the resonance frequency of the physical mode,
	\begin{equation}
		\omega_0 = \sqrt{\frac{-\mathcal{B} + \sqrt{\mathcal{B}^2 - 4\mathcal{A}(\kappa)\mathcal{C}}}{2\mathcal{A}(\kappa)}},
		\label{EqSuppMatOmega0Dimer}
	\end{equation}
	which is real only for $\kappa \geq \kappa_\mathrm{th}$, with
	\begin{equation}
		\kappa_\mathrm{th}^2 = \frac{Lc_\Sigma}{R^2} - \frac{L^2}{4R^4}.
		\label{EqSuppMatKappaTh}
	\end{equation}
	For our parameters, $\kappa_\mathrm{th} \approx 2.97$\,nF coincides with the EP of the dimer, $\kappa = 1/(\omega R)$ (Eq.~\eqref{Eq2NodePTLaplacian}). In the ideal lossless limit, the impedance diverges at $\omega_0$, which would imply a vanishing linewidth and an infinite Q factor. The leakage admittance $Y_{\mathrm{leak}}$, however, broadens the response. At resonance, the slope of the denominator and the numerator are
	\begin{equation}
		\mathcal{D}'_0=4\mathcal{A}(\kappa)\,\omega_0^3+2\mathcal{B}\,\omega_0=2\omega_0\sqrt{\mathcal{B}^2-4\mathcal{A}\mathcal{C}},\qquad \mathcal{N}=-2\omega_0LR^2\left[\omega_0^2L(c_\Sigma-\kappa)-1\right],
		\label{EqSuppMatDimerSlope}
	\end{equation}
	and substituting them into Eq.~\eqref{EqSuppMatQLeak} gives
	\begin{equation}
		\mathrm{Q}(\kappa)=\frac{\omega_0\sqrt{\kappa^2-\kappa_\mathrm{th}^2}}{Y_\mathrm{leak}\left|\omega_0^2L(c_\Sigma-\kappa)-1\right|},
		\label{EqSuppMatQ2NodePT}
	\end{equation}
	where we used $\sqrt{\mathcal{B}^2-4\mathcal{A}\mathcal{C}}=2LR^2\sqrt{\kappa^2-\kappa_\mathrm{th}^2}$. This expression shows how a nonzero $Y_{\mathrm{leak}}$ bounds the resonance peak and prevents an unphysical divergence. It holds except within about $0.05$\,nF of $\kappa_\mathrm{th}$, where the two resonance peaks of the dimer overlap. Fig.~\ref{FigSuppMatLeakageQ} shows the Q obtained from Eq.~\eqref{EqSuppMatQ2NodePT} for $Y_{\mathrm{leak}}=5\times10^{-5}$\,S, which gives Q $\approx1.7\times10^4$ at $\kappa=4$\,nF and $3.5\times10^4$ at 5\,nF, of the order of the measured Q $=1.82\times10^4$ of the dimer (Fig.~\ref{figMain_Experiment}c).

	\section{Eigenvalue condition and exceptional points of the topological PT lattice}
	\label{SuppMatEPTheory}

	\subsection{Mid-gap spectral relation of the coupled lattices}
	\label{SuppMatMidGap}

	We consider a two-ring topological system comprising a gain ring (`g') and a lossy ring (`l'), each of which contains $n$ unit cells and is described in its bulk by the non-Bloch Laplacian~[S9,\,S10]
	\begin{equation}
		J_\gamma = \begin{pmatrix} s_\gamma \frac{1}{R} & i\omega \left(c_1 + c_2\beta^{-1}\right) \\ i\omega (c_1 + c_2\beta) & s_\gamma \frac{1}{R} \end{pmatrix},\ \gamma \in (\mathrm{l},\mathrm{g})
	\end{equation}
	in the A, B sublattice basis. Here, $\gamma \in (\mathrm{l},\mathrm{g})$ labels the lossy and gain rings, respectively, with $s_{\mathrm{l}}=1$ for the lossy Laplacian and $s_{\mathrm{g}}=-1$ for the gain Laplacian. The other diagonal term, $Y_0=i\omega c_\Sigma+1/(i\omega L)$, is the same on every node and shifts all eigenvalues by $Y_0$. It is omitted for simplicity, so that $\lambda$ is measured from $Y_0$, and it vanishes at the resonance frequency. The sign of the off-diagonal terms differs from that of Eq.~\eqref{EqSuppMatBlochLaplacian} by a sign change of the B amplitudes, which does not affect the eigenvalues.

	The wavefunctions of an eigenstate within the individual rings have the generic form
	\begin{equation}
		\psi_\gamma(x_\gamma) = \sum_{i=1}^{2} c_{\gamma; i} \beta_{\gamma; i}^x \begin{pmatrix} 1 \\ \chi_{\gamma; i} \end{pmatrix}, \label{EqSuppMatPsiX}
	\end{equation}
	where $x_\gamma$ is the unit cell index within the $\gamma$th ring and $|\beta_{\gamma; 1}| \leq |\beta_{\gamma; 2}|$. For an eigenvalue $\lambda$, the $\beta_{\gamma; i}$ and $\chi_{\gamma; i}$ take the explicit forms
	\begin{align}
		\beta_{\gamma;\pm} =& -\frac{1}{2 c_1 c_2 \omega^2} \Bigg( \lambda^2 + \frac{1}{(s_{\gamma} R)^2} - \frac{2 \lambda}{s_{\gamma} R} + c_1^2 \omega^2 + c_2^2 \omega^2 \nonumber \\ &\pm \left( -4 c_1^2 c_2^2 \omega^4 + \left( \frac{1}{(s_{\gamma} R)^2} ( -1 + \lambda s_{\gamma} R )^2 + ( c_1^2 + c_2^2 ) \omega^2 \right)^2 \right)^{1/2} \Bigg) \label{EqSuppMatBeta}, \\
		\chi_{\gamma;\pm} =&  \mp \frac{i}{2 c_1 s_{\gamma} R ( -1 + \lambda s_{\gamma} R ) \omega} \Bigg( 1 - 2 \lambda s_{\gamma} R + \lambda^2 R^2 - c_1^2 R^2 \omega^2 + c_2^2 R^2 \omega^2 \nonumber  \\
		&+ R^2 \left( -4 c_1^2 c_2^2 \omega^4 + \left( \frac{1}{R^2} ( -1 + \lambda s_{\gamma} R )^2 + ( c_1^2 + c_2^2 ) \omega^2 \right)^2 \right)^{1/2} \Bigg) \label{EqSuppMatChi}.
	\end{align}
	For the values of $c_1$, $c_2$, $R$ and $\omega$ in the experiment, $\beta_1$ with the smaller modulus corresponds to $\beta_-$, and $\beta_2$ with the larger modulus to $\beta_+$, since $\beta_+\beta_-=1$.

	We assume that the B node in the $x=n$th unit cell of the lossy ring (node 16) is coupled to the A node in the $x=1$st unit cell of the gain ring (node 17) through a capacitive coupling $\kappa$. The boundary equations due to this coupling and to the closure of each SSH lattice into a ring are then
	\begin{align}
		&c_2 \langle \mathrm{A}|\psi_{\mathrm{l}}(1) + \kappa \langle \mathrm{A}|\psi_{\mathrm{g}}(1) - c_2 \langle \mathrm{A}|\psi_{\mathrm{l}}(n+1) = 0 \label{EqSuppMatBC1}, \\
		& \langle \mathrm{B}|\psi_{\mathrm{l}}(n) - \langle \mathrm{B}|\psi_{\mathrm{l}}(0) = 0, \\
		& c_2 \langle \mathrm{B}|\psi_{\mathrm{g}}(n) + \kappa \langle \mathrm{B}|\psi_{\mathrm{l}}(n) - c_2 \langle \mathrm{B}|\psi_{\mathrm{g}}(0) = 0, \\
		& \langle \mathrm{A}|\psi_{\mathrm{g}}(1) - \langle\mathrm{A}|\psi_{\mathrm{g}}(n+1) = 0. \label{EqSuppMatBC4}
	\end{align}

	Substituting Eq.~\eqref{EqSuppMatPsiX} into the boundary conditions, Eqs.~\eqref{EqSuppMatBC1} to~\eqref{EqSuppMatBC4}, gives a system of four homogeneous linear equations in the four unknowns $c_{\mathrm{l};1}$, $c_{\mathrm{l};2}$, $c_{\mathrm{g};1}$ and $c_{\mathrm{g};2}$. A nontrivial solution exists only when the determinant of this system vanishes, which gives the following condition, satisfied at every eigenvalue of the topological PT system:
	\begin{align}
		&c_2^2 \left(1 - \beta_{\mathrm{l};1}^{n}\right) \left(1 - \beta_{\mathrm{g};1}^{n}\right)
		\left(\chi_{\mathrm{l};2} \beta_{\mathrm{l};1} - \chi_{\mathrm{l};1} \beta_{\mathrm{l};2}\right) \left(1 - \beta_{\mathrm{l};2}^{n}\right)
		\left(\chi_{\mathrm{g};2} \beta_{\mathrm{g};1} - \chi_{\mathrm{g};1} \beta_{\mathrm{g};2}\right) \left(1 - \beta_{\mathrm{g};2}^{n}\right) \nonumber \\
		&\quad - \chi_{\mathrm{l};1} \chi_{\mathrm{l};2} \beta_{\mathrm{g};1} \left(\beta_{\mathrm{l};1}^{n} - \beta_{\mathrm{l};2}^{n}\right)
		\beta_{\mathrm{g};2} \left(\beta_{\mathrm{g};1}^{n} - \beta_{\mathrm{g};2}^{n}\right) \kappa^2 = 0. \label{EqSuppMatConsCon}
	\end{align}

	For the parameters of our experiment, $|\beta_{\gamma;1}| < 1 < |\beta_{\gamma;2}|$ for the mid-gap states. Therefore, in the large-$n$ limit, $\beta_{\gamma,1}^n\rightarrow 0$ while $(1-\beta^n_{\gamma;2})\approx - \beta^n_{\gamma;2}$, and the consistency condition of Eq.~\eqref{EqSuppMatConsCon} simplifies to
	\begin{equation}
		c_2^2 \left( \chi_{\mathrm{l};2} \beta_{\mathrm{l};1} - \chi_{\mathrm{l};1} \beta_{\mathrm{l};2} \right) \left( \chi_{\mathrm{g};2} \beta_{\mathrm{g};1} - \chi_{\mathrm{g};1} \beta_{\mathrm{g};2} \right) - \chi_{\mathrm{l};1} \chi_{\mathrm{l};2} \beta_{\mathrm{g};1} \beta_{\mathrm{g};2} \kappa^2 = 0. \label{EqSuppMatConsConLarge}
	\end{equation}

	Substituting $\beta_{\gamma;2}=\beta_{\gamma, +}$ and $\beta_{\gamma;1}=\beta_{\gamma, -}$ from Eq.~\eqref{EqSuppMatBeta} and the corresponding expressions for $\chi_{\gamma, i}$ from Eq.~\eqref{EqSuppMatChi} into Eq.~\eqref{EqSuppMatConsConLarge} gives the following implicit expression for the eigenvalues $\lambda$ of the mid-gap states:
	\begin{align}
		(1 - \lambda^2 R^2)\, \kappa^2 \omega^2 - \Bigg(& R^2 \sqrt{ -4 c_1^2 c_2^2 \omega^4 + \left( \frac{(-1 + \lambda R)^2}{R^2} + (c_1^2 + c_2^2) \omega^2 \right)^2 } \nonumber  \\
		&\quad \times \sqrt{ -4 c_1^2 c_2^2 \omega^4 + \left( \frac{(1 + \lambda R)^2}{R^2} + (c_1^2 + c_2^2) \omega^2 \right)^2 } \Bigg) = 0. \label{EqSuppMatMidGap}
	\end{align}

	\subsection{Reduction to the two coupling nodes}
	\label{SuppMatTwoNodeReduction}

	The mid-gap relation of Eq.~\eqref{EqSuppMatMidGap} takes a compact form that makes its connection with the PT dimer explicit. We write the circuit Laplacian as $J=J_0+J_1$, with the coupling $J_1$ of Eq.~\eqref{EqSuppMatJ1}, and denote by $G_0(\lambda)=(J_0-\lambda)^{-1}$ the Green's function of the two uncoupled rings. Then
	\begin{equation}
		\det(J-\lambda)=\det(J_0-\lambda)\,\det\!\left(\mathds{I}+G_0(\lambda)J_1\right).
		\label{EqSuppMatDetFactor}
	\end{equation}
	Since $J_1$ couples only the nodes $a$ and $b$, the matrix $\mathds{I}+G_0J_1$ differs from the identity only in the columns $a$ and $b$, and its determinant reduces to a $2\times2$ determinant,
	\begin{equation}
		\det\!\left(\mathds{I}+G_0J_1\right)=1+\omega^2\kappa^2\,G_{+;a,a}(\lambda)\,G_{-;b,b}(\lambda).
		\label{EqSuppMatDet2x2}
	\end{equation}
	Away from the bulk eigenvalues of the uncoupled rings, the eigenvalues of $J$ therefore satisfy $1+\omega^2\kappa^2G_{+;a,a}G_{-;b,b}=0$. The local Green's functions are the one-point impedances of Supplementary Section~\ref{SuppMatOnePointZ} with the onsite admittance shifted by the eigenvalue, $Y_\pm\rightarrow Y_\pm-\lambda$. We write them as $G_{\pm;\alpha,\alpha}=1/\mathcal{Y}(Y_\pm-\lambda)$, which defines the input admittance of a ring
	\begin{equation}
		\mathcal{Y}(y)=\frac{s\sqrt{\left(y^2+D_+\right)^2-4\omega^4c_1^2c_2^2}}{y}=\frac{\sqrt{(y^2+Y_\mathrm{gap}^2)(y^2+Y_\mathrm{band}^2)}}{y},
		\label{EqSuppMatDressedY}
	\end{equation}
	with $D_+=(c_1^2+c_2^2)\omega^2$, the gap edge $Y_\mathrm{gap}=\omega|c_2-c_1|$ and the band edge $Y_\mathrm{band}=\omega(c_1+c_2)$ (Supplementary Section~\ref{SuppMatOnePointZ}). The sign function $s$ of Eq.~\eqref{EqSuppMatSignFunction} selects the branch with $\mathcal{Y}(y)\rightarrow y$ for $|y|\rightarrow\infty$, so that $\mathcal{Y}$ is odd, real for real $y$, and equal to the node admittance $y$ for an isolated node ($c_1=c_2=0$). Inserting $\mathcal{Y}$ into Eq.~\eqref{EqSuppMatDet2x2} gives
	\begin{equation}
		\mathcal{Y}(Y_+-\lambda)\,\mathcal{Y}(Y_--\lambda)+\omega^2\kappa^2=0,
		\label{EqSuppMatDressedDimer}
	\end{equation}
	which is Eq.~\eqref{EqDressedDimer} of the main text. At $\lambda=0$, Eq.~\eqref{EqSuppMatDressedDimer} is the pole condition of the two-point impedance, Eq.~\eqref{EqMethodsFinalZ}. For the dimer, $\mathcal{Y}(y)=y$, and Eq.~\eqref{EqSuppMatDressedDimer} is the characteristic equation of Eq.~\eqref{Eq2NodePTLaplacian}.

	Equation~\eqref{EqSuppMatDressedDimer} is equivalent to the mid-gap relation. At the resonance frequency, $Y_\pm=\pm1/R$, and the two square roots in Eq.~\eqref{EqSuppMatMidGap} are the bulk factors $f_\gamma$ of the lossy and gain rings,
	\begin{equation}
		f_\gamma^2=\left[\frac{(s_\gamma+\lambda R)^2}{R^2}+Y_\mathrm{gap}^2\right]\left[\frac{(s_\gamma+\lambda R)^2}{R^2}+Y_\mathrm{band}^2\right].
		\label{EqSuppMatBulkFactor}
	\end{equation}
	Comparing with Eq.~\eqref{EqSuppMatDressedY}, each bulk factor is the input admittance multiplied by its argument,
	\begin{equation}
		f_\mathrm{l}=\left(\tfrac{1}{R}+\lambda\right)\mathcal{Y}\!\left(\tfrac{1}{R}+\lambda\right),\qquad f_\mathrm{g}=\left(\tfrac{1}{R}-\lambda\right)\mathcal{Y}\!\left(\tfrac{1}{R}-\lambda\right),
		\label{EqSuppMatBulkFactorY}
	\end{equation}
	and Eq.~\eqref{EqSuppMatMidGap} reads
	\begin{equation}
		\left(\frac{1}{R^2}-\lambda^2\right)\left[\omega^2\kappa^2-\mathcal{Y}\!\left(\tfrac{1}{R}+\lambda\right)\mathcal{Y}\!\left(\tfrac{1}{R}-\lambda\right)\right]=0,
		\label{EqSuppMatMidgapDressed}
	\end{equation}
	which, since $\mathcal{Y}$ is odd, is Eq.~\eqref{EqSuppMatDressedDimer} apart from the trivial factor $1/R^2-\lambda^2$. For a finite ring, $G_{\pm;\alpha,\alpha}$ is the finite Bloch sum of Eq.~\eqref{EqSuppMatOnePointZSum}, which differs from the closed form by terms that decay exponentially with $N$ for mid-gap eigenvalues. For rings of $N=16$ nodes, the solutions of Eq.~\eqref{EqSuppMatDressedDimer} agree with the eigenvalues of the full 32-node Laplacian to better than $10^{-4}$. For example, at $\kappa=4.6$\,nF both give the purely imaginary eigenvalues $\pm1.4928$ and $\pm4.0980$\,mS to within $10^{-4}$\,mS.

	This form also gives the effect of a local perturbation exactly. If the onsite admittance of the gain-side coupling node $b$ changes by $\delta$, the Dyson equation of the isolated gain ring gives
	\begin{equation}
		G'_{-;b,b}=\frac{G_{-;b,b}}{1+\delta\,G_{-;b,b}},\qquad\text{so that}\qquad \mathcal{Y}(Y_--\lambda)\rightarrow\mathcal{Y}(Y_--\lambda)+\delta.
		\label{EqSuppMatLocalShift}
	\end{equation}
	The same holds for node $a$ of the lossy ring. A gain fluctuation at a coupling node therefore enters the lattice through the input admittance of its ring in the same way as it enters the node admittance of the dimer.

	\subsection{Couplings of the central and outer exceptional points}
	\label{SuppMatEPCouplings}

	Figure~\ref{FigSuppMatCleanLimitSpectra}a shows that, at each frequency, the mid-gap eigenvalues are purely real at large values of $\kappa$, purely imaginary at intermediate values of $\kappa$ and complex at small values of $\kappa$. Here we derive from Eq.~\eqref{EqSuppMatMidGap} the couplings at which these transitions occur.

	\textbf{Central EP.} The eigenvalues come in pairs $\pm\lambda$, so a pair changes from purely real to purely imaginary where $\lambda^2$ changes sign, that is, at $\lambda=0$. Substituting $\lambda=0$ into Eq.~\eqref{EqSuppMatMidGap} and solving for $\kappa$ gives this critical coupling,
	\begin{equation}
		\kappa_{\mathrm{EP_c}} = \frac{1}{\omega R} \sqrt{1 + 2 (c_1^2 + c_2^2)\omega^2 R^2 + (c_2^2 - c_1^2)^2 \omega^4 R^4}.
		\label{EqSuppMatKappaEPc}
	\end{equation}
	The expression under the square root factorizes with the gap edge $Y_\mathrm{gap}=\omega|c_2-c_1|$ and the band edge $Y_\mathrm{band}=\omega(c_1+c_2)$, which gives
	\begin{equation}
		\kappa_{\mathrm{EP_c}}=\frac{\mathcal{Y}(1/R)}{\omega}=\frac{1}{\omega R}\sqrt{\left(1+Y_\mathrm{gap}^2R^2\right)\left(1+Y_\mathrm{band}^2R^2\right)},
		\label{EqSuppMatKappaEPcY}
	\end{equation}
	Eq.~\eqref{EqMainKappaEPc} of the main text. For our parameters, $\kappa_{\mathrm{EP_c}}=4.727$\,nF.

	\textbf{Away from the resonance frequency.} Here $Y_\pm=Y_0\pm1/R$ with an imaginary $Y_0$. An eigenvalue at the origin requires $\mathcal{Y}(Y_+)\,\mathcal{Y}(Y_-)+\omega^2\kappa^2=0$, and since $\mathcal{Y}$ is odd and $\mathcal{Y}(z^*)=\mathcal{Y}(z)^*$, this condition becomes
	\begin{equation}
		\omega\kappa=\left|\mathcal{Y}\!\left(\tfrac{1}{R}+Y_0(\omega)\right)\right|.
		\label{EqSuppMatKappaOffResonance}
	\end{equation}
	This is the curve of Eq.~\eqref{EqHump} with $\lambda=-Y_0$. It gives, for each drive frequency, the coupling at which an eigenvalue sits at the origin, that is, the ridges of the impedance map in Fig.~\ref{figMain_CleanLimit}c, and it reduces to Eq.~\eqref{EqSuppMatKappaEPcY} at the resonance frequency.

	\textbf{Outer EPs.} The transition between purely imaginary and complex eigenvalues occurs at a turning point of $\lambda^2$ as a function of $\kappa$,
	\begin{equation}
		\frac{\partial\lambda^2}{\partial\kappa}=0.
		\label{EqSuppMatTurningPoint}
	\end{equation}
	Implicit differentiation of Eq.~\eqref{EqSuppMatMidGap} turns this condition into a quartic equation for $\lambda^2$. For our parameters, its solution is $\lambda=\pm2.926i$\,mS at $\kappa_{\mathrm{EP_o}}=4.482$\,nF, which coincides with the two minima of the curve $|\mathcal{Y}(1/R+i\mu)|$ of Supplementary Section~\ref{SuppMatThreeEPs}. These minima give the outer EPs more directly. At the resonance frequency, the spectrum is symmetric under $\lambda\rightarrow-\lambda$ and $\lambda\rightarrow\lambda^*$, so EP$_\mathrm{o1}$ and EP$_\mathrm{o2}$ form at the same coupling, $\kappa\approx4.48$\,nF, as mirror images on the imaginary axis.

	\subsection{Three exceptional points from the input admittance}
	\label{SuppMatThreeEPs}

	\textbf{Lossless and real eigenvalues.} At the resonance frequency, $Y_\pm=\pm1/R$, and Eq.~\eqref{EqSuppMatDressedDimer} reads
	\begin{equation}
		\mathcal{Y}\!\left(\tfrac{1}{R}-\lambda\right)\mathcal{Y}\!\left(-\tfrac{1}{R}-\lambda\right)+\omega^2\kappa^2=0.
		\label{EqSuppMatResonanceCondition}
	\end{equation}
	Since $\mathcal{Y}$ is odd and $\mathcal{Y}(z^*)=\mathcal{Y}(z)^*$, a purely imaginary eigenvalue $\lambda=i\mu$, with $\mu=\mathrm{Im}\,\lambda$, describes a lossless mode and satisfies
	\begin{equation}
		\left|\mathcal{Y}\!\left(\tfrac{1}{R}+i\mu\right)\right|=\omega\kappa,
		\label{EqSuppMatLossless}
	\end{equation}
	which is Eq.~\eqref{EqHump} of the main text. A purely real eigenvalue $\lambda=\nu$ describes a mode with net loss or net gain and satisfies
	\begin{equation}
		\mathcal{Y}\!\left(\tfrac{1}{R}-\nu\right)\mathcal{Y}\!\left(\tfrac{1}{R}+\nu\right)=\omega^2\kappa^2.
		\label{EqSuppMatRealPair}
	\end{equation}
	The eigenvalues come in pairs $\pm\lambda$ and $\pm\lambda^*$. Divided by $\omega$, the left-hand side of Eq.~\eqref{EqSuppMatLossless} is the curve of Fig.~\ref{figMain_CleanLimit}d, and for a given $\kappa$, every crossing of the horizontal line $\omega\kappa$ with this curve gives one lossless eigenvalue.

	\textbf{Exceptional-point condition.} An EP of the interface pair is a double root of
	\begin{equation}
		F(\lambda)=\mathcal{Y}(Y_+-\lambda)\,\mathcal{Y}(Y_--\lambda)+\omega^2\kappa^2,
		\label{EqSuppMatFlambda}
	\end{equation}
	that is, $F=\partial_\lambda F=0$, at which the eigenvectors also coalesce (see below). For $\lambda=i\mu$, $F=\omega^2\kappa^2-|\mathcal{Y}(1/R+i\mu)|^2$ is real, and $\partial_\mu F=0$ requires $|\mathcal{Y}(1/R+i\mu)|$ to be stationary in $\mu$. The EPs on the imaginary axis are therefore the extrema of the curve of Eq.~\eqref{EqSuppMatLossless}, and the coupling of each EP is the value of the curve at the extremum divided by $\omega$. At $\mu=0$, the derivative
	\begin{equation}
		\partial_\lambda F\big|_{\lambda=0}=-\mathcal{Y}'\!\left(\tfrac{1}{R}\right)\mathcal{Y}\!\left(-\tfrac{1}{R}\right)-\mathcal{Y}\!\left(\tfrac{1}{R}\right)\mathcal{Y}'\!\left(-\tfrac{1}{R}\right)=0
		\label{EqSuppMatDoubleRoot}
	\end{equation}
	vanishes identically, because $\mathcal{Y}$ is odd and $\mathcal{Y}'$ is even. The eigenvalue $\lambda=0$ is therefore a double root whenever $\omega\kappa=\mathcal{Y}(1/R)$, which gives Eq.~\eqref{EqMainKappaEPc}.

	\textbf{Shape of the curve.} Whether the centre of the curve is a maximum or a minimum decides how many EPs form. With $u=\ln\mathcal{Y}$, which is real on the real axis, the curve near its centre is
	\begin{equation}
		\ln\left|\mathcal{Y}\!\left(\tfrac{1}{R}+i\mu\right)\right|=\mathrm{Re}\,u\!\left(\tfrac{1}{R}+i\mu\right)=u\!\left(\tfrac{1}{R}\right)-\frac{\mu^2}{2}u''\!\left(\tfrac{1}{R}\right)+\frac{\mu^4}{24}u''''\!\left(\tfrac{1}{R}\right)+\cdots.
		\label{EqSuppMatHumpExpansion}
	\end{equation}
	The curve has a maximum at $\mu=0$ if $u''(1/R)>0$ and a minimum if $u''(1/R)<0$. For large $|\mu|$, $|\mathcal{Y}(1/R+i\mu)|\simeq|\mu|$ grows, so that a maximum at $\mu=0$ is always flanked by two minima at $\pm\mu^*$. We apply this criterion to the three models in turn.

	\textbf{PT dimer.} Here $\mathcal{Y}(y)=y$, so that
	\begin{equation}
		u''\!\left(\tfrac{1}{R}\right)=-R^2<0.
		\label{EqSuppMatDimerCurvature}
	\end{equation}
	The curve has a single minimum at $\mu=0$, which gives a single EP at $\omega\kappa=1/R$. The pair is real below and imaginary above this coupling (Re$\lambda$-to-Im$\lambda$).

	\textbf{Gapless lattice.} Here $Y_\mathrm{gap}=0$ and $\mathcal{Y}(y)=\sqrt{y^2+Y_\mathrm{band}^2}$, so that
	\begin{equation}
		u''(y)=\frac{Y_\mathrm{band}^2-y^2}{\left(y^2+Y_\mathrm{band}^2\right)^2}.
		\label{EqSuppMatGaplessCurvature}
	\end{equation}
	The curve therefore has a minimum at $\mu=0$, and a single EP, when $1/R>Y_\mathrm{band}$. For $c_1=c_2=1.005$\,nF, $Y_\mathrm{band}=2.51$\,mS is smaller than $1/R=3.70$\,mS, and the gapless lattice of Supplementary Section~\ref{SuppMatDisorderSpectra} has a single EP, like the dimer. For weaker gain and loss, $1/R<Y_\mathrm{band}$, which means $R>398\,\Omega$, the centre of the curve lies inside the bulk band, $|\mathrm{Im}\,\lambda|<Y_\mathrm{band}$. This regime lies outside the operating range considered here.

	\textbf{SSH rings.} With a band gap, $Y_\mathrm{gap}\neq0$, the input admittance behaves as
	\begin{equation}
		\mathcal{Y}(y)\simeq\frac{Y_\mathrm{gap}Y_\mathrm{band}}{y}\quad(y\rightarrow0),\qquad \mathcal{Y}(y)\simeq y\quad(y\rightarrow\infty),
		\label{EqSuppMatSSHLimits}
	\end{equation}
	so that $\mathcal{Y}$ has a minimum on the real axis at $y^2=Y_\mathrm{gap}Y_\mathrm{band}$, with $\mathcal{Y}_\mathrm{min}=Y_\mathrm{gap}+Y_\mathrm{band}=2\omega c_2$. To find the sign of $u''$, we write
	\begin{equation}
		u(y)=\frac{1}{2}\ln\left(y^2+Y_\mathrm{gap}^2+Y_\mathrm{band}^2+\frac{Y_\mathrm{gap}^2Y_\mathrm{band}^2}{y^2}\right)
		\label{EqSuppMatSSHlog}
	\end{equation}
	and differentiate twice. This shows that $u''(y)>0$ if and only if $P(y^2)>0$, with
	\begin{equation}
		P(x)=-x^4+(Y_\mathrm{gap}^2+Y_\mathrm{band}^2)\,x^3+8Y_\mathrm{gap}^2Y_\mathrm{band}^2x^2+3(Y_\mathrm{gap}^2+Y_\mathrm{band}^2)Y_\mathrm{gap}^2Y_\mathrm{band}^2x+Y_\mathrm{gap}^4Y_\mathrm{band}^4.
		\label{EqSuppMatHumpPolynomial}
	\end{equation}
	Since $P(0)=Y_\mathrm{gap}^4Y_\mathrm{band}^4>0$, the maximum at $\mu=0$, and with it the three EPs, exists for any band gap when the gain and loss are weak enough. It disappears when $1/R^2$ exceeds the positive root of $P$. For our $c_1$ and $c_2$, this happens at $R=182\,\Omega$, and at $R=270\,\Omega$ it happens when the gap is reduced to $c_2=1.53$\,nF. At this boundary, $u''(1/R)=0$, EP$_\mathrm{c}$ merges with EP$_\mathrm{o1}$ and EP$_\mathrm{o2}$, and the broadening coefficient $\eta_\mathrm{c}$ of Eq.~\eqref{EqEPcShare} diverges. For $R=270\,\Omega$, the maximum gives $\kappa_{\mathrm{EP_c}}=4.727$\,nF, and the minima lie at $\mu^*=\pm2.926$\,mS with $\kappa=4.482$\,nF, in agreement with Fig.~\ref{FigSuppMatCleanLimitSpectra} and Fig.~\ref{figMain_CleanLimit}b,d of the main text. The curves of Fig.~\ref{figMain_CleanLimit}e use this closed form for long rings. For rings of $N=16$ nodes, the full Laplacian gives the same three EPs from $182\,\Omega$ to about $1.3$\,k$\Omega$, a range that contains $R^*$ and our operating point.

	\textbf{Counting the lossless eigenvalues.} For $\kappa$ between the outer EPs and EP$_\mathrm{c}$, the line $\omega\kappa$ crosses the curve at four values of $\mu$, and all four interface eigenvalues are purely imaginary, which gives the four impedance peaks of Fig.~\ref{figMain_CleanLimit}f. For $\kappa>\kappa_{\mathrm{EP_c}}$, two crossings remain, and the pair that met at EP$_\mathrm{c}$ becomes real, because Eq.~\eqref{EqSuppMatRealPair} can now be satisfied. Near $\nu=0$, its left-hand side grows as
	\begin{equation}
		\mathcal{Y}\!\left(\tfrac{1}{R}-\nu\right)\mathcal{Y}\!\left(\tfrac{1}{R}+\nu\right)\simeq\mathcal{Y}^2\!\left(\tfrac{1}{R}\right)\left[1+u''\!\left(\tfrac{1}{R}\right)\nu^2\right]
		\label{EqSuppMatRealGrowth}
	\end{equation}
	when $u''>0$, so that a real pair appears as soon as $\omega\kappa$ exceeds $\mathcal{Y}(1/R)$. For $\kappa$ below the outer EPs, the line does not cross the curve, and all four interface eigenvalues are complex. Since EP$_\mathrm{c}$ sits on a maximum, its transition (Im$\lambda$-to-Re$\lambda$) is the reverse of the transitions at the outer EPs and at the dimer EP.

	\textbf{Coalescence of the eigenvectors.} The eigenvector $\psi$ of $J$ at the eigenvalue $\lambda$ satisfies $\psi=-G_0J_1\psi$, and since $J_1$ acts only on the coupling nodes, its components there, $\phi=(\psi_a,\psi_b)^\mathsf{T}$, satisfy $M(\lambda)\,\phi=0$ with
	\begin{equation}
		M(\lambda)=\begin{pmatrix}\mathcal{Y}(Y_+-\lambda) & -i\omega\kappa\\ -i\omega\kappa & \mathcal{Y}(Y_--\lambda)\end{pmatrix},\qquad \det M=F(\lambda).
		\label{EqSuppMatInterfaceBlock}
	\end{equation}
	All other components follow from those of the coupling nodes,
	\begin{equation}
		\psi_j=i\omega\kappa\,G_{+;j,a}\,\psi_b\quad(\text{lossy ring}),\qquad \psi_j=i\omega\kappa\,G_{-;j,b}\,\psi_a\quad(\text{gain ring}).
		\label{EqSuppMatEigenvectorRest}
	\end{equation}
	At EP$_\mathrm{c}$, where $\lambda=0$, $Y_\pm=\pm1/R$ and $\omega\kappa=\mathcal{Y}(1/R)$, the block becomes
	\begin{equation}
		M=\mathcal{Y}(1/R)\begin{pmatrix}1&-i\\-i&-1\end{pmatrix},\qquad M^2=0.
		\label{EqSuppMatJordan}
	\end{equation}
	It is a nonzero nilpotent matrix, a $2\times2$ Jordan block, with the single null vector $\phi=(1,-i)^\mathsf{T}$, which is self-orthogonal, $\phi^\mathsf{T}\phi=0$. The double root of $F$ is therefore accompanied by a single eigenvector, and EP$_\mathrm{c}$ is a genuine EP of second order rather than only a zero eigenvalue. The eigenvectors of the full 32-node Laplacian confirm this. As $\kappa$ approaches $\kappa_{\mathrm{EP_c}}$, the normalized overlap of the two eigenvectors of the mid-gap pair rises to 0.870, 0.974, 0.993, 0.998 and 0.99995 at $\kappa=4.600$, 4.700, 4.720, 4.725 and 4.7270\,nF, the ratio $|\psi^\mathsf{T}\psi|/\psi^\dagger\psi$ falls to $3\times10^{-3}$, and $\psi_b/\psi_a$ approaches $-i$, as Eq.~\eqref{EqSuppMatJordan} predicts. The outer EPs coalesce in the same way. Just above $\kappa=4.4817$\,nF, the overlap of the two eigenvectors that merge at EP$_\mathrm{o1}$ (or EP$_\mathrm{o2}$) rises to 0.962, 0.9925, 0.9985, 0.99962 and 0.99992 at distances of 0.05, 0.01, 0.002, 0.0005 and 0.0001\,nF from the EP, and $|\psi^\mathsf{T}\psi|/\psi^\dagger\psi$ falls to $5\times10^{-3}$.

	\subsection{Response of the resonant eigenvalue to local gain fluctuations}
	\label{SuppMatGainFluctuations}

	Here we calculate how far the resonant eigenvalue moves when the gain at the coupling node fluctuates. The real part of this shift is the broadening coefficient $\eta$ of Eq.~\eqref{EqGainShare} of the main text. We use only the input admittance $\mathcal{Y}$ of Supplementary Section~\ref{SuppMatTwoNodeReduction} and its logarithm $u=\ln\mathcal{Y}$ of Supplementary Section~\ref{SuppMatThreeEPs}, whose derivative $u'=\mathcal{Y}'/\mathcal{Y}$ is the relative change of $\mathcal{Y}$. The calculation has four steps.

	\textbf{Step 1, shift of the eigenvalue.} By Eq.~\eqref{EqSuppMatLocalShift}, a change $\delta$ of the gain admittance at node $b$ adds $\delta$ to the input admittance of the gain ring. The eigenvalue condition, Eq.~\eqref{EqSuppMatDressedDimer}, then reads
	\begin{equation}
		F(\lambda,\delta)=\mathcal{Y}(Y_+-\lambda)\left[\mathcal{Y}(Y_--\lambda)+\delta\right]+\omega^2\kappa^2=0.
		\label{EqSuppMatFdelta}
	\end{equation}
	When $\delta$ changes, $\lambda$ must change so that $F$ stays zero. Implicit differentiation at $\delta=0$, $d\lambda/d\delta=-(\partial F/\partial\delta)/(\partial F/\partial\lambda)$, gives the shift of the eigenvalue per unit change of the gain,
	\begin{equation}
		p_b=\frac{d\lambda}{d\delta}=\frac{\mathcal{Y}(Y_+-\lambda)}{\mathcal{Y}'(Y_+-\lambda)\,\mathcal{Y}(Y_--\lambda)+\mathcal{Y}(Y_+-\lambda)\,\mathcal{Y}'(Y_--\lambda)}=\frac{1}{\mathcal{Y}(Y_--\lambda)\left[u'(Y_+-\lambda)+u'(Y_--\lambda)\right]},
		\label{EqSuppMatPbGeneral}
	\end{equation}
	where the second form follows by dividing the numerator and the denominator by $\mathcal{Y}(Y_+-\lambda)\,\mathcal{Y}(Y_--\lambda)$. The quantity $p_b$ is the weight $\psi_b^2/\psi^\mathsf{T}\psi$ of the resonant mode on node $b$ (Methods). Its real part changes the net loss and is the broadening coefficient $\eta$, whereas its imaginary part only shifts the resonance frequency.

	\textbf{Step 2, operating point.} We now place one lossless eigenvalue at the origin, $\lambda=0$, by tuning the drive frequency. The detuned drive adds the same imaginary admittance $i\varepsilon$ to every node (Supplementary Section~\ref{SuppMatFreqShift}), so that
	\begin{equation}
		Y_\pm=\pm\frac{1}{R}+i\varepsilon.
		\label{EqSuppMatOperatingY}
	\end{equation}
	The detuning $\varepsilon$ measures how far the resonant mode lies from the gap centre. At the resonance frequency, the same mode has the eigenvalue $\lambda=-i\varepsilon$, so $|\varepsilon|$ is its position along the horizontal axis of Fig.~\ref{figMain_CleanLimit}d, and $\varepsilon=0$ at EP$_\mathrm{c}$. (The small frequency dependence of the coupling admittances is neglected here and included in the numerical checks.) Two properties of $\mathcal{Y}$ simplify Eq.~\eqref{EqSuppMatPbGeneral}. Since $\mathcal{Y}$ is odd and real for real arguments, the terms of the gain ring are mirror images of those of the lossy ring,
	\begin{equation}
		\mathcal{Y}(Y_-)=-\mathcal{Y}(Y_+)^*,\qquad u'(Y_-)=-u'(Y_+)^*,
		\label{EqSuppMatMirror}
	\end{equation}
	and the eigenvalue condition at $\lambda=0$ fixes the magnitude $|\mathcal{Y}(Y_+)|=\omega\kappa$. The bracket in Eq.~\eqref{EqSuppMatPbGeneral} becomes $u'(Y_+)-u'(Y_+)^*=2i\,\mathrm{Im}\,u'(Y_+)$, and
	\begin{equation}
		p_b=\frac{i\,\mathcal{Y}(Y_+)}{2\,\omega^2\kappa^2\,\mathrm{Im}\,u'(Y_+)}.
		\label{EqSuppMatPbOperating}
	\end{equation}
	The denominator contains $\mathrm{Im}\,u'(Y_+)$, which vanishes at every extremum of the curve of Fig.~\ref{figMain_CleanLimit}d, that is, at every EP (Supplementary Section~\ref{SuppMatThreeEPs}). This is why $p_b$ diverges at the EPs. Whether this divergence broadens the line depends on the phase of the numerator, $\mathcal{Y}(Y_+)$, which we examine next.

	\textbf{Step 3, near EP$_\mathrm{c}$.} Near EP$_\mathrm{c}$, $\varepsilon$ is small, and we expand around the real point $1/R$, where $\mathcal{Y}$ and all derivatives of $u$ are real. To the order needed,
	\begin{equation}
		\mathcal{Y}\!\left(\tfrac{1}{R}+i\varepsilon\right)=\mathcal{Y}\exp\left[i\varepsilon u'-\frac{\varepsilon^2}{2}u''+O(\varepsilon^3)\right],\qquad \mathrm{Im}\,u'\!\left(\tfrac{1}{R}+i\varepsilon\right)=\varepsilon u''+O(\varepsilon^3),
		\label{EqSuppMatCentralExpansion}
	\end{equation}
	where $\mathcal{Y}$, $u'$ and $u''$ without an argument are taken at $1/R$. Inserting these into Eq.~\eqref{EqSuppMatPbOperating}, with $\omega^2\kappa^2=|\mathcal{Y}(1/R+i\varepsilon)|^2$, gives
	\begin{equation}
		p_b=\frac{i}{2\,\mathcal{Y}u''\,\varepsilon}-\frac{u'}{2\,\mathcal{Y}u''}+O(\varepsilon).
		\label{EqSuppMatPbExpansion}
	\end{equation}
	The two terms have a clear meaning. The first term diverges as $1/\varepsilon$ at EP$_\mathrm{c}$, but it is purely imaginary, so the strong EP response only shifts the resonance frequency. The second term is real and finite, and it is the broadening coefficient at EP$_\mathrm{c}$,
	\begin{equation}
		\eta_\mathrm{c}=-\frac{u'}{2\,\mathcal{Y}u''}.
		\label{EqSuppMatEtac}
	\end{equation}
	To connect this result with Eq.~\eqref{EqEPcShare} of the main text, we express $u'$ and $u''$ through the two curves of Fig.~\ref{figMain_CleanLimit}. Since $\mathcal{Y}(1/R)=\omega\kappa_{\mathrm{EP_c}}$ (Eq.~\eqref{EqSuppMatKappaEPcY}), the slope of the EP$_\mathrm{c}$ curve in Fig.~\ref{figMain_CleanLimit}e gives $u'=\kappa_{\mathrm{EP_c}}^{-1}\,d\kappa_{\mathrm{EP_c}}/d(1/R)$, and Eq.~\eqref{EqSuppMatHumpExpansion} gives the curvature of the curve in Fig.~\ref{figMain_CleanLimit}d at its maximum, $d^2\kappa/d\mu^2=-\kappa_{\mathrm{EP_c}}u''$. Inserting both into Eq.~\eqref{EqSuppMatEtac} gives
	\begin{equation}
		\eta_\mathrm{c}=\frac{d\kappa_{\mathrm{EP_c}}/d(1/R)}{2\,\omega\,\kappa_{\mathrm{EP_c}}\,d^2\kappa/d\mu^2},
		\label{EqSuppMatEtacCurves}
	\end{equation}
	which is Eq.~\eqref{EqEPcShare}. For the dimer, $\mathcal{Y}=y$, $u'=1/y$ and $u''=-1/y^2$, and Eq.~\eqref{EqSuppMatPbOperating} gives, without approximation,
	\begin{equation}
		p_b=\frac{1}{2}-\frac{i}{2\varepsilon R},
		\label{EqSuppMatPbDimer}
	\end{equation}
	in agreement with Eq.~\eqref{EqDimerShare}. Its real part is exactly $1/2$ at every $\varepsilon$. Table~\ref{TabSuppMatEta} compares $\eta_\mathrm{c}$ with the weight computed from the eigenvectors of the full 32-node Laplacian near EP$_\mathrm{c}$. At the operating point of the disorder simulations ($\kappa=4.688$\,nF, $\varepsilon R=0.19$), the full Laplacian gives $p_b=-0.278+4.19i$, and Eq.~\eqref{EqSuppMatPbOperating} gives $-0.275+4.38i$.

	\begin{table}[h]
		\centering
		\caption{\textbf{Calculated broadening coefficient at EP$_\mathrm{c}$, the part of a gain fluctuation at the coupling node that reaches the linewidth.} The broadening coefficient vanishes at $R^*$, and the closed form agrees with the full Laplacian. The values are obtained from the closed form, Eq.~\eqref{EqSuppMatPbExpansion}, and from the full 32-node Laplacian on the EP$_\mathrm{c}$ branch at $\varepsilon=0.1/R$, for $c_1=1.005$\,nF and $c_2=2.212$\,nF. The dimer value holds at any $\kappa$, and the gapless lattice has $c_1=c_2=1.005$\,nF.}
		\label{TabSuppMatEta}
		\begin{tabular}{lcccccc|cc}
			\hline\hline
			$R$ ($\Omega$) & 230 & 270 & 330 & 407.2 ($R^*$) & 500 & 700 & dimer & gapless (270\,$\Omega$)\\
			\hline
			$\eta_\mathrm{c}$, Eq.~\eqref{EqSuppMatPbExpansion} & $-0.593$ & $-0.238$ & $-0.073$ & $0.000$ & $+0.031$ & $+0.043$ & $0.5$ & $1.11$\\
			$\eta$, full Laplacian & $-0.623$ & $-0.248$ & $-0.076$ & $-0.001$ & $+0.030$ & $+0.043$ & $0.500$ & $1.10$\\
			\hline\hline
		\end{tabular}
	\end{table}

	\textbf{Step 4, near the outer EPs.} The outer EPs lie at the minima $\mu=\pm\mu^*$ of the curve, inside the frequency range of the bulk bands. We repeat Step 3 with $Y_+=1/R+i(\mu^*+\varepsilon)$, where $\varepsilon$ now measures the distance from the outer EP. At the minimum, $\mathrm{Im}\,u'(1/R+i\mu^*)=0$, which is the extremum condition of Eq.~\eqref{EqSuppMatHumpExpansion}, so that
	\begin{equation}
		\mathrm{Im}\,u'(Y_+)\simeq\varepsilon\,\mathrm{Re}\,u''\!\left(\tfrac{1}{R}+i\mu^*\right).
		\label{EqSuppMatOuterExpansion}
	\end{equation}
	Unlike at EP$_\mathrm{c}$, $\mathcal{Y}$ is complex here, because the bulk bands no longer lie symmetrically above and below the mode in frequency. Writing its value at the outer EP as $\mathcal{Y}(1/R+i\mu^*)=|\mathcal{Y}(1/R+i\mu^*)|\,e^{i\varphi}$, Eq.~\eqref{EqSuppMatPbOperating} gives
	\begin{equation}
		p_b\simeq\frac{i\,e^{i\varphi}}{2\left|\mathcal{Y}\!\left(\tfrac{1}{R}+i\mu^*\right)\right|\mathrm{Re}\,u''\!\left(\tfrac{1}{R}+i\mu^*\right)\varepsilon},\qquad \mathrm{Re}\,p_b\simeq-\frac{\sin\varphi}{2\left|\mathcal{Y}\!\left(\tfrac{1}{R}+i\mu^*\right)\right|\mathrm{Re}\,u''\!\left(\tfrac{1}{R}+i\mu^*\right)\varepsilon}.
		\label{EqSuppMatPbOuter}
	\end{equation}
	The phase $\varphi$ rotates the divergent response away from the imaginary axis. For our circuit, $\varphi\approx20^\circ$, so the real part of the divergent term is $-\tan\varphi$ times its imaginary part, and a part $\tan20^\circ\approx0.36$ of the divergent frequency response reaches the linewidth. The real part of $p_b$, and with it the response of the linewidth to gain fluctuations, therefore diverges as $1/\varepsilon$ at the outer EPs. At the resonance frequency, the full Laplacian gives $\mathrm{Re}\,p_b=1.00$, 1.54 and 2.11 at $\kappa=4.60$, 4.52 and 4.50\,nF on the outer branch towards an outer EP, whereas $\mathrm{Re}\,p_b=-0.45$, $-0.29$ and $-0.27$ at $\kappa=4.60$, 4.68 and 4.70\,nF on the branch towards EP$_\mathrm{c}$. This explains why the outer EPs broaden like or more than the dimer EP in the disorder simulations.

	\textbf{Gain-insensitive point.} By Eq.~\eqref{EqSuppMatEtac}, $\eta_\mathrm{c}$ vanishes where $u'=\mathcal{Y}'/\mathcal{Y}=0$, that is, at a minimum of $\mathcal{Y}$ on the real axis. For the SSH rings,
	\begin{equation}
		\mathcal{Y}^2=y^2+Y_\mathrm{gap}^2+Y_\mathrm{band}^2+\frac{Y_\mathrm{gap}^2Y_\mathrm{band}^2}{y^2}
		\label{EqSuppMatYsquared}
	\end{equation}
	is stationary at $y^2=Y_\mathrm{gap}Y_\mathrm{band}$, where $\mathcal{Y}=Y_\mathrm{gap}+Y_\mathrm{band}=2\omega c_2$. This gives
	\begin{equation}
		\frac{1}{R^*}=\sqrt{Y_\mathrm{gap}Y_\mathrm{band}}=\omega\sqrt{c_2^2-c_1^2},\qquad \kappa_{\mathrm{EP_c}}(R^*)=2c_2,
		\label{EqSuppMatRstar}
	\end{equation}
	as stated in the main text, that is, $R^*=407.2\,\Omega$ and $\kappa_{\mathrm{EP_c}}(R^*)=4.424$\,nF for our circuit. Since $\omega\,d\kappa_{\mathrm{EP_c}}/d(1/R)=\mathcal{Y}'(1/R)$, the zero of $\eta_\mathrm{c}$ coincides with the minimum of $\kappa_{\mathrm{EP_c}}$ as a function of the gain--loss strength, the flat bottom of the EP$_\mathrm{c}$ curve in Fig.~\ref{figMain_CleanLimit}e, and since $u''>0$ at a minimum of $\mathcal{Y}$, it always lies inside the three-EP regime. This zero holds to first order in the fluctuation, and at $R^*$ the second-order terms set the remaining broadening. The band gap decides where the zero lies. A wider gap moves $R^*$ towards stronger gain and loss, and at fixed $R=270\,\Omega$ it lowers $|\eta_\mathrm{c}|$. Equation~\eqref{EqSuppMatEtac} gives $\eta_\mathrm{c}=-0.46$, $-0.24$, $-0.08$, $0$ and $+0.04$ for $c_2=2.0$, 2.212, 2.6, 3.137 and 4.0\,nF (full Laplacian, $-0.48$, $-0.25$, $-0.08$, $-0.002$ and $+0.04$), and the three-EP regime ends below $c_2=1.53$\,nF, where $\eta_\mathrm{c}$ diverges. For a gapless lattice, $d\mathcal{Y}/dy=y/\sqrt{y^2+Y_\mathrm{band}^2}>0$, and $\eta_\mathrm{c}$ has no zero.

	\textbf{Scope.} The weights of the resonant mode on all nodes $j$ and on the gain nodes alone satisfy
	\begin{equation}
		\sum_jp_j=1,\qquad \sum_{j\in\mathrm{gain}}\mathrm{Re}\,p_j=\frac{1}{2}
		\label{EqSuppMatSumRule}
	\end{equation}
	for the PT-symmetric models at their operating points (the second to $10^{-13}$ numerically). A uniform change of all gain elements therefore shifts $\mathrm{Re}\,\lambda$ by half its value in every model, and the suppression discussed here concerns local fluctuations at the coupling interface, where the measured disorder is concentrated (Supplementary Section~\ref{SuppMatDisorderProfile}). The weight is not confined to node $b$. At the operating point of the disorder simulations, the resonant mode has $\mathrm{Re}\,p_j=-0.28$ on node 17 but $+0.80$ on node 32, its neighbour across $c_2$, and $+0.05$ on node 18. Independent fluctuations of these neighbours therefore also reach the linewidth (Supplementary Section~\ref{SuppMatDisorderSpectra}). A reactive change $i\delta_m$ of the admittance of node $b$ gives $\mathrm{Re}\,\delta\lambda=-\mathrm{Im}(p_b)\,\delta_m$, which grows as $1/\varepsilon$ near EP$_\mathrm{c}$ as it does near the dimer EP. The resonance is therefore operated on the EP$_\mathrm{c}$ branch at a finite distance from the EP, and the static reactive mismatch of the nodes is reduced by fine tuning (Supplementary Section~\ref{SuppMatNodeTuning}).

	\section*{Supplementary references}
	\begin{list}{}{\setlength{\leftmargin}{2.5em}\setlength{\labelwidth}{2em}\setlength{\itemsep}{2pt}}
		\item[{[S1]}] Hofmann, T., Helbig, T., Lee, C. H., Greiter, M. \& Thomale, R. Chiral voltage propagation and calibration in a topolectrical Chern circuit. \textit{Phys. Rev. Lett.} \textbf{122}, 247702 (2019).
		\item[{[S2]}] Helbig, T. \textit{et al.} Generalized bulk--boundary correspondence in non-Hermitian topolectrical circuits. \textit{Nat. Phys.} \textbf{16}, 747--750 (2020).
		\item[{[S3]}] Gregory, A. P. \textit{Q-factor measurement by using a Vector Network Analyser} DOI: 10.47120/npl.MAT58 (2021).
		\item[{[S4]}] Asb\'oth, J. K., Oroszl\'any, L. \& P\'alyi, A. \textit{A Short Course on Topological Insulators}, Lecture Notes in Physics Vol.~919 (Springer, 2016).
		\item[{[S5]}] Wang, H., Lai, Y.-H., Yuan, Z., Suh, M.-G. \& Vahala, K. Petermann-factor sensitivity limit near an exceptional point in a Brillouin ring laser gyroscope. \textit{Nat. Commun.} \textbf{11}, 1610 (2020).
		\item[{[S6]}] Lau, H.-K. \& Clerk, A. A. Fundamental limits and non-reciprocal approaches in non-Hermitian quantum sensing. \textit{Nat. Commun.} \textbf{9}, 4320 (2018).
		\item[{[S7]}] Economou, E. N. \textit{Green's Functions in Quantum Physics} 3rd edn (Springer, 2006).
		\item[{[S8]}] Cserti, J., D\'avid, G. \& Pir\'oth, A. Perturbation of infinite networks of resistors. \textit{Am. J. Phys.} \textbf{70}, 153--159 (2002).
		\item[{[S9]}] Yao, S. \& Wang, Z. Edge states and topological invariants of non-Hermitian systems. \textit{Phys. Rev. Lett.} \textbf{121}, 086803 (2018).
		\item[{[S10]}] Yokomizo, K. \& Murakami, S. Non-Bloch band theory of non-Hermitian systems. \textit{Phys. Rev. Lett.} \textbf{123}, 066404 (2019).
		\item[{[S11]}] Teo, J. C. Y. \& Kane, C. L. Topological defects and gapless modes in insulators and superconductors. \textit{Phys. Rev. B} \textbf{82}, 115120 (2010).
		\item[{[S12]}] Smirnova, D., Leykam, D., Chong, Y. \& Kivshar, Y. Nonlinear topological photonics. \textit{Appl. Phys. Rev.} \textbf{7}, 021306 (2020).
		\item[{[S13]}] Sahin, H. \textit{et al.} Protected chaos in a topological lattice. \textit{Adv. Sci.} \textbf{12}, e03216 (2025).
	\end{list}

\end{document}